\documentclass[12pt]{article}
\usepackage{amssymb}
\usepackage{pstricks}
\usepackage{color}
\usepackage{pst-node}
\usepackage{graphics}
\usepackage{epic}
\usepackage{curves}
\catcode`\@=11
\def\marginnote#1{}

\newcount\hour
\newcount\minute
\newtoks\amorpm
\hour=\time\divide\hour by60 \minute=\time{\multiply\hour by60
\global\advance\minute by-\hour}
\edef\standardtime{{\ifnum\hour<12 \global\amorpm={am}%
        \else\global\amorpm={pm}\advance\hour by-12 \fi
        \ifnum\hour=0 \hour=12 \fi
        \number\hour:\ifnum\minute<10 0\fi\number\minute\the\amorpm}}
\edef\militarytime{\number\hour:\ifnum\minute<10 0\fi\number\minute}

\def\draftlabel#1{{\@bsphack\if@filesw {\let\thepage\relax
      \xdef\@gtempa{\write\@auxout{\string
          \newlabel{#1}{{\@currentlabel}{\thepage}}}}}\@gtempa \if@nobreak
    \ifvmode\nobreak\fi\fi\fi\@esphack} \gdef\@eqnlabel{#1}}
    \def\@eqnlabel{}
\def\@vacuum{}
\def\draftmarginnote#1{\marginpar{\raggedright\scriptsize\tt#1}}

\def\draft{
  \oddsidemargin -.5truein
  \def\@oddfoot{\footnotesize \sl preliminary draft \hfil
    \rm\thepage\hfil\sl\today\quad\militarytime}
  \let\@evenfoot\@oddfoot \overfullrule 3pt
    \let\label=\draftlabel
    \let\marginnote=\draftmarginnote
  \def\@eqnnum{(\theequation)\rlap{\kern\marginparsep\tt\@eqnlabel}%
    \global\let\@eqnlabel\@vacuum}

  }

\draft

\newtheorem{theorem}{Theorem}
\newtheorem{proposition}[theorem]{Proposition}
\newtheorem{lemma}[theorem]{Lemma}
\newtheorem{cor}[theorem]{Corollary}

\newtheorem{defin}{Definition}

\newtheorem{example}{Example}

\newtheorem{definition}{Definition}
\let\Bbb=\mathbb
\let\wtd=\widetilde
\def\XX{{\hbox{\scriptsize${{\hbox{\tiny$\times$}}\atop{
\hbox{\tiny$\times$}}}$}}}
\newcommand{\ORD}[1]{\XX{#1}\XX}

\renewcommand{\theequation}{\thesection.\arabic{equation}}

\newcommand{\newsection}{
\setcounter{equation}{0} \setcounter{theorem}{0} \setcounter{con}{0}
\setcounter{defin}{0} \setcounter{remark}{0} \setcounter{example}{0}
\section}
\def\appendix#1{
\addtocounter{section}{1} \setcounter{equation}{0}
\renewcommand{\thesection}{\Alph{section}}
\section*{Appendix \thesection\protect\indent
#1}
}
\newcommand{\tr}{\,{\rm tr}\,}

\def\e{e}

\def\be{\begin{equation}}
\def\ee{\end{equation}}
\def\bea{\begin{eqnarray}}
\def\eea{\end{eqnarray}}

\def\HH{{\Bbb H}}
\def\RR{{\Bbb R}}
\def\ZZ{{\Bbb Z}}

\def\mod{\mathop{\mbox{mod}}\nolimits}

\begin{document}
\setlength{\unitlength}{1.5mm}


\begin{center}
\hfill ITEP/TH-??/09\\
\end{center}


\begin{center}
{\Large Orbifold Riemann surfaces and geodesic algebras}\\
\vspace{8mm}
L.~O.~Chekhov\footnote{E-mail:
chekhov@mi.ras.ru.}
\\
\vspace{18pt}
{\small{\it Steklov Mathematical
Institute, Moscow, Russia,}}
\\
{\small{\it Institute for Theoretical and Experimental Physics,
Moscow, Russia}},
\\
{\small{\it Poncelet Laboratoire International Franco--Russe,
Moscow, Russia}}.
\end{center}
\vskip 0.9 cm
{\flushright {\em To the memory of Alesha Zamolodchikov}\\}
\vskip 0.9 cm
\begin{abstract}
We study the Teichm\"uller theory of Riemann surfaces
with orbifold points of order two using the fat graph technique.
The previously developed technique of
quantization, classical and quantum mapping-class group transformations,
and Poisson and quantum algebras of geodesic
functions is applicable to the surfaces with orbifold points.
We describe classical and quantum braid group relations for particular
sets of geodesic functions corresponding to $A_n$ and $D_n$ algebras and
describe their central elements for the Poisson and quantum algebras.
\end{abstract}

\renewcommand{\thefootnote}{\arabic{footnote}}
\setcounter{footnote}{0}

\newsection{Introduction}

Algebraic structures that arise in studies of Teichm\"uller spaces are an
interesting object deserving a deeper investigation and understanding.
Particular cases of these algebras happen to be related to algebras of
monodromies in Fuchsian systems \cite{DM,Ugaglia} and to algebras of
groupoid of upper triangular matrices \cite{Bondal}. In this paper, we review
these cases and present another case of these (closed) Poisson algebras \cite{Ch1}.
We also present a new result obtained in collaboration with M.~Mazzocco (see~\cite{ChM})
concerning constructing central elements for this new algebra.

Teichm\"uller spaces of hyperbolic structures on Riemann surfaces admit a fruitful
graph (combinatorial) descriptions~\cite{Penn1,Fock1}. These structures
proved to be especially useful when describing sets of geodesic
functions and the related Poisson and quantum structures \cite{ChF}.
These Riemann surfaces necessarily contain holes. A generalization of this
construction to bordered Riemann surfaces \cite{KP,FST,FG,Ch1}, or to the
Riemann surfaces with ${\ZZ_2}$-orbifold points \cite{Ch2} was developed.
First, Kaufmann and Penner~\cite{KP} have demonstrated the relation between
the Thurston theory of measured foliations and a combinatorial description of open/closed
string diagrammatic. The original description in terms of the
Teichm\"uller space coordinates was proposed in \cite{FG} for the
{\em ciliated} Riemann surfaces (a {\em cilium} was there a
marked point on the boundary). The algebraic structures behind this geometry are
cluster algebras (originated in \cite{FZ} and applied to bordered surfaces in
\cite{FST}). Then, the Teichm\"uller theory of bordered surfaces
in the shear-coordinate pattern was reconstructed and the
corresponding algebras of geodesic functions were investigated in \cite{Ch1}.

In~\cite{Ch2}, we gave a detailed geometrical accounting for the orbifold Riemann
surfaces in the graph description; in the present paper, we briefly recall this
description, but our main goal is to describe algebras of geodesic functions,
the mapping class group transformations, and the corresponding braid group
transformations on the level of these algebras (classical and quantum)
and to obtain central elements of these algebras in the special cases of
algebras of $A_n$ and $D_n$ types.

Riemann surfaces obtained as quotients of the hyperbolic upper half-plane under the
action of a Fuchsian subgroup of $PSL(2,{\mathbb R})$ with elliptic elements $e_i$ of a
fixed order $m_i$ \ $(e_i^{m_i}=1)$ (a crystallographic subgroup) has been studied for long (see,
e.g., \cite{EM} and the references therein). A class of Fuchsian groups generated by
half-turns (with all $m_i=2$) was of a special importance, and a difficult
problems was to characterize moduli spaces of such surfaces. It was shown in~\cite{Ch2}
that the fat graph description provides such a characterization.

In Section~\ref{s:hyp}, we first recall the structure of geodesic functions in the fat graph
technique for the Riemann surfaces with holes but without orbifold points.
A fat graph is a nondirected graph with marking on the edges and with a
prescribed cyclic order of edges entering each vertex. We call
a fat graph a {\em spine} of the corresponding
Riemann surface with holes if we can draw this graph without self intersections on this surface and
cutting along all the edges of the graph decomposes the surface into a disjoint set of faces:
each face contains exactly one hole and becomes simply connected polygon upon gluing this hole.

We then generalize this structures
to the case of ${\mathbb Z}_2$-orbifold points (half-turns) using that (see \cite{Ch2})
the space of all regular (metrizable) Riemann surfaces with $|\delta|$ ${\mathbb Z}_2$-orbifold points
is covered by the Teichm\"uller space of fat
graphs with real parameters $Z_\alpha$ on edges and with one- and three-valent vertices; the number
of one-valent vertices (endpoints of ``pending edges'') is $|\delta|$ whereas the total number of
edges is $6g-6+3s+2|\delta|$ and it coincides with the dimension of the corresponding Teichm\"uller
space. Our main examples are
the genus zero Riemann surface with $n$ ${\mathbb Z}_2$-orbifold points
and {\em one} hole, which corresponds to the case of $A_n$ algebra, and the genus zero
Riemann surface with $n$ ${\mathbb Z}_2$-orbifold points
and {\em two} holes (the annulus), which corresponds to the case of $D_n$ algebra \cite{Ch1}.

In Section~\ref{s:mcg} we recall the
Poisson brackets for coordinates of the Teichm\"uller space \cite{FG},
\cite{Ch1}. We then consider all possible mapping class group transformations
generated by flips of edges (internal and pending) of the spine graph
and by changing the directions of spiraling to the hole perimeters for lines
of an ideal triangle decomposition of
the Riemann surface. We prove the invariance of both the set of the
geodesic functions and their Poisson relations (the Goldman bracket \cite{Gold})
under all these transformations thus proving that {\em any} choice of the spine
with {\em arbitrary} marking of edges provides the same geodesic algebra.

In Section~\ref{s:braid}, we first describe the Poisson algebras of geodesic
functions in the $A_n$ and $D_n$ cases, then
present the braid-group transformations that generate
the whole group of modular transformations in the $A_n$ and $D_n$ cases
and leaves these Poisson algebras invariant and, third, describe the
central elements of these two algebras ($A_n$ and $D_n$). These are the
main results of this paper.

We quantize in Section~\ref{s:q}. We briefly recall the
quantization procedure from \cite{ChF} coming then to the quantum
geodesic operators and to the corresponding quantum
geodesic algebras of $A_n$ and $D_n$ type. We also describe the
quantum counterparts of the braid group transformations and the
corresponding representations for quantum geodesic functions in
the $A_n$ and $D_n$ cases.

\newsection{Graph description and hyperbolic geometry\label{s:hyp}}

\subsection{Hyperbolic geometry and inversions\label{ss:graph}}

\subsubsection{Graph description for Riemann surfaces with holes}

Recall the fat graph combinatorial description of the Fuchsian group $\Delta_{g,s}$, which is a
discrete finitely generated subgroup of $PSL(2,\RR)$.
In the case without orbifold points, we assume
a fat graph to be a three-valent graph (each vertex is incident to three terminal points of
nonoriented edges; two terminal points of the same edge can be incident to the same vertex)
with edges labeled by distinct integers $\alpha=1,2,\dots,6g-6+3s$
and with the prescribed cyclic ordering of edges entering each vertex.
A natural way to represent such a graph is to draw it without self-intersections on an orientable
Riemann surface with holes.

\begin{definition}\label{def-spine}
{\rm
We call a three-valent fat graph $\Gamma_{g,s}$ a {\em spine} of the Riemann surface $\Sigma_{g,s}$
of genus $g$ with $s$ holes ($s\ge1$, $2g-2+s>0$) if the fat graph $\Gamma_{g,s}$ can be embed without
self-intersections into this Riemann surface in such a way that the complement of a graph is
a disjoint set of {\em faces}, each face being a polygon with exactly one
hole inside and gluing this hole makes the face simply-connected.
}
\end{definition}

We consider a
spine $\Gamma_{g,s}$ corresponding to the Riemann surface
$\Sigma_{g,s}$ with $g$ handles and $s$ boundary components (holes).
The first homotopy groups $\pi_1(\Sigma_{g,s})$ and $\pi_1(\Gamma_{g,s})$ coincide because
each closed path in $\Sigma_{g,s}$ can be homotopically transformed to a closed path in $\Gamma_{g,s}$
in a unique way. The standard statement in hyperbolic geometry is that conjugate classes of elements of
a Fuchsian group $\Delta_{g,s}$ are in the 1-1 correspondence with homotopy
classes of closed paths in the Riemann surface $\Sigma_{g,s}=\HH^2_+/\Delta_{g,s}$ and that the
``actual'' length $\ell_\gamma$
of a hyperbolic element $\gamma\in\Delta_{g,s}$ coincides with the minimum length of
curves from the corresponding homotopy class: it is then the length of a unique closed
geodesic line belonging to this class.

The standard set of generators of the first homotopy group $\pi_1(\Sigma_{g,s})$ comprises
$2g$ elements
$A_i$, $B_i$, $i=1,\dots,g$, corresponding to going around cycles $a_i$, $b_i$ in the Riemann surface
(with the standard intersection conditions $a_i\circ b_j=\delta_{ij}$, $a_i\circ a_j=b_i\circ b_j=0$)
and $s$ generators $P_j$, $j=1,\dots, s$, corresponding to going around holes (in one and the same
direction w.r.t. the orientation of the Riemann surface) with a single restriction that
$$
A_1B_1A_1^{-1}B_1^{-1}A_2B_2A_2^{-1}B_2^{-1}\cdots A_gB_gA_g^{-1}B_g^{-1}P_1\cdots P_s=I,
$$
from which the total number of parameters is $6g+3s-6$ (three parameters fixed by the constraint and
three extra fixed by the general conjugation freedom).

The combinatorial description of moduli spaces uses the above 1-1 correspondence
between conjugacy classes of the Fuchsian group and closed paths in the spine.
We set the real number $Z_\alpha$ into the correspondence to the edge with the label $\alpha$ and
insert~\cite{Fock1} the matrix of the M\"obius transformation
\be
\label{XZ} X_{Z_\alpha}=\left(
\begin{array}{cc} 0 & -\e^{Z_\alpha/2}\\
                \e^{-Z_\alpha/2} & 0\end{array}\right)
\ee
each time the path homeomorphic to a geodesic $\gamma$ passes through the $\alpha$th edge.

We also introduce the ``right'' and ``left'' turn matrices
to be set in the proper place when a path makes the corresponding turn,
\be
\label{R}
R=\left(\begin{array}{cc} 1 & 1\\ -1 & 0\end{array}\right), \qquad
L= R^2=\left(\begin{array}{cc} 0 & 1\\ -1 &
-1\end{array}\right),
\ee
and define the related operators $R_Z$ and $L_Z$,
\bea
\label{Rz}
R_Z\equiv RX_Z&=&\left(\begin{array}{cc}
                \e^{-Z/2}&-\e^{Z/2}\\
                     0   &\e^{Z/2}
                     \end{array}\right),\\
\label{Lz}
L_Z\equiv LX_Z&=&\left(\begin{array}{cc}
                \e^{-Z/2}&   0\\
                 -\e^{-Z/2}&\e^{Z/2}
                     \end{array}\right).
\eea

An element of a Fuchsian group has then the structure
$$
P_{\gamma}=LX_{Z_n}RX_{Z_{n-1}}\cdots RX_{Z_2}RX_{Z_1},
$$
and the corresponding {\em geodesic function}
\be
\label{G}
G_{\gamma}\equiv \tr P_\gamma=2\cosh(\ell_\gamma/2)
\ee
is expressed via the actual length $\ell_\gamma$ of the closed
geodesic on the Riemann surface.

The total number of parameters here equals the number of edges of $\Gamma_{g,s}$, which, by the
Euler formula, is exactly the desired number $6g-6+3s$.

\subsubsection{Generalization to surfaces with ${\mathbb Z}_2$ orbifold points}

New generators of the Fuchsian group are
rotations through the angle $\pi$ at a finite set $\delta$ of points on the Riemann
surface.\footnote{We let $|\,\cdot\,|$ denote the cardinality of a set.} \
All these generators $F_i$, $i=1,\dots,|\delta|$, are conjugates of the same matrix
\be
\label{F}
F_i=U_iFU_i^{-1},\qquad F=\left(\begin{array}{cc} 0 & 1\\ -1 & 0\end{array}\right).
\ee
Adding this set of generators to the standard one (translations along $A$- and $B$-cycles
and around holes) not necessarily result in a regular (metrizable) surface because the action of the
resulting group is not necessarily discrete. The necessary and
sufficient conditions for this generators to result in a {\em regular} surface can be
formulated in terms of graphs \cite{Ch2}.\footnote{In what follows, we call a Riemann surface
regular if it is locally a smooth
constant-curvature surface everywhere except exactly $|\delta|$ ${\mathbb Z}_2$-orbifold points.}

We now extend the notion of the fat graph to the case of orbifold Riemann surfaces. First,
we label all the orbifold points by distinct integers $\beta=1,\dots,n$. Second, we arbitrarily
split the set of orbifold points $\delta$ into $s$ nonintersecting (may be empty) subsets $\delta_k$,
$|\delta_k|\ge0$, \ $\cup_{k=1}^s\delta_k=\delta$,
and we then associate (in the set-theoretical sense)
the $k$th hole to the subset $\delta_k$.
We also introduce the cyclic ordering in every subset $\delta_k$.

\begin{definition}\label{def-pend}
{\rm
We call a fat graph $\Gamma_{g,s,|\delta|}$ a spine of the Riemann surface $\Sigma_{g,s,\delta}$
with $g$ handles, $s$ holes, and $|\delta|$ ${\mathbb Z}_2$-orbifold points if
\begin{itemize}
\item[(a)] this graph can be embed without self-intersections in $\Sigma_{g,s,\delta}$,
\item[(b)] all vertices of $\Gamma_{g,s,|\delta|}$ are three-valent except exactly $|\delta|$
one-valent vertices (endpoints of ``pending'' edges), which are placed at the corresponding
orbifold points,
\item[(c)] upon cutting along all edges of $\Gamma_{g,s,|\delta|}$ the Riemann surface
$\Sigma_{g,s,\delta}$ splits into $s$ polygons each containing exactly one hole and being
simply connected upon gluing this hole.
\end{itemize}
}
\end{definition}

We therefore
set into correspondence to each orbifold point the special one-valence vertex (the end of a
{\em pending edge}) indicated by a dot in figures below.

\begin{figure}[tb]
\setlength{\unitlength}{0.7mm}%
\begin{picture}(80,40)(-100,-35)
\thicklines
\put(-40,0){\line(1,0){80}}
\put(-40,-16){\line(1,0){32}}
\put(8,-16){\line(1,0){32}}
\put(-8,-16){\line(0,-1){16}}
\put(8,-16){\line(0,-1){16}}
\put(0,-30){\circle*{2}}
\qbezier(-2,-30)(-4,-9)(-12,-9)
\qbezier(2,-30)(4,-9)(12,-9)
\put(-12,-9){\line(-1,0){16}}
\put(12,-9){\line(1,0){16}}
\put(-32,-9.1){\makebox(0,0){$\cdots$}}
\put(32,-9.1){\makebox(0,0){$\cdots$}}
\put(-28,-7){\line(1,0){56}}
\put(-32,-7.1){\makebox(0,0){$\cdots$}}
\put(32,-7.1){\makebox(0,0){$\cdots$}}
\put(4,-32){\makebox(0,0){$F$}}
\put(-11,-30){\makebox(0,0){$Z$}}
\put(-30,4){\makebox(0,0){$Y_1$}}
\put(30,4){\makebox(0,0){$Y_2$}}
\color[rgb]{0,0,1}
\qbezier(-2,-30)(-2,-32)(0,-32)
\qbezier(2,-30)(2,-32)(0,-32)
\end{picture}
\caption{\small Part of the graph with the pending edge.
Its endpoint with the orbifold point is directed to the interior
of the boundary component this point is associated with.
The variable $Z$ corresponds to the respective pending edge. Two
types of geodesic lines are shown in the figure: one that does not
come to the edge $Z$ is parameterized in the standard way, the
other undergoes the inversion with the matrix $F$ (\ref{F}).
The corresponding geodesic line then goes around the dot vertex representing the
orbifold point.}
\label{fi:corner}
\end{figure}
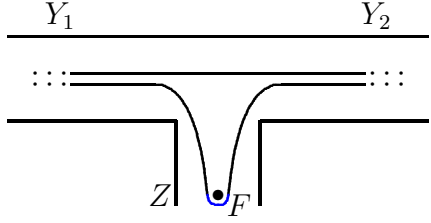

We now construct a spine $\Gamma_{g,s,|\delta|}$ that represents the above splitting of $\delta$
into $\delta_k$. For this, note that every pending edge terminates at an orbifold point belonging to
some subset $\delta_k$ and points towards some boundary component (hole)
(see Fig.~\ref{fi:corner} and the example in Fig.~\ref{fi:center}). The spine $\Gamma_{g,s,|\delta|}$
must be such that this is just the hole associated with the set $\delta_k$. A natural w.r.t. the
surface orientation cyclic ordering
of pending vertices of the spine $\Gamma_{g,s,|\delta|}$ associated to the $k$th boundary component
must coincide with the prescribed cyclic ordering in the subset $\delta_k$.

As the result, we obtain a fat graph of a given genus,
a given number of holes (punctures), and a given number of pending edges. We now endow {\em each}
edge of this graph including the pending edges
(the total number of edges is $6g-6+3s+2|\delta|$) with the real number $Z_\alpha$ (the subscript
$\alpha$ enumerates the edges of the graph). This pattern was first
proposed by Fock and Goncharov \cite{FG}.

\subsubsection{Geodesic functions for Riemann surfaces with ${\mathbb Z}_2$ orbifold points}

There exists a convenient parametrization of the geodesic lines corresponding to elements of the
group generated by the complete set of generators.
A geodesic line undergoes the {\em inversion} when it goes around the dot-vertex:
we then insert the matrix $F$ (\ref{F})
into the corresponding string of $2\times2$-matrices. For example, a part of geodesic function
in Fig.~\ref{fi:corner} that is inverted reads
$$
\dots X_{Y_1}LX_Z FX_Z LX_{Y_2}\dots\,,
$$
whereas the other geodesic that does not go around the dot vertex reads merely
$$
\dots X_{Y_1} RX_{Y_2}\dots\,.
$$

Note the simple relation,
\be
\label{2Z}
X_{Z}FX_{Z}=X_{2Z}.
\ee

Together with the explicit form of $R_Z$ and $L_Z$ (formulas (\ref{Rz}) and (\ref{Lz})), Eq.~(\ref{2Z})
implies that {\em any} product of $R_Z$ and $L_Z$ (represented by a matrix
$\left({a\ b\atop c\ d}\right)$)
always has strictly positive elements on the main diagonal
and nonpositive elements on the antidiagonal and, since it has
the unit determinant $ad-bc=1$, its trace $a+d\ge2$, that is,
such an element is almost always hyperbolic; it may be parabolic only when either
$b=0$ or $c=0$, which happens only when all the matrices in the product are either
$L_Z$ or $R_Z$, and this corresponds to
passing around a hole (puncture), and even in this case the trace is two
only if the hole is actually a puncture. The only
elliptic elements are conjugates of $F_i$  ($\tr F_i=0$).

We have therefore a metrizable Riemann surface for {\em any} choice of real numbers
$Z_\alpha$, associated to the edges of a spine; the main lemma in \cite{Ch2} states that
the converse is also true: {\em any} metrizable Riemann surface can be obtained this way.
So, in what follows, we identify the Teichm\"uller space of Riemann surfaces
with orbifold points with the space $\RR^{6g-6+3s+2|\delta|}$ of real parameters
on the edges of a spine $\Gamma_{g,s,|\delta|}$.

We also have the statement concerning the polynomiality of geodesic functions.

\begin{proposition}\label{prop-Laurent}
All $G_{\gamma}$ constructed by {\rm(\ref{G})}
are Laurent polynomials in $e^{Z_i}$ and $e^{Y_j/2}$ with positive integer coefficients, that is, we have
the Laurent property, which holds, e.g., in cluster algebras~\cite{FZ}. Here we let
$Z_i$ denote the variables of pending edges and $Y_j$ denote those of internal edges of the graph.
All these geodesic functions preserve their
polynomial structures upon Whitehead moves on inner edges (Fig.~\ref{fi:flip}) and upon
mapping class group transformations in Fig.~\ref{fi:mcg-pending}, in (\ref{braid1}), and in
(\ref{loopinvert}). All these geodesic functions correspond to hyperbolic elements
($G_\gamma>2$), the only exception where $G_\gamma=2$ are paths homotopic to going around holes
of zero length
(punctures).
\end{proposition}

\begin{example}\label{ex-geod}
{\rm
For the closed path drawn in Fig.~\ref{fi:treegraph}, we have
\bea
G_{24}&=&\tr L_{Y_2}R_{Y_3}LX_{Z_4}FX_{Z_4}R_{Y_3}L_{Y_2}RX_{Z_2}FX_{Z_2}
\nonumber
\\
&=&\tr L_{Y_2}R_{Y_3}L_{2Z_4}R_{Y_3}L_{Y_2}R_{2Z_2}.
\nonumber
\eea
}
\end{example}

Before we switch to purely algebraic part of this paper, let us remind the geometric
pattern underlying these algebraic constructions. For this, we recall the construction of
{\em ideal triangle decomposition} for orbifold Riemann surfaces.

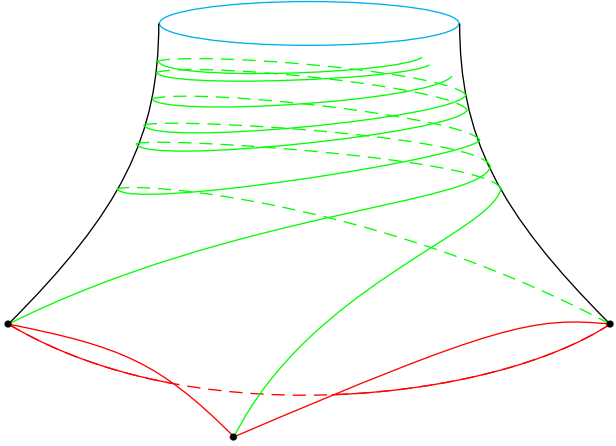
\begin{figure}[tb]
{\psset{unit=1}
\begin{pspicture}(-7,0)(7,-6)
\psellipse[linecolor=cyan, linewidth=0.5pt](0,0)(2,0.3)
\psbezier[linewidth=0.5pt](-4,-4)(-2.5,-2.5)(-2,-1.5)(-2,0)
\psbezier[linewidth=0.5pt](4,-4)(2.5,-2.5)(2,-1.5)(2,0)
\psbezier[linecolor=red, linewidth=0.5pt](-4,-4)(-1.5,-5.5)(2.5,-5)(4,-4)
\psframe[linecolor=white, fillstyle=solid, fillcolor=white](-1.8,-5.3)(0.3,-4)
\psbezier[linecolor=red, linewidth=0.5pt](-4,-4)(-2.5,-4.3)(-2,-4.5)(-1,-5.5)
\psbezier[linecolor=red, linewidth=0.5pt](4,-4)(3,-3.9)(2.5,-4)(-1,-5.5)
\psbezier[linecolor=red, linestyle=dashed, linewidth=0.5pt](-4,-4)(-1.5,-5.5)(2.5,-5)(4,-4)
\psbezier[linecolor=green, linewidth=0.5pt](-1,-5.5)(0,-3.5)(2.4,-2.8)(2.55,-2.2)
\psbezier[linecolor=green, linewidth=0.5pt](-4,-4)(-1,-2.5)(2.3,-2.4)(2.41,-1.9)
\psbezier[linecolor=green, linestyle=dashed, linewidth=0.5pt](4,-4)(1,-2.4)(-2.2,-2.1)(-2.55,-2.2)
\psbezier[linecolor=green, linestyle=dashed, linewidth=0.5pt](-2.3,-1.6)(-2,-1.5)(2.4,-1.8)(2.55,-2.2)
\psbezier[linecolor=green, linestyle=dashed, linewidth=0.5pt](-2.2,-1.35)(-1.8,-1.25)(2.3,-1.5)(2.41,-1.9)
\psbezier[linecolor=green, linewidth=0.5pt](-2.55,-2.2)(-2.45,-2.5)(2.1,-1.8)(2.27,-1.55)
\psbezier[linecolor=green, linewidth=0.5pt](-2.3,-1.6)(-2.2,-1.9)(2,-1.45)(2.1,-1.15)
\psbezier[linecolor=green, linewidth=0.5pt](-2.2,-1.35)(-2.1,-1.65)(2,-1.25)(2.08,-.95)
\psbezier[linecolor=green, linestyle=dashed, linewidth=0.5pt](-2.09,-1)(-1.45,-.85)(2.05,-1.25)(2.27,-1.55)
\psbezier[linecolor=green, linestyle=dashed, linewidth=0.5pt](-2.03,-.65)(-1.4,-0.5)(1.9,-.85)(2.1,-1.15)
\psbezier[linecolor=green, linestyle=dashed, linewidth=0.5pt](-2.03,-.5)(-1.4,-.37)(1.86,-.65)(2.08,-.95)
\psbezier[linecolor=green, linewidth=0.5pt](-2.09,-1)(-2,-1.25)(1.6,-1.05)(1.9,-.7)
\psbezier[linecolor=green, linewidth=0.5pt](-2.03,-.65)(-1.95,-0.87)(1.3,-.75)(1.6,-.55)
\psbezier[linecolor=green, linewidth=0.5pt](-2.01,-.5)(-1.92,-0.78)(1.2,-.67)(1.5,-.45)
\rput(-4,-4){\makebox(0,0){\tiny $\bullet$}}
\rput(4,-4){\makebox(0,0){\tiny $\bullet$}}
\rput(-1,-5.5){\makebox(0,0){\tiny $\bullet$}}
\end{pspicture}
}
\caption{\small The example of a regular genus zero
Riemann surface with three ${\mathbb Z}_2$-orbifold points ${\mathfrak s}_i$ and one hole.
The orbifold points are connected by finite-length geodesic lines, whereas
the geodesic lines of the partition start at the corresponding point
${\mathfrak s}_i$ and spiral asymptotically to the
closed geodesic that is the boundary of the hole with the perimeter $\ell_P=|Z_1+Z_2+Z_3|$,
where $Z_i$ are the Teichm\"uller space variables (from the graph description).}
\label{fi:saucer-pan}
\end{figure}

Given a regular Riemann surface of genus $g$ with $s$ holes and with $|\delta_k|$
${\mathbb Z}_2$-orbifold points assigned to the $k$th hole and cyclically ordered in every set
$\delta_k$, we issue from each of these
orbifold points a geodesic line (which is unique in a given homotopy class) that spirals in a given direction
(one and the same for all
lines spiraling to a given hole) to the perimeter line (the horizon) of this hole.
The example of a genus zero Riemann surface with three orbifold points and one hole is depicted in
Fig.~\ref{fi:saucer-pan}.

For pre-images of the spiraling geodesic lines starting at the orbifold points on the
Riemann surface we obtain that, for a ${\mathbb Z}_2$-orbifold point,
the pre-image of such a line in a Poincar\'e disc
consists of {\em two half-lines} originating
at this point and pointing in
opposite directions. We can then represent it as a {\em single} infinite geodesic line
passing through a preimage of the orbifold point. As the result, we have a pattern
like the one depicted in
Fig.~\ref{fi:disc}, that is, as in the case of Riemann surfaces without orbifold points, the
{\em fundamental domain} is a union of {\em ideal triangles}. Boundary geodesic curves of this fundamental
domain can be of two sorts: either (as in the standard case) an infinite
boundary curve is to be identified with
another boundary curve of this domain
or it contains a unique preimage of a ${\mathbb Z}_2$-orbifold point, and we then identify
its two halves separated by this point. It follows immediately from this consideration that lines
containing preimages of
${\mathbb Z}_2$-orbifold points must necessarily lie on the boundary of a fundamental domain.

\begin{figure}[tb]
{\psset{unit=0.7}
\begin{pspicture}(-7,-5)(8,5)
\pscircle[linewidth=0.5pt](0,0){4}
\psarc[linecolor=green, linewidth=0.5pt](-4,1.64){1.64}{-90}{45}
\psarc[linecolor=green, linewidth=0.5pt](0,5.66){4}{-135}{-45}
\psarc[linecolor=green, linewidth=0.5pt](4,1.64){1.64}{135}{270}
\psarc[linecolor=green, linewidth=0.5pt](4,-2.31){2.31}{90}{210}
\psarc[linecolor=green, linewidth=0.5pt](-.67,-4.99){3.06}{30}{135}
\psarc[linecolor=green, linewidth=0.5pt](-4,-1.64){1.64}{-45}{90}
\rput{37}(0,0){\psarc[linestyle=dashed, linecolor=red, linewidth=0.5pt](0,5.66){4}{-135}{-45}}
\rput{43.5}(0,0){\psarc[linestyle=dashed, linecolor=green, linewidth=0.5pt](0,5.8){4.2}{-133}{-47}}
\rput{-24}(0,0){\psarc[linestyle=dashed, linecolor=green, linewidth=0.5pt](0,4.3){1.6}{202}{338}}
\rput(-2.75,0.6){\makebox(0,0){\tiny $\bullet$}}
\rput(-0.6,1.7){\makebox(0,0){\tiny $\bullet$}}
\rput(2.35,1.45){\makebox(0,0){\tiny $\bullet$}}
\rput(2.6,-0.45){\makebox(0,0){\tiny $\bullet$}}
\rput(-1.1,-1.9){\makebox(0,0){\tiny $\bullet$}}
\rput(-2.8,-0.45){\makebox(0,0){\tiny $\bullet$}}
\rput(0.34,3.16){\makebox(0,0){\tiny $\bullet$}}
\rput(-2.85,0.7){\makebox(0,0)[rb]{$s_{\alpha_i}$}}
\rput(.45,3.2){\makebox(0,0)[lc]{$s'_{\alpha_i}$}}
\rput(-0.3,1.4){\makebox(0,0)[ct]{$s_{\alpha_{i+1}}$}}
\rput(-4.1,0){\makebox(0,0)[rc]{$-d$}}
\rput(-4,0.7){\makebox(0,0)[rc]{$\infty$}}
\rput(-3,3){\makebox(0,0)[rb]{$e$}}
\rput(0,4.1){\makebox(0,0)[cb]{$f$}}
\rput(0.7,4){\makebox(0,0)[cb]{$0$}}
\rput(2.5,3){\makebox(0,0)[lb]{$-c$}}
\end{pspicture}
\begin{pspicture}(-2,-5)(0,0)
\pscircle[linewidth=0.5pt](0,0){4}
\pcline[linestyle=dashed, linecolor=red, linewidth=0.5pt]{<-}(2.8,-2.8)(-2.8,2.8)
\pcline[linecolor=green, linewidth=0.5pt](2,-3.46)(-2,3.46)
\pcline[linestyle=dashed, linecolor=green, linewidth=0.5pt](-2,-3.46)(2,3.46)
\psarc[linecolor=green, linewidth=0.5pt](-8,0){6.92}{-30}{30}
\psarc[linestyle=dashed, linecolor=green, linewidth=0.5pt](8,0){6.92}{150}{210}
\rput{247.5}(0,0){\psarc[linecolor=green, linewidth=0.5pt](4.035,0){.52}{97.5}{262.5}}
\rput{247.5}(0,0){\rput(3.515,0){\makebox(0,0){\tiny $\bullet$}}}
\rput{262.5}(0,0){\psarc[linecolor=green, linewidth=0.5pt](4.035,0){.52}{97.5}{262.5}}
\rput{262.5}(0,0){\rput(3.515,0){\makebox(0,0){\tiny $\bullet$}}}
\rput{277.5}(0,0){\psarc[linecolor=green, linewidth=0.5pt](4.035,0){.52}{97.5}{262.5}}
\rput{277.5}(0,0){\rput(3.515,0){\makebox(0,0){\tiny $\bullet$}}}
\rput{292.5}(0,0){\psarc[linecolor=green, linewidth=0.5pt](4.035,0){.52}{97.5}{262.5}}
\rput{292.5}(0,0){\rput(3.515,0){\makebox(0,0){\tiny $\bullet$}}}
\rput(0,0){\makebox(0,0){\tiny $\bullet$}}
\rput(-1.2,1.2){\makebox(0,0){\tiny $\bullet$}}
\rput(1.2,-1.2){\makebox(0,0){\tiny $\bullet$}}
\rput(-1.3,1.2){\makebox(0,0)[rt]{$s_{\alpha_i}$}}
\rput(0,-1){\makebox(0,0)[ct]{$s_{\alpha_{i+1}}$}}
\rput(1.3,-1.2){\makebox(0,0)[lb]{$s'_{\alpha_i}$}}
\rput(-2.1,-3.6){\makebox(0,0)[rt]{$-d$}}
\rput(2.1,-3.6){\makebox(0,0)[lt]{$-c$}}
\rput(2.9,-2.9){\makebox(0,0)[lt]{$0$}}
\rput(-2.9,2.9){\makebox(0,0)[rb]{$\infty$}}
\rput(2.1,3.6){\makebox(0,0)[lb]{$f$}}
\rput(-2.1,3.6){\makebox(0,0)[rb]{$e$}}
\end{pspicture}
}
\caption{\small
Changing the pattern of ``natural'' ordering when replacing the preimage $s_{\alpha_i}$ of an
orbifold point by another preimage $s'_{\alpha_i}$ obtained upon rotation through the angle
$\pi$ about the neighbor preimage $s_{\alpha_{i+1}}$
of another orbifold point. Two bounding curves are replaces by two new bounding curves
(dashed lines), the third dashed line is the geodesic line connecting $s_{\alpha_i}$
and $s_{\alpha_{i+1}}$ (and also $s'_{\alpha_i}$). The equivalent form is presented in the right side
where the point $s_{\alpha_{i+1}}$ is at the center of the disc; from the picture it is obvious that
$\hbox{dist\,}(s_{\alpha_i},s_{\alpha_{i+1}})=\hbox{dist\,}(s_{\alpha_{i+1}},s'_{\alpha_i})$.
}
\label{fi:interchange}
\end{figure}
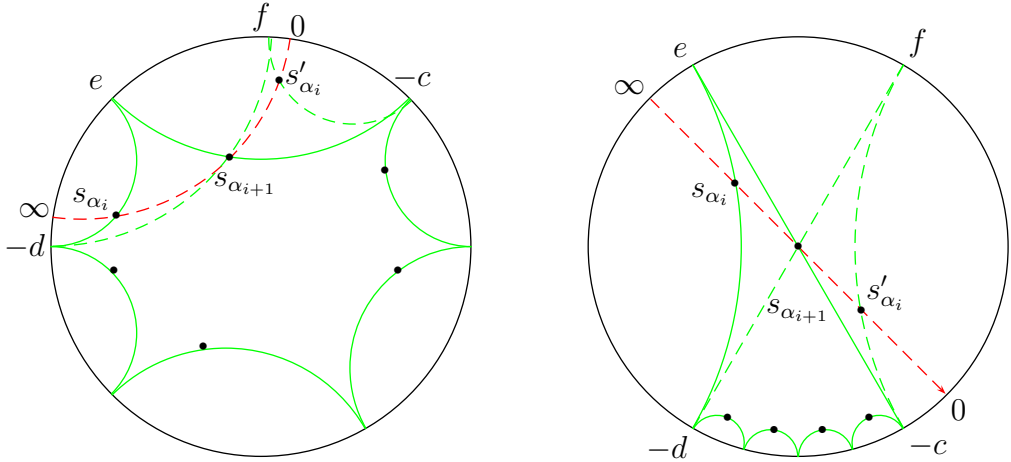

Note that choosing other representatives of the orbifold points, we obtain different
fundamental domains with different cyclic ordering of the (preimages) of the orbifold points
$s_i$ $(i=1,\dots, |\delta_k|)$. In Fig.~\ref{fi:interchange} we describe the changing of the
fundamental domain when we replace the preimage $s_{\alpha_i}$ by another preimage $s'_{\alpha_i}$
of the same orbifold point obtained upon
rotation about the neighbor (in the sense of the natural ordering inherited from the structure of
a fundamental domain)
point $s_{\alpha_{i+1}}$. As the result, the natural ordering changes: instead of
$\{\dots, s_{\alpha_i}, s_{\alpha_{i+1}}, \dots \}$
we obtain $\{\dots, s_{\alpha_{i+1}}, s'_{\alpha_{i}}, \dots \}$. Performing a series of such elementary
interchange operations, we can obtain any ordering starting from a given one.

The transformation in Fig.~\ref{fi:interchange} is an example of the braid
group transformation from Section~\ref{s:braid}.

After constructing a pattern with splitting of the fundamental domain into ideal triangles with vertices
at the absolute (see Fig.~\ref{fi:treegraph}),
we can apply the above graph technique (cf. \cite{Fock1},~\cite{ChF2}).
We then have the spine with $|\delta|$ new pending edges pointed outward the
fundamental domain and passing through the preimages of the
${\mathbb Z}_2$-orbifold points. The corresponding
Fuchsian group is then parameterized by real numbers $Z_\alpha$ associated to all the
edges of the constructed graph.

We consider below the example of the Riemann surface with
$n$ ${\mathbb Z}_2$-orbifold points and with one hole (puncture).

\subsection{The group  ${\mathfrak G}_n$}

We consider the Poincar\'e disc with $n$ different marked points $s_i$, $i=1,\dots,n$  inside it.
At each point $s_i$ we introduce the element $F_i$ of the rotation through $\pi$; each
$F_i=U_iFU^{-1}_i$ is a conjugate of the matrix (\ref{F}).

We are interested in the group ${\mathfrak G}_n$ generated by all the $F_i$. Observe, first, that the element
$\gamma_{ij}=F_iF_j$ is always a hyperbolic element whose {\em invariant axis} is a unique
geodesic that passes through the points $s_i$ and $s_j$ and its {\em length} is exactly
the double geodesic distance between $s_i$ and $s_j$. The fundamental domain necessarily has the
form (an ideal polygon) depicted in Fig.~\ref{fi:disc}; finite-length geodesic lines between the
points $s_i$ in Fig.~\ref{fi:disc} represent the above elements $\gamma_{ij}$.

\begin{figure}[tb]
{\psset{unit=0.7}
\begin{pspicture}(-10,-5)(8,5)
\pscircle[linewidth=0.5pt](0,0){4}
\psarc[linecolor=green, linewidth=0.5pt](-4,1.64){1.64}{-90}{45}
\psarc[linecolor=green, linewidth=0.5pt](0,5.66){4}{-135}{-45}
\psarc[linecolor=green, linewidth=0.5pt](4,1.64){1.64}{135}{270}
\psarc[linecolor=green, linewidth=0.5pt](4,-2.31){2.31}{90}{210}
\psarc[linecolor=green, linewidth=0.5pt](-.67,-4.99){3.06}{30}{135}
\psarc[linecolor=green, linewidth=0.5pt](-4,-1.64){1.64}{-45}{90}
\rput{20}(0,0){\psarc[linestyle=dashed, linecolor=red, linewidth=0.5pt](-4,1.64){1.64}{-90}{45}}
\rput{20}(0,0){\psarc[linecolor=red, linewidth=0.5pt](-4,1.64){1.64}{-40}{-5}}
\rput{37}(0,0){\psarc[linestyle=dashed, linecolor=red, linewidth=0.5pt](0,5.66){4}{-135}{-45}}
\rput{37}(0,0){\psarc[linecolor=red, linewidth=0.5pt](0,5.66){4}{-120}{-80}}
\rput{-12}(0,0){\psarc[linestyle=dashed, linecolor=red, linewidth=0.5pt](0,6.22){4.77}{-130}{-50}}
\rput{-12}(0,0){\psarc[linecolor=red, linewidth=0.5pt](0,6.22){4.77}{-100}{-65}}
\rput{39}(0,0){\psarc[linestyle=dashed, linecolor=red, linewidth=0.5pt](4,-2.31){2.31}{90}{210}}
\rput{39}(0,0){\psarc[linecolor=red, linewidth=0.5pt](4,-2.31){2.31}{120}{175}}
\rput{-157}(0,0){\psarc[linestyle=dashed, linecolor=red, linewidth=0.5pt](0,8){6.92}{-120}{-60}}
\rput{-157}(0,0){\psarc[linecolor=red, linewidth=0.5pt](0,8){6.92}{-110}{-75}}
\rput{-41}(0,0){\psarc[linestyle=dashed, linecolor=red, linewidth=0.5pt](-.67,-4.99){3.06}{30}{135}}
\rput{-41}(0,0){\psarc[linecolor=red, linewidth=0.5pt](-.67,-4.99){3.06}{70}{110}}
\rput(-2.75,0.6){\makebox(0,0){\tiny $\bullet$}}
\rput(-0.55,1.7){\makebox(0,0){\tiny $\bullet$}}
\rput(2.35,1.45){\makebox(0,0){\tiny $\bullet$}}
\rput(2.6,-0.45){\makebox(0,0){\tiny $\bullet$}}
\rput(-1.1,-1.9){\makebox(0,0){\tiny $\bullet$}}
\rput(-2.8,-0.45){\makebox(0,0){\tiny $\bullet$}}
\rput(-2.65,0.5){\makebox(0,0)[lt]{$s_1$}}
\rput(-2.7,-0.45){\makebox(0,0)[lb]{$s_2$}}
\rput(-1.2,-1.6){\makebox(0,0)[lb]{$s_3$}}
\rput(2.45,-0.45){\makebox(0,0)[rb]{$s_4$}}
\rput(2.2,1.3){\makebox(0,0)[rt]{$s_5$}}
\rput(-0.55,1.5){\makebox(0,0)[ct]{$s_6$}}
\end{pspicture}
}
\caption{\small
The Poincar\'e disc with $n=6$ preimages $s_i$ of orbifold points (marked by {\tiny$\bullet$}).
These points lie on sides of an ideal polygon, and
complementary geodesic lines in the figure are
the invariant axes of elements $\gamma_{i,i+1}=F_iF_{i+1}$, the part of an axis that lies
in the fundamental domain is drawn as the solid line and the outer part is drawn as a dashed line.}
\label{fi:disc}
\end{figure}
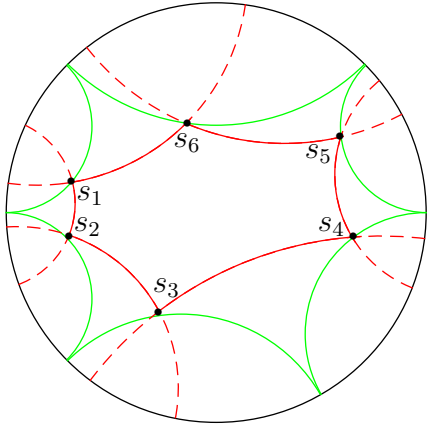

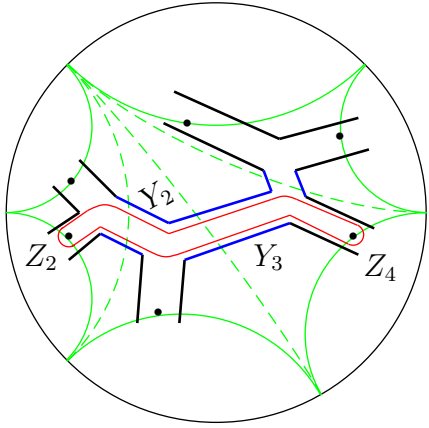
\begin{figure}[tb]
{\psset{unit=0.7}
\begin{pspicture}(-10,-5)(8,5)
\pscircle[linewidth=0.5pt](0,0){4}
\psarc[linecolor=green, linewidth=0.5pt](-4,1.64){1.64}{-90}{45}
\psarc[linecolor=green, linewidth=0.5pt](0,5.66){4}{-135}{-45}
\psarc[linecolor=green, linewidth=0.5pt](4,1.64){1.64}{135}{270}
\psarc[linecolor=green, linewidth=0.5pt](4,-2.31){2.31}{90}{210}
\psarc[linecolor=green, linewidth=0.5pt](-.67,-4.99){3.06}{30}{135}
\psarc[linecolor=green, linewidth=0.5pt](-4,-1.64){1.64}{-45}{90}
\psbezier[linecolor=green, linestyle=dashed, linewidth=0.5pt](-2.83,2.83)(-1.4,1.4)(2,0)(4,0)
\psbezier[linecolor=green, linestyle=dashed, linewidth=0.5pt](-2.83,2.83)(-1.4,1.4)(1,-1.732)(2,-3.464)
\rput{90}(0,0){\psarc[linecolor=green, linestyle=dashed, linewidth=0.5pt](0,5.66){4}{-135}{-45}}
\pcline[linewidth=1pt](-3.1,0.5)(-2.6,0)
\pcline[linewidth=1pt](-2.6,1)(-1.9,0.3)
\pcline[linewidth=1pt](-3.2,-0.4)(-2.6,0)
\pcline[linewidth=1pt](-2.8,-0.9)(-2.2,-0.4)
\pcline[linewidth=1pt](-1.5,-2.1)(-1.4,-0.8)
\pcline[linewidth=1pt](-.7,-2.1)(-.6,-0.9)
\pcline[linewidth=1pt](1.7,.3)(3,-.3)
\pcline[linewidth=1pt](1.4,-.2)(2.7,-.8)
\pcline[linewidth=1pt](1.2,1.4)(2.7,1.8)
\pcline[linewidth=1pt](1.5,.8)(2.9,1.2)
\pcline[linewidth=1pt](-1,1.7)(.9,.8)
\pcline[linewidth=1pt](-.8,2.3)(1.2,1.4)
\pcline[linecolor=blue, linewidth=1pt](-2.2,-0.4)(-1.4,-0.8)
\pcline[linecolor=blue, linewidth=1pt](-1.9,0.3)(-.9,-0.2)
\pcline[linecolor=blue, linewidth=1pt](-.6,-0.9)(1.4,-.2)
\pcline[linecolor=blue, linewidth=1pt](1.05,.4)(-.9,-0.2)
\pcline[linecolor=blue, linewidth=1pt](1.5,.8)(1.7,.3)
\pcline[linecolor=blue, linewidth=1pt](.9,.8)(1.05,.4)
\rput(-2.75,0.6){\makebox(0,0){\tiny $\bullet$}}
\rput(-0.55,1.7){\makebox(0,0){\tiny $\bullet$}}
\rput(2.35,1.45){\makebox(0,0){\tiny $\bullet$}}
\rput(2.6,-0.45){\makebox(0,0){\tiny $\bullet$}}
\rput(-1.1,-1.9){\makebox(0,0){\tiny $\bullet$}}
\rput(-2.8,-0.45){\makebox(0,0){\tiny $\bullet$}}
\psarc[linecolor=red, linewidth=0.5pt](-2.8,-.45){.2}{125}{305}
\pcline[linecolor=red, linewidth=0.5pt](-2.94,-0.31)(-2.44,0.04)
\pcline[linecolor=red, linewidth=0.5pt](-2.75,-0.65)(-2.2,-0.27)
\pcline[linecolor=red, linewidth=0.5pt](-2.2,-0.27)(-1.4,-0.67)
\pcline[linecolor=red, linewidth=0.5pt](-1.9,0.1)(-.9,-0.4)
\pcline[linecolor=red, linewidth=0.5pt](-.9,-0.4)(1.05,.25)
\pcline[linecolor=red, linewidth=0.5pt](-.6,-0.75)(1.4,-.05)
\pcline[linecolor=red, linewidth=0.5pt](1.4,-.05)(2.51,-0.6)
\pcline[linecolor=red, linewidth=0.5pt](1.53,.25)(2.68,-0.24)
\psarc[linecolor=red, linewidth=0.5pt](2.6,-.42){.2}{-115}{65}
\psbezier[linecolor=red, linewidth=0.5pt](-2.44,0.04)(-2.14,.19)(-2.1,0.16)(-1.9,0.1)
\psbezier[linecolor=red, linewidth=0.5pt](-1.4,-0.67)(-1,-.87)(-1,-0.89)(-.6,-0.75)
\psbezier[linecolor=red, linewidth=0.5pt](1.05,.25)(1.25,.315)(1.29,0.35)(1.53,.25)
\rput{30}(-1,0.1){\makebox(0,0)[cb]{$Y_2$}}
\rput(1,-0.6){\makebox(0,0)[ct]{$Y_3$}}
\rput(-3,-0.6){\makebox(0,0)[rt]{$Z_2$}}
\rput(2.8,-0.8){\makebox(0,0)[lt]{$Z_4$}}
\end{pspicture}
}
\caption{\small
The Poincar\'e disc with $n=6$ preimages ${\mathbb Z}_2$-orbifold points $s_i$ (marked by {\tiny$\bullet$})
with the associated fat graph that is dual to the ideal triangle decomposition of the fundamental
domain (additional geodesics of ideal triangle partition inside the fundamental domain
are drawn by dashed infinite lines).
We associate real numbers $Z_i$, $i=1,\dots,6$, to the pending edges and real numbers
$Y_2$, $Y_3$, and $Y_4$ to the inner edges (some of these parameters are indicated in the figure).
The closed curve in the graph corresponds to
$\tr F_2F_4$.}
\label{fi:treegraph}
\end{figure}

We now set in the correspondence to the Poincar\'e disc
with $n$ orbifold points the tree fat graph with $n$ pending edges.
We present the result in Fig.~\ref{fi:treegraph}.

\subsection{Structure of geodesic lines and multicurves (laminations)}

\subsubsection{Geodesic functions corresponding to paths in graphs}

To each closed path in a fat graph $\Gamma_{g,s,|\delta|}$, which is a spine of a genus $g$ Riemann surface
with $s$ holes and $|\delta|$ \ ${\mathbb Z}_2$-orbifold points, we set into a correspondence a closed
path in the Riemann surface. This closed path is a closed geodesic being an image of the invariant
axis of the corresponding hyperbolic element of the
Fuchsian group. In the orbifold case, we
have however a new class of finite-length geodesic
paths. Namely, let us consider a path in the graph connecting two dot-vertices
$s_i$ and $s_j$
(may be the same dot-vertex $s_i$) and going by
exactly the same sequence of edges of the graph in the both directions
(see the example in Fig.~\ref{fi:treegraph}). The corresponding element of the almost-hyperbolic
Fuchsian group
then has the form
\be
A^{-1}_{ij}F_iA_{ij}F_j=\tilde F_i F_j
\label{AFAF}
\ee
and, by the same reason as above, the invariant
axis of this element passes exactly through the orbifold points $s_i$ and $s_j$ (and its length is again the
doubled length of the corresponding path between these two points). The corresponding path in
the orbifold Riemann surface has then $s_i$ and $s_j$ as its terminal points.
We must therefore add to
the set of smooth closed geodesic lines in the Riemann surface the set of all geodesic lines that start and
terminate at the orbifold points (including cases where it is the same orbifold point).

\begin{definition}\label{def-lam}
{\rm
The geodesic multicurve (GM), or lamination, for an orbifold Riemann surface $\Sigma_{g,s,|\delta|}$ is a set
of non(self)intersecting geodesic lines (with multiplicities) including lines that terminate at the
orbifold points. In the latter case, only one line (with the multiplicity) that terminates at a
point $s_i$ is allowed in a GM.
}
\end{definition}

The algebraic counterpart of a GM is the $GM$ function (we use the same notation as we believe it does not
lead to a confusion)
\be
GM:=\prod_{\gamma\in \mathop{GM}} G_\gamma^{m_\gamma},
\label{GM}
\ee
where the product is over all geodesics $\gamma$ entering the GM with the multiplicities $m_\gamma$.

\newsection{Mapping class group transformations}\label{s:mcg}

\subsection{Poisson structure}\label{ss:Poisson}

One of the most attractive properties of the graph description is a very simple Poisson algebra on the set
of parameters $Z_\alpha$. Namely, we have the following theorem. It was formulated for surfaces
without marked points in~\cite{Fock1} and was extended to {\em arbitrary graphs} with pending
vertices in~\cite{FG} (see also \cite{Ch1}).

\begin{theorem}\label{th-WP} In the coordinates $Z_\alpha $ on any fixed spine
corresponding to a surface with orbifold points,
the Weil--Petersson bracket $B_{{\mbox{\tiny WP}}}$ is given by
\be
\label{WP-PB}
B_{{\mbox{\tiny WP}}}
= \sum_{v} \sum_{i=1}^{3}\frac{\partial}{\partial Z_{v_i}}\wedge
\frac{\partial}{\partial Z_{v_{i+1}}},
\ee
where the sum is taken over all three-valent {\rm(}i.e., not pending\/{\rm)} vertices~$v$ and $v_i$, \
$i=1,2,3\ \hbox{\rm mod}\ 3$, are the labels of the cyclically ordered
edges incident on this vertex irrespectively on whether they are internal or
pending edges of the graph.
\end{theorem}

The center of this Poisson algebra is provided by the proposition.

\begin{proposition}\label{prop12}
The center of the Poisson algebra {\rm(\ref{WP-PB})} is generated by
elements of the form $\sum Z_\alpha$, where the sum ranges all edges
of $\Gamma_{g,\delta} $ belonging to the same boundary component
taken with multiplicities.
This means, in particular, that each pending edge contributes twice to such sums.
The dimension of this center is obviously $s$.
\end{proposition}

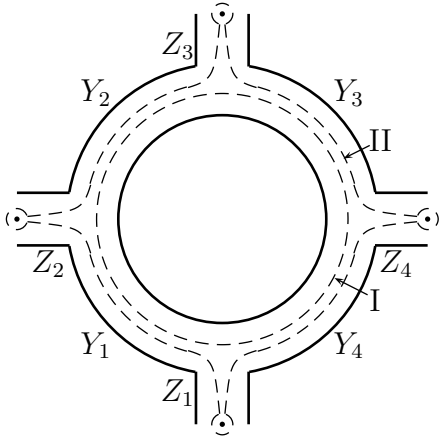
\begin{figure}[tb]
{\psset{unit=0.7}
\begin{pspicture}(-12,-4)(7,4)
\psarc[linewidth=1pt](0,0){2.95}{-80}{-10}
\psarc[linewidth=1pt](0,0){2.95}{10}{80}
\psarc[linewidth=1pt](0,0){2.95}{100}{170}
\psarc[linewidth=1pt](0,0){2.95}{190}{260}
\pscircle[linewidth=1pt](0,0){2}
\pcline[linewidth=1pt](-0.5,-2.9)(-0.5,-3.9)
\pcline[linewidth=1pt](0.5,-2.9)(0.5,-3.9)
\pcline[linewidth=1pt](-0.5,2.9)(-0.5,3.9)
\pcline[linewidth=1pt](0.5,2.9)(0.5,3.9)
\pcline[linewidth=1pt](-2.9,-0.5)(-3.9,-0.5)
\pcline[linewidth=1pt](-2.9,0.5)(-3.9,0.5)
\pcline[linewidth=1pt](2.9,-0.5)(3.9,-0.5)
\pcline[linewidth=1pt](2.9,0.5)(3.9,0.5)
\rput(-0.85,-3.3){\makebox(0,0){$Z_1$}}
\rput(-3.3,-0.85){\makebox(0,0){$Z_2$}}
\rput(-0.85,3.3){\makebox(0,0){$Z_3$}}
\rput(3.3,-0.85){\makebox(0,0){$Z_4$}}
\rput(-2.4,-2.4){\makebox(0,0){$Y_1$}}
\rput(-2.4,2.4){\makebox(0,0){$Y_2$}}
\rput(2.4,2.4){\makebox(0,0){$Y_3$}}
\rput(2.4,-2.4){\makebox(0,0){$Y_4$}}
\pscircle[linestyle=dashed, linewidth=0.5pt](0,0){2.4}
\psarc[linestyle=dashed, linewidth=0.5pt](0,0){2.6}{-75}{-15}
\psarc[linestyle=dashed, linewidth=0.5pt](0,0){2.6}{15}{75}
\psarc[linestyle=dashed, linewidth=0.5pt](0,0){2.6}{105}{165}
\psarc[linestyle=dashed, linewidth=0.5pt](0,0){2.6}{195}{255}
\psbezier[linestyle=dashed, linewidth=0.5pt](-0.65,-2.5)(-0.15,-2.65)(-0.15,-2.75)(-0.05,-3.7)
\psbezier[linestyle=dashed, linewidth=0.5pt](0.65,-2.5)(0.15,-2.65)(0.15,-2.75)(0.05,-3.7)
\psbezier[linestyle=dashed, linewidth=0.5pt](-0.65,2.5)(-0.15,2.65)(-0.15,2.75)(-0.05,3.7)
\psbezier[linestyle=dashed, linewidth=0.5pt](0.65,2.5)(0.15,2.65)(0.15,2.75)(0.05,3.7)
\psbezier[linestyle=dashed, linewidth=0.5pt](-2.5,-0.65)(-2.65,-0.15)(-2.75,-0.15)(-3.7,-0.05)
\psbezier[linestyle=dashed, linewidth=0.5pt](-2.5,0.65)(-2.65,0.15)(-2.75,0.15)(-3.7,0.05)
\psbezier[linestyle=dashed, linewidth=0.5pt](2.5,-0.65)(2.65,-0.15)(2.75,-0.15)(3.7,-0.05)
\psbezier[linestyle=dashed, linewidth=0.5pt](2.5,0.65)(2.65,0.15)(2.75,0.15)(3.7,0.05)
\psarc[linestyle=dashed, linewidth=0.5pt](0,-3.9){.2}{-250}{70}
\psarc[linestyle=dashed, linewidth=0.5pt](0,3.9){.2}{-70}{250}
\psarc[linestyle=dashed, linewidth=0.5pt](-3.9,0){.2}{20}{340}
\psarc[linestyle=dashed, linewidth=0.5pt](3.9,0){.2}{-160}{160}
\pscircle*[linestyle=dashed, linewidth=0.5pt](0,-3.9){.05}
\pscircle*[linestyle=dashed, linewidth=0.5pt](0,3.9){.05}
\pscircle*[linestyle=dashed, linewidth=0.5pt](-3.9,0){.05}
\pscircle*[linestyle=dashed, linewidth=0.5pt](3.9,0){.05}
\rput(2.9,-1.55){\makebox(0,0){I}}
\rput(3.0,1.5){\makebox(0,0){II}}
\pcline[linewidth=0.5pt]{->}(2.76,-1.44)(2.13,-1.12)
\pcline[linewidth=0.5pt]{->}(2.8,1.4)(2.3,1.15)
\end{pspicture}
}
\caption{\small An example of geodesics whose geodesic functions $G_{\mathrm I}$ and $G_{\mathrm II}$
are in the center of the Poisson algebra (dashed lines).}
\label{fi:center}
\end{figure}

\begin{example}\label{Ex1}
{\rm
Let us consider the graph in Fig.~\ref{fi:center}. It has two boundary components and two
corresponding geodesic lines. Their lengths, $\sum_{i=1}^4 Y_i$ and $\sum_{i=1}^4 (Y_i+2Z_i)$,
are the two Casimirs of the Poisson algebra with the defining relations
$$
\{Y_i,Y_{i-1}\}=1\quad\mod\ 4,\qquad \{Z_i,Y_i\}=-\{Z_i,Y_{i-1}\}=1\quad \mod 4,
$$
and with all other brackets equal to zero.
}
\end{example}

\subsection{Flip morphisms of fat graphs}\label{ss:flip}

In this section, we present the complete list of mapping class group transformations
that enable us to change numbers $|\delta_k|$ of orbifold points associated with the $k$th
hole, to change the cyclic ordering inside any of the sets $\delta_k$, to flip any inner edge
of the graph and, eventually, change the orientation of the geodesic
spiraling to the hole perimeter (in the case where we have more than one hole).\footnote{On the language
of cluster algebras \cite{FZ,FG}, this means that we are able to mutate {\em any} edge of any fat
graph possibly with pending edges and, possibly, with inner edges starting and terminating
at the same vertex.}
We can therefore
establish a morphism between any two of the graphs belonging to the same class $\Gamma_{g,s,|\delta|}$.

\subsubsection{Whitehead moves on inner edges}\label{sss:mcg}

The $Z_\alpha$-coordinates (which are the logarithms of cross ratios) are called {\it (Thurston) shear
coordinates} \cite{ThSh},\cite{Bon2} in the case of punctured Riemann surface (without boundary components).
We preserve this notation and this term also in the case of orbifold surfaces.

In the case of surfaces with holes, $Z_\alpha$ are the coordinates on the Teichm\"uller space
${\mathcal T}_{g,s}^H$, which is the $2^s$-fold covering of the standard Teichm\"uller space ramified over
surfaces with punctures (when a hole perimeter becomes zero, see~\cite{Fock2}).
We set $Z_\alpha$
to be the coordinates of the corresponding spaces
${\mathcal T}^H_{g,|\delta_1|,|\delta_2|,\dots,|\delta_s|}$ in the orbifold case, where, as above,
we let $|\delta_i|$ denote the number of orbifold points (may be zero) associated to the $i$th hole.

\begin{figure}[tb]
\setlength{\unitlength}{1.5mm}%
\begin{picture}(50,27)(-12,48)
\thicklines
\put(32,62){\line( -1,2){ 4}}
\put(36,64){\line( -1,2){ 4}}
\put(32,62){\line(-1,-2){ 4}}
\put(36,60){\line(-1,-2){ 4}}
\put(36,60){\line( 1, 0){20}}
\put(36,64){\line( 1, 0){20}}
\put(60,62){\line( 1, 2){ 4}}
\put(60,62){\line( 1,-2){ 4}}
\put(56,64){\line( 1, 2){ 4}}
\put(56,60){\line( 1,-2){ 4}}
\thinlines
\put(21,62){\vector(-1, 0){  0}}
\put(21,62){\vector( 1, 0){ 5}}
\thicklines
\put(10,54){\line( -2,-1){ 8}}
\put(8,58){\line( -2,-1){ 8}}
\put(8,58){\line( 0,1){8}}
\put(12,58){\line( 0,1){8}}
\put(10,54){\line(2,-1){ 8}}
\put(12,58){\line(2,-1){ 8}}
\put(10,70){\line( -2,1){ 8}}
\put(8,66){\line( -2,1){ 8}}
\put(10,70){\line(2,1){ 8}}
\put(12,66){\line(2,1){ 8}}
\put( 4,74){\makebox(0,0)[lb]{$A$}}
\put(16,74){\makebox(0,0)[rb]{$B$}}
\put(14,62){\makebox(0,0)[lc]{$Z$}}
\put(16,50){\makebox(0,0)[rt]{$C$}}
\put( 4,50){\makebox(0,0)[lt]{$D$}}
\put(32,52){\makebox(0,0)[lt]{$D - \phi(-Z)$}}
\put(60,52){\makebox(0,0)[rt]{$C+\phi(Z)$}}
\put(60,73){\makebox(0,0)[rb]{$B-\phi(-Z)$}}
\put(32,73){\makebox(0,0)[lb]{$A+\phi(Z)$}}
\put(47,66){\makebox(0,0)[cb]{$-Z$}}
\color[rgb]{1,0,0}
\thinlines
\put(0.5,53){\line( 2,1){ 6}}
\put(1,52){\line( 2,1){ 6}}
\put(1.5,51){\line( 2,1){ 6}}
\put(0.5,71){\line( 2,-1){ 6}}
\put(19,72){\line( -2,-1){ 6}}
\put(18.5,51){\line( -2,1){ 6}}
\qbezier(6.5,56)(9,57.75)(9,60)
\qbezier(6.5,68)(9,66.25)(9,64)
\put(9,60){\line( 0,1){ 4}}
\qbezier(7,55)(10,56.5)(10,62)
\qbezier(13,69)(10,67.5)(10,62)
\qbezier(7.5,54)(10,55.75)(12.5,54)
\put(-1.5,72){\makebox(0,0)[cc]{\hbox{\small$1$}}}
\put(-1.5,72){\circle{3}}
\put(21,73){\makebox(0,0)[cc]{\hbox{\small$2$}}}
\put(21,73){\circle{3}}
\put(20.5,50){\makebox(0,0)[cc]{\hbox{\small$3$}}}
\put(20.5,50){\circle{3}}
\put(29,53.5){\line( 1,2){ 3}}
\put(30,53){\line( 1,2){ 3}}
\put(31,52.5){\line( 1,2){ 3}}
\put(29,70.5){\line( 1,-2){ 3}}
\put(62,71){\line( -1,-2){ 3}}
\put(61,52.5){\line( -1,2){ 3}}
\qbezier(34,58.5)(35.25,61)(38,61)
\qbezier(58,58.5)(56.75,61)(54,61)
\put(38,61){\line( 1,0){ 16}}
\qbezier(33,59)(34.5,62)(38,62)
\qbezier(59,65)(57.5,62)(54,62)
\put(38,62){\line( 1,0){ 16}}
\qbezier(32,59.5)(33.25,62)(32,64.5)
\put(28,72.5){\makebox(0,0)[cc]{\hbox{\small$1$}}}
\put(28,72.5){\circle{3}}
\put(63,73){\makebox(0,0)[cc]{\hbox{\small$2$}}}
\put(63,73){\circle{3}}
\put(62,50.5){\makebox(0,0)[cc]{\hbox{\small$3$}}}
\put(62,50.5){\circle{3}}
\end{picture}
\caption{\small Flip, or Whitehead move on the shear coordinates $Z_\alpha$. The outer edges can be
pending, but the edge with respect to which the morphism is performed must be an internal
edge.}
\label{fi:flip}
\end{figure}
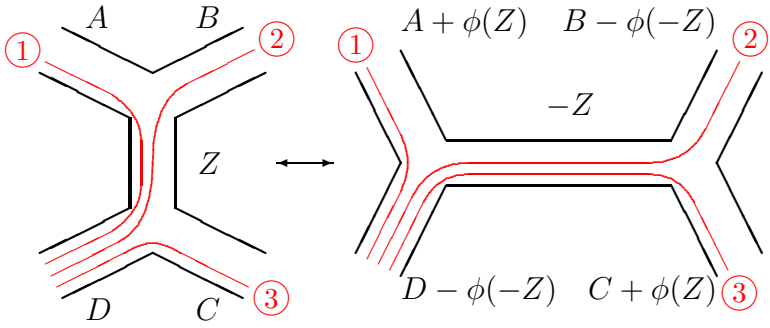

Given an enumeration of the edges of the spine $\Gamma$ of $\Sigma$ and assuming the
edge $\alpha$ to have distinct endpoints, we may produce
another spine $\Gamma _\alpha$ of $\Sigma$ by contracting and expanding edge $\alpha$ of
$\Gamma $, the edge labeled $Z$ in Figure~\ref{fi:flip}, to produce $\Gamma _\alpha$ as in the
figure. Furthermore, an enumeration of the edges of
$\Gamma $ induces an enumeration of the edges of $\Gamma _\alpha$ in the natural
way, where the vertical edge labelled $Z$ in Figure~\ref{fi:flip} corresponds to the horizontal
edge labelled $-Z$.  We say that $\Gamma _\alpha$ arises from $\Gamma$ by a
{\it Whitehead move} (or flip) along the edge $\alpha$.
A labeling of edges of the spine $\Gamma$ implies a natural labeling of edges of the
spine $\Gamma_\alpha$; we then obtain a morphism between the spines $\Gamma$ and $\Gamma_\alpha$.

\begin{proposition} {\rm \cite{ChF}}\label{propcase}
Setting $\phi (Z)={\rm log}(e^Z+1)$ and adopting the notation of Fig.~\ref{fi:flip}
for shear coordinates of nearby edges, the effect of a
Whitehead move is as follows:
\be
W_Z\,:\ (A,B,C,D,Z)\to (A+\phi(Z), B-\phi(-Z), C+\phi(Z), D-\phi(-Z), -Z)
\label{abc}
\ee
In the various cases where the edges are not distinct
and identifying an edge with its shear coordinate in the obvious notation we have:
if $A=C$, then $A'=A+2\phi(Z)$;
if $B=D$, then $B'=B-2\phi(-Z)$;
if $A=B$ (or $C=D$), then $A'=A+Z$ (or $C'=C+Z$);
if $A=D$ (or $B=C$), then $A'=A+Z$ (or $B'=B+Z$).
Any variety of edges among $A$, $B$, $C$, and $D$ can be pending edges of the graph.
\end{proposition}

We also have two simple but important lemmas establishing the properties of
invariance w.r.t. the flip morphisms~\cite{ChF}.

\begin{lemma} \label{lem-abc}
Transformation~{\rm(\ref{abc})} preserves
the traces of products over paths {\rm(\ref{G})}.
\end{lemma}

\begin{lemma} \label{lem-Poisson}
Transformation~{\rm(\ref{abc})} preserves
Poisson structure {\rm(\ref{WP-PB})} on the shear coordinates.
\end{lemma}

That the Poisson algebra for the orbifold Riemann surfaces case
is invariant under the flip transformations follows immediately
because we flip here inner, not pending, edges of a graph, which reduces
the situation to the ``old" statement for surfaces without orbifold points.

\subsubsection{Whitehead moves on pending edges}\label{sss:pending}

In the case of orbifold surfaces we encounter a new phenomenon, namely,
we can construct morphisms relating {\em any} two of the Teichm\"uller spaces
${\mathcal T}_{g,\delta^1}^H$ and ${\mathcal T}_{g,\delta^2}^H$ with
$\delta^1=\{|\delta_1^1|,\dots, |\delta_{s_1}^1|\}$
and $\delta^2=\{|\delta_1^2|,\dots, |\delta_{s_2}^2|\}$ providing $s_1=s_2=s$ and
$\sum_{i=1}^{s_1}|\delta^1_i|=\sum_{i=1}^{s_2}|\delta^2_i|$, that is, we explicitly construct
morphisms relating
any two of algebras corresponding to orbifold surfaces of the same genus, same number of boundary components, and
with the same total number of orbifold points whose
distributions into the holes can be arbitrary.

This new morphism corresponds in a sense to flipping a pending edge.

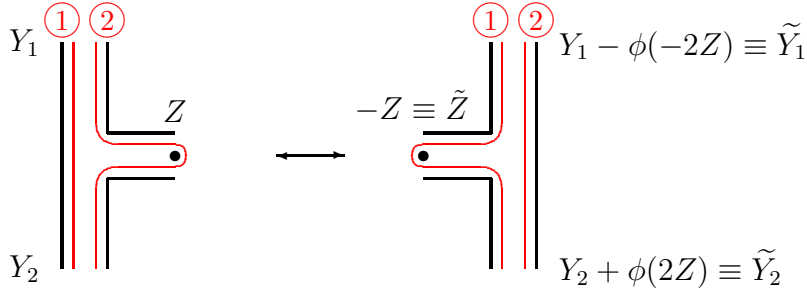
\begin{figure}[tb]
\setlength{\unitlength}{1.5mm}%
\begin{picture}(50,27)(-12,48)
\thicklines
\put(50,70){\line( 0,-1){20}}
\put(46,70){\line( 0,-1){8}}
\put(46,58){\line( 0,-1){8}}
\put(46,62){\line(-1,0){ 6}}
\put(46,58){\line(-1,0){ 6}}
\put(40,60){\circle*{1}}
\put( 52,70){\makebox(0,0)[lc]{$Y_1-\phi(-2Z)\equiv {\widetilde{Y_1}}$}}
\put(39,63){\makebox(0,0)[cb]{$-Z\equiv\tilde Z$}}
\put( 52,50){\makebox(0,0)[lc]{$Y_2+\phi(2Z)\equiv {\widetilde{Y_2}}$}}
\thinlines
\put(27,60){\vector(-1, 0){  0}}
\put(27,60){\vector( 1, 0){ 6}}
\thicklines
\put(8,70){\line( 0,-1){20}}
\put(12,70){\line( 0,-1){8}}
\put(12,58){\line( 0,-1){8}}
\put(12,62){\line(1,0){ 6}}
\put(12,58){\line(1,0){ 6}}
\put(18,60){\circle*{1}}
\put( 6,70){\makebox(0,0)[rc]{$Y_1$}}
\put(18,63){\makebox(0,0)[cb]{$Z$}}
\put( 6,50){\makebox(0,0)[rc]{$Y_2$}}
\color[rgb]{1,0,0}
\thinlines
\put(9,70){\line( 0,-1){20}}
\put(11,70){\line( 0,-1){ 7}}
\put(11,50){\line( 0,1){ 7}}
\put(13,61){\line( 1,0){ 5}}
\put(13,59){\line( 1,0){ 5}}
\qbezier(11,63)(11,61)(13,61)
\qbezier(11,57)(11,59)(13,59)
\qbezier(19,60)(19,61)(18,61)
\qbezier(19,60)(19,59)(18,59)
\put(8,72){\makebox(0,0)[cc]{\hbox{\small$1$}}}
\put(8,72){\circle{3}}
\put(12,72){\makebox(0,0)[cc]{\hbox{\small$2$}}}
\put(12,72){\circle{3}}
\put(49,70){\line( 0,-1){20}}
\put(47,70){\line( 0,-1){ 7}}
\put(47,50){\line( 0,1){ 7}}
\put(45,61){\line( -1,0){ 5}}
\put(45,59){\line( -1,0){ 5}}
\qbezier(47,63)(47,61)(45,61)
\qbezier(47,57)(47,59)(45,59)
\qbezier(39,60)(39,61)(40,61)
\qbezier(39,60)(39,59)(40,59)
\put(46,72){\makebox(0,0)[cc]{\hbox{\small$1$}}}
\put(46,72){\circle{3}}
\put(50,72){\makebox(0,0)[cc]{\hbox{\small$2$}}}
\put(50,72){\circle{3}}
\end{picture}
\caption{\small Flip, or Whitehead move on the shear coordinates when flipping the pending
edge $Z$ (indicated by bullet). Any (or both) of edges $Y_1$ and $Y_2$ can be pending.}
\label{fi:mcg-pending}
\end{figure}

\begin{lemma} \label{lem-pending}
Transformation~in Fig.~\ref{fi:mcg-pending} is the morphism between the spaces
${\mathcal T}_{g,\delta^1}^H$ and ${\mathcal T}_{g,\delta^2}^H$. These
morphisms preserve both Poisson structures {\rm(\ref{WP-PB})} and the geodesic
functions. In Fig.~\ref{fi:mcg-pending} any (or both) of $Y$-variables can be
variables of pending edges (the transformation formula is insensitive to it).
\end{lemma}

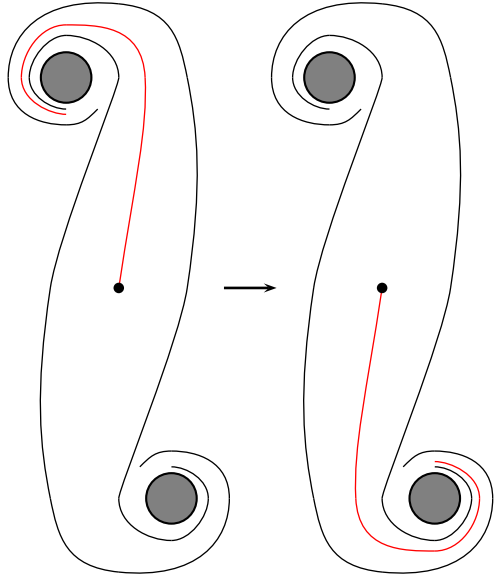
\begin{figure}[tb]
{\psset{unit=0.7}
\begin{pspicture}(-12,-6)(7,6)
\newcommand{\BASE}{%
\pscircle[fillstyle=solid, fillcolor=gray](-1,4){0.5}
\psbezier[linewidth=0.5pt](-1,3.4)(-1.3,3.4)(-1.7,3.65)(-1.7,4)
\psbezier[linewidth=0.5pt](-1.7,4)(-1.7,4.35)(-1.35,4.8)(-1,4.8)
\psbezier[linewidth=0.5pt](-1,4.8)(-.55,4.8)(0,4.5)(0,4)
\psbezier[linewidth=0.5pt](0,4)(0,3.8)(-1.15,1)(-1.3,0)
\psbezier[linewidth=0.5pt](1.3,4)(1.5,3)(1.6,2)(1.3,0)
\psbezier[linewidth=0.5pt](1.3,4)(1.15,4.75)(1,5.3)(0,5.4)
\psbezier[linewidth=0.5pt](0,5.4)(-1.5,5.55)(-2.1,4.8)(-2.1,4)
\psbezier[linewidth=0.5pt](-2.1,4)(-2.1,3.3)(-1.5,3.1)(-1,3.1)
\psbezier[linewidth=0.5pt](-1,3.1)(-.7,3.1)(-.6,3.2)(-.4,3.4)
}
\newcommand{\ORBI}{%
\psbezier[linecolor=red, linewidth=0.5pt](-1,3.3)(-1.3,3.3)(-1.85,3.55)(-1.85,4)
\psbezier[linecolor=red, linewidth=0.5pt](-1.85,4)(-1.85,4.45)(-1.5,5)(-1,5)
\psbezier[linecolor=red, linewidth=0.5pt](-1,5)(-0.4,5)(0.45,5)(0.5,4)
\psbezier[linecolor=red, linewidth=0.5pt](0.5,4)(0.55,3)(0.3,2)(0,0)
}
\rput(-2.5,0){\BASE}
\rput{180}(-2.5,0){\BASE}
\rput(-2.5,0){\ORBI}
\rput(-2.5,0){\pscircle*{0.1}}
\pcline[linewidth=1pt]{->}(-0.5,0)(0.5,0)
\rput(2.5,0){\BASE}
\rput{180}(2.5,0){\BASE}
\rput{180}(2.5,0){\ORBI}
\rput(2.5,0){\pscircle*{0.1}}
\end{pspicture}
}
\caption{\small Changing ideal triangular decomposition corresponding to transformation
in Fig.~\ref{fi:mcg-pending}.}
\label{fi:snail}
\end{figure}

In the ideal triangular decomposition of the original Riemann surface, this transformation
reduces to changing the boundary geodesic line that starts at the corresponding orbifold
point as shown in Fig.~\ref{fi:snail}. There we assume $Y_1$ and $Y_2$ to be variables
of internal edges. The two holes in the figure can be the same hole if
the two boundary lines in Fig.~\ref{fi:mcg-pending} belong to the same boundary component.

{\bf Proof.} Verifying the preservation of Poisson relations (\ref{WP-PB}) is simple, whereas
for traces over paths we have four cases, and in each of these cases we have the following
$2\times2$-{\em matrix} equalities (each can be verified directly)
\bea
X_{Y_2}LX_ZFX_ZLX_{Y_1}&=&X_{{\tilde Y}_2}LX_{{\tilde Y}_1},\nonumber\\
X_{Y_1}RX_ZFX_ZRX_{Y_1}&=&X_{{\tilde Y}_1}LX_{\tilde Z}FX_{\tilde Z}RX_{{\tilde Y}_1},\nonumber\\
X_{Y_2}RX_{Y_1}&=&X_{{\tilde Y}_2}RX_{\tilde Z}FX_{\tilde Z}RX_{{\tilde Y}_1},\nonumber\\
X_{Y_2}LX_ZFX_ZRX_{Y_2}&=&X_{{\tilde Y}_2}RX_{\tilde Z}FX_{\tilde Z}LX_{{\tilde Y}_2},\nonumber
\eea
where (in the exponentiated form)
\be
\label{exp-p}
e^{{\tilde Y}_1}=e^{Y_1}\bigl(1+e^{-2Z}\bigr)^{-1},\qquad e^{{\tilde Y}_2}=e^{Y_1}\bigl(1+e^{2Z}\bigr),
\qquad
e^{{\tilde Z}}=e^{-Z}.\qquad\square
\ee

From the technical standpoint, all these equalities follow from flip transformation (\ref{abc})
upon the substitution $A=C=Y_2$, $B=D=Y_1$, and $Z=2Z$. The above four cases of geodesic functions are
then exactly four possible cases of geodesic arrangement in the (omitted) proof of Lemma~\ref{lem-abc}.

Using flip morphisms in Fig.~\ref{fi:mcg-pending} and in formula
(\ref{abc}), we establish a morphism between any two algebras
corresponding to surfaces of the same genus, same number of boundary
components, and same total number of marked points on these
components. And it is again a standard tool that if, after a
series of morphisms, we come to a graph of the same combinatorial
type as the initial one (disregarding labeling of edges), we
associate a {\em mapping class group} operation to this morphism
therefore passing from the groupoid of morphisms to the group of
modular transformations.

\begin{example}\label{Ex2}
{\rm
The flip morphism w.r.t. the edge $Z_1$ in the pattern in (\ref{braid1}),
\be
\label{braid1}
{\psset{unit=0.7}
\begin{pspicture}(-5,-3)(7,1)
\pcline[linewidth=1pt](-6,0.5)(-1,0.5)
\pcline[linewidth=1pt](-6,-0.5)(-4,-0.5)
\pcline[linewidth=1pt](-3,-0.5)(-1,-0.5)
\pcline[linewidth=1pt](-4,-0.5)(-4,-2.5)
\pcline[linewidth=1pt](-3,-0.5)(-3,-2.5)
\pscircle*(-6,0){0.1}
\pscircle*(-3.5,-2.5){0.1}
\rput(-5.5,1){\makebox(0,0){$Z_{1}$}}
\rput(-4.6,-2){\makebox(0,0){$Z_{2}$}}
\rput(-1.5,1){\makebox(0,0){$Y$}}
\pcline[linewidth=1pt]{<->}(0,0)(2,0)
\pcline[linewidth=1pt](3,0.5)(8,0.5)
\pcline[linewidth=1pt](3,-0.5)(5,-0.5)
\pcline[linewidth=1pt](6,-0.5)(8,-0.5)
\pcline[linewidth=1pt](5,-0.5)(5,-2.5)
\pcline[linewidth=1pt](6,-0.5)(6,-2.5)
\rput(3.5,1){\makebox(0,0){$Z_{2}-\phi(-2Z_1)$}}
\rput(7.5,1){\makebox(0,0){$Y+\phi(2Z_1)$}}
\rput(4.4,-2){\makebox(0,0){$-Z_{1}$}}
\pscircle*(3,0){0.1}
\pscircle*(5.5,-2.5){0.1}
\rput(10,0){\makebox(0,0){,}}
\psarc[linecolor=green, linewidth=1pt]{<-}(-6,-10){10}{79}{88}
\psarc[linecolor=green, linewidth=1pt]{->}(-9.5,-2.5){6}{2}{14}
\pcline[linecolor=red, linewidth=0.5pt](-1,0.3)(-3,0.3)
\pcline[linecolor=red, linewidth=0.5pt](-1,-0.3)(-3,-0.3)
\pcline[linecolor=red, linewidth=0.5pt](-3.8,-0.5)(-3.8,-2.5)
\pcline[linecolor=red, linewidth=0.5pt](-3.2,-0.5)(-3.2,-2.5)
\psarc[linecolor=red, linewidth=0.5pt](-3,-0.5){.2}{90}{180}
\psarc[linecolor=red, linewidth=0.5pt](-3,-0.5){.8}{90}{180}
\psarc[linecolor=red, linewidth=0.5pt](-3.5,-2.5){.3}{180}{360}
\pcline[linecolor=red, linewidth=0.5pt](8,0.3)(6,0.3)
\pcline[linecolor=red, linewidth=0.5pt](8,-0.3)(6,-0.3)
\pcline[linecolor=red, linewidth=0.5pt](5.7,-2.5)(5.7,0)
\pcline[linecolor=red, linewidth=0.5pt](5.9,-2.5)(5.9,-0.4)
\psarc[linecolor=red, linewidth=0.5pt](6,-0.4){.1}{90}{180}
\psarc[linecolor=red, linewidth=0.5pt](6,0){.3}{90}{180}
\pcline[linecolor=red, linewidth=0.5pt](3,0.3)(5,0.3)
\pcline[linecolor=red, linewidth=0.5pt](3,-0.3)(5,-0.3)
\pcline[linecolor=red, linewidth=0.5pt](5.3,-2.5)(5.3,0)
\pcline[linecolor=red, linewidth=0.5pt](5.1,-2.5)(5.1,-0.4)
\psarc[linecolor=red, linewidth=0.5pt](5,-0.4){.1}{0}{90}
\psarc[linecolor=red, linewidth=0.5pt](5,0){.3}{0}{90}
\psarc[linecolor=red, linewidth=0.5pt](3,0){.3}{90}{270}
\psarc[linecolor=red, linewidth=0.5pt](5.5,-2.5){.2}{180}{360}
\psarc[linecolor=red, linewidth=0.5pt](5.5,-2.5){.4}{180}{360}
\end{pspicture}
}
\ee
where $Z_1$ and $Z_2$ are the pending edges, generates the (unitary) mapping class group transformation
\be
\label{braid2}
e^{Z_2}\to e^{-Z_1},\qquad e^{Z_1}\to e^{Z_2}\bigl(1+e^{-2Z_1}\bigr)^{-1},\qquad e^Y\to e^Y
\bigl(1+e^{2Z_1}\bigr)
\ee
on the corresponding Teichm\"uller space ${\mathcal T}_{g,\delta}^H$. This is one of the generators of
the {\em braid group} \cite{Ch1}.
}
\end{example}

\subsubsection{Changing the spiraling direction}

We introduce a new mapping class group transformation that change the sign of the hole perimeter:
\be
\label{loopinvert}
{\psset{unit=0.7}
\begin{pspicture}(-5,-3)(7,1)
\pcline[linewidth=1pt](-6,-0.5)(-4,-0.5)
\pcline[linewidth=1pt](-6,-1.5)(-4,-1.5)
\psbezier[linewidth=1pt](-4,-0.5)(-3,1)(-1,1)(-1,-1)
\psbezier[linewidth=1pt](-4,-1.5)(-3,-3)(-1,-3)(-1,-1)
\psbezier[linewidth=1pt](-3.2,-1)(-2.4,-0.2)(-2,-0.3)(-2,-1)
\psbezier[linewidth=1pt](-3.2,-1)(-2.4,-1.8)(-2,-1.7)(-2,-1)
\rput(-5.5,0.2){\makebox(0,0){$Y$}}
\rput(-1.5,1){\makebox(0,0){$X$}}
\pcline[linewidth=1pt]{<->}(0,-1)(2,-1)
\pcline[linewidth=1pt](3,-0.5)(5,-0.5)
\pcline[linewidth=1pt](3,-1.5)(5,-1.5)
\psbezier[linewidth=1pt](5,-0.5)(6,1)(8,1)(8,-1)
\psbezier[linewidth=1pt](5,-1.5)(6,-3)(8,-3)(8,-1)
\psbezier[linewidth=1pt](5.8,-1)(6.6,-0.2)(7,-0.3)(7,-1)
\psbezier[linewidth=1pt](5.8,-1)(6.6,-1.8)(7,-1.7)(7,-1)
\rput(3.5,0.2){\makebox(0,0){$Y+X$}}
\rput(7.5,1){\makebox(0,0){$-X$}}
\rput(10,0){\makebox(0,0){.}}
\pcline[linecolor=red, linewidth=0.5pt]{->}(-6,-.7)(-3.9,-.7)
\pcline[linecolor=red, linewidth=0.5pt](-6,-1.3)(-3.9,-1.3)
\psbezier[linecolor=red, linewidth=0.5pt](-3.9,-.7)(-3,.7)(-1.3,.8)(-1.3,-1)
\psbezier[linecolor=red, linewidth=0.5pt](-3.9,-1.3)(-3,-2.7)(-1.3,-2.8)(-1.3,-1)
\pcline[linecolor=red, linewidth=0.5pt]{->}(3,-.7)(5.1,-.7)
\pcline[linecolor=red, linewidth=0.5pt](3,-1.3)(5.1,-1.3)
\psbezier[linecolor=red, linewidth=0.5pt](5.1,-.7)(6,.7)(7.7,.8)(7.7,-1)
\psbezier[linecolor=red, linewidth=0.5pt](5.1,-1.3)(6,-2.7)(7.7,-2.8)(7.7,-1)
\end{pspicture}
}
\ee
That this transformation preserves geodesic functions follows from two matrix equalities:
\bea
&&X_YLX_XLX_Y=X_{Y+X}LX_{-X}LX_{Y+X},
\nonumber
\\
&&X_YRX_XRX_Y=X_{Y+X}RX_{-X}RX_{Y+X},
\nonumber
\eea
and we can correspondingly enlarge the mapping class group of ${\cal T}^H_{g,\delta}$ by adding
symmetries between sheets of the $2^s$-ramified covering of the ``genuine'' (nondecorated)
Teichm\"uller space ${\cal T}_{g,\delta}$.

The geometrical meaning of this transformation is clear: we change the direction of spiraling to the
hole perimeter line for all lines of the ideal triangle decomposition that spiral to a given hole.

\subsection{The graphical representation}\label{ss:new}

In the case of usual geodesic functions, there exists a very convenient representation in which
one can apply classical skein and Poisson relations in classical case or the quantum skein
relation in the quantum case and ensure the Reidemeister moves when ``disentangling" the products of
geodesic function representing them as linear combinations of multicurve functions.
However, in our case, it is still obscure what happens when geodesic lines
intersect in some way at the dot vertex. In fact, we can propose the comprehensive graphical
representation in this case as well! For this,
we turn to Fig.~\ref{fi:corner} and assume that the inversion
matrix $F$ corresponds to actual winding around this dot-vertex as shown in the figure.

We now formulate the rules for geodesic algebra that follow from relations (\ref{WP-PB}) and
classical skein relations. They coincide with the rules in the case of surfaces with holes except
the one new nontrivial case depicted in Fig.~\ref{fi:dot-skein}. Note that all claims below follow from
direct and explicit calculations involving representations from Sec.~\ref{s:hyp}.

\subsubsection{Classical skein relation}\label{sss:skein}
The trace relation $\tr(AB)+\tr(AB^{-1})-\tr A\cdot\tr B=0$ for
arbitrary $2\times 2$ matrices~$A$ and~$B$ with unit determinant allows one to
``disentangle'' any product of geodesic functions $G_A\cdot G_B$,
i.e., express it uniquely as a finite linear combination of generalized multicurves
(see Definition~\ref{def-lam}).
This relation corresponds to resolving the crossing between two geodesics $A$ and $B$
as indicated in the formula below and it is referred to as the {\em skein relation.}
\be
{\psset{unit=0.5}
\begin{pspicture}(-14,-4)(28,4)
\newcommand{\LOOPS}{%
\psarc[linestyle=dashed, linewidth=1pt](2,0){1.42}{-45}{135}
\pcline[linestyle=dashed, linewidth=1pt](3,-1)(1,-3)
\psarc[linestyle=dashed, linewidth=1pt](0,-2){1.42}{135}{315}
\psarc[linestyle=dashed, linewidth=1pt](0.2,-1.8){1.13}{-135}{45}
\pcline[linestyle=dashed, linewidth=1pt](-0.6,-2.6)(-2.6,-0.6)
\psarc[linestyle=dashed, linewidth=1pt](-1.8,0.2){1.13}{45}{225}
}
\rput(-10,0){\LOOPS}
\rput(-10,0){
\pcline[linewidth=1pt](-1,1)(1,-1)
\pcline[linewidth=1pt](-1,-1)(-0.2,-0.2)
\pcline[linewidth=1pt](1,1)(0.2,0.2)
\rput(-1.8,1.6){\makebox(0,0)[cb]{$G_A$}}
\rput(2,1.8){\makebox(0,0)[cb]{$G_B$}}
}
\rput(-4.8,0){\makebox(0,0)[cc]{$=$}}
\rput(0,0){\LOOPS}
\rput(0,0){
\psarc[linewidth=1pt](-2,0){1.42}{-45}{45}
\psarc[linewidth=1pt](2,0){1.42}{135}{225}
\rput(0,1.3){\makebox(0,0)[cb]{$G_{AB}$}}
}
\rput(5.2,0){\makebox(0,0)[cc]{$+$}}
\rput(10,0){\LOOPS}
\rput(10,0){
\psarc[linewidth=1pt](0,2){1.42}{225}{315}
\psarc[linewidth=1pt](0,-2){1.42}{45}{135}
\rput(0,1.3){\makebox(0,0)[cb]{$G_{AB^{-1}}$}}
}
\end{pspicture}
}
\label{skeinclass}
\ee

\subsubsection{Poisson brackets for geodesic functions.}

We first mention that two geodesic functions Poisson
commute if the underlying geodesics are disjointly embedded in the sense of the new
graph technique involving dot-vertices. Because of
the Leibnitz rule for the Poisson bracket, it suffices to consider only ``simple'' intersections of
pairs of geodesics when the respective geodesic functions
$G_1$ and $G_2$ are
\bea
\label{G1}
G_1&=&\tr\dots X_CRX_ZLX_A\dots,
\\
\label{G2}
G_2&=&\tr\dots X_B L X_ZRX_D\dots.
\eea
The positions of edges $A,B,C,D,$ and $Z$ are as in Fig.~\ref{fi:flip}.
Ellipses in (\ref{G1}), (\ref{G2}) refer to arbitrary sequences of matrices $R$, $L$,
$X_{Z_i}$, and $F$; $G_1$ and $G_2$ must correspond
to closed geodesic lines, but are otherwise arbitrary.

Direct calculations then give
\be
\label{Goldman}
\{G_1,G_2\}=\frac{1}{2}(G_{AB^{-1}}-G_{AB}),
\ee
in the notation of (\ref{skeinclass}) where we set $G_1=G_A$ and $G_2=G_B$.
Then $G_{AB^{-1}}$ corresponds to the geodesic that
passes over the edge~$Z$ twice, so it has the form
$\tr \dots X_CR_ZR_D\dots$ $\dots X_BL_ZL_A\dots$.
These relations were first obtained
by Goldman~\cite{Gold} in the continuous parametrization (the classical Turaev--Viro
algebra).

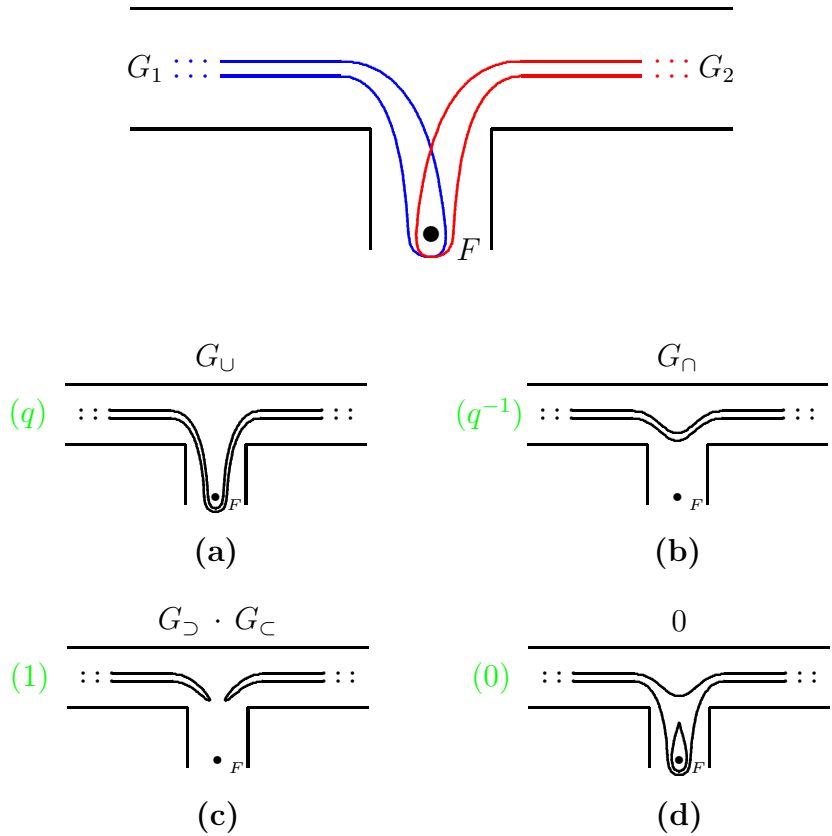
\begin{figure}[tb]
\setlength{\unitlength}{1.0mm}%
\begin{picture}(80,110)(-70,-110)
\thicklines
\put(-40,0){\line(1,0){80}}
\put(-40,-16){\line(1,0){32}}
\put(8,-16){\line(1,0){32}}
\put(-8,-16){\line(0,-1){16}}
\put(8,-16){\line(0,-1){16}}
\put(0,-30){\circle*{2}}
\put(5,-32){\makebox(0,0){$F$}}
\put(-38,-8){\makebox(0,0){$G_1$}}
\put(38,-8){\makebox(0,0){$G_2$}}
\color[rgb]{0,0,1}
\qbezier(-3,-30)(-4,-9)(-12,-9)
\qbezier(2,-30)(0,-7)(-12,-7)
\put(-32,-9.1){\makebox(0,0){$\cdots$}}
\put(-32,-7.1){\makebox(0,0){$\cdots$}}
\put(-12,-9){\line(-1,0){16}}
\put(-28,-7){\line(1,0){16}}
\qbezier(-3,-30)(-2.8,-33.5)(0.6,-33)
\qbezier(2,-30)(2,-32.8)(0.6,-33)
\color[rgb]{1,0,0}
\qbezier(3,-30)(4,-9)(12,-9)
\qbezier(-2,-30)(0,-7)(12,-7)
\put(32,-9.1){\makebox(0,0){$\cdots$}}
\put(32,-7.1){\makebox(0,0){$\cdots$}}
\put(12,-9){\line(1,0){16}}
\put(28,-7){\line(-1,0){16}}
\qbezier(3,-30)(2.8,-33.5)(-0.6,-33)
\qbezier(-2,-30)(-2,-32.8)(-0.6,-33)
\end{picture}
\setlength{\unitlength}{.5mm}%
\begin{picture}(0,0)(80,-120)
\thicklines
\put(-40,0){\line(1,0){80}}
\put(-40,-16){\line(1,0){32}}
\put(8,-16){\line(1,0){32}}
\put(-8,-16){\line(0,-1){16}}
\put(8,-16){\line(0,-1){16}}
\put(0,-30){\circle*{2}}
\put(5,-32){\makebox(0,0){\tiny $F$}}
\qbezier(-3,-30)(-4,-9)(-12,-9)
\qbezier(-2,-30)(-3,-7)(-12,-7)
\put(-32,-9.1){\makebox(0,0){$\cdots$}}
\put(-32,-7.1){\makebox(0,0){$\cdots$}}
\put(-12,-9){\line(-1,0){16}}
\put(-28,-7){\line(1,0){16}}
\qbezier(3,-30)(2.8,-34)(0,-34)
\qbezier(2,-30)(1.8,-33)(0,-33)
\qbezier(3,-30)(4,-9)(12,-9)
\qbezier(2,-30)(3,-7)(12,-7)
\put(32,-9.1){\makebox(0,0){$\cdots$}}
\put(32,-7.1){\makebox(0,0){$\cdots$}}
\put(12,-9){\line(1,0){16}}
\put(28,-7){\line(-1,0){16}}
\qbezier(-3,-30)(-2.8,-34)(0,-34)
\qbezier(-2,-30)(-1.8,-33)(0,-33)
\put(0,7){\makebox(0,0){$G_{\cup}$}}
\put(0,-45){\makebox(0,0){\bf (a)}}
\color[rgb]{0,1,0}
\put(-50,-8){\makebox(0,0){$(q)$}}
\end{picture}
\begin{picture}(0,0)(-40,-120)
\thicklines
\put(-40,0){\line(1,0){80}}
\put(-40,-16){\line(1,0){32}}
\put(8,-16){\line(1,0){32}}
\put(-8,-16){\line(0,-1){16}}
\put(8,-16){\line(0,-1){16}}
\put(0,-30){\circle*{2}}
\put(5,-32){\makebox(0,0){\tiny $F$}}
\qbezier(-4,-13)(-8,-9)(-12,-9)
\qbezier(-4,-11)(-8,-7)(-12,-7)
\qbezier(4,-13)(8,-9)(12,-9)
\qbezier(4,-11)(8,-7)(12,-7)
\qbezier(-4,-13)(0,-17)(4,-13)
\qbezier(-4,-11)(0,-15)(4,-11)
\put(-32,-9.1){\makebox(0,0){$\cdots$}}
\put(-32,-7.1){\makebox(0,0){$\cdots$}}
\put(-12,-9){\line(-1,0){16}}
\put(-28,-7){\line(1,0){16}}
\put(32,-9.1){\makebox(0,0){$\cdots$}}
\put(32,-7.1){\makebox(0,0){$\cdots$}}
\put(12,-9){\line(1,0){16}}
\put(28,-7){\line(-1,0){16}}
\put(0,7){\makebox(0,0){$G_{\cap}$}}
\put(0,-45){\makebox(0,0){\bf (b)}}
\color[rgb]{0,1,0}
\put(-50,-8){\makebox(0,0){$(q^{-1})$}}
\end{picture}
\begin{picture}(0,0)(85,-50)
\thicklines
\put(-40,0){\line(1,0){80}}
\put(-40,-16){\line(1,0){32}}
\put(8,-16){\line(1,0){32}}
\put(-8,-16){\line(0,-1){16}}
\put(8,-16){\line(0,-1){16}}
\put(0,-30){\circle*{2}}
\put(5,-32){\makebox(0,0){\tiny $F$}}
\qbezier(-4,-13)(-8,-9)(-12,-9)
\qbezier(-4,-11)(-8,-7)(-12,-7)
\qbezier(4,-13)(8,-9)(12,-9)
\qbezier(4,-11)(8,-7)(12,-7)
\qbezier(-4,-13)(0,-16)(-4,-11)
\qbezier(4,-13)(0,-16)(4,-11)
\put(-32,-9.1){\makebox(0,0){$\cdots$}}
\put(-32,-7.1){\makebox(0,0){$\cdots$}}
\put(-12,-9){\line(-1,0){16}}
\put(-28,-7){\line(1,0){16}}
\put(32,-9.1){\makebox(0,0){$\cdots$}}
\put(32,-7.1){\makebox(0,0){$\cdots$}}
\put(12,-9){\line(1,0){16}}
\put(28,-7){\line(-1,0){16}}
\put(0,7){\makebox(0,0){$G_{\supset}\,\cdot\,G_{\subset}$}}
\put(0,-45){\makebox(0,0){\bf (c)}}
\color[rgb]{0,1,0}
\put(-50,-8){\makebox(0,0){$(1)$}}
\end{picture}
\begin{picture}(0,0)(-35,-50)
\thicklines
\put(-40,0){\line(1,0){80}}
\put(-40,-16){\line(1,0){32}}
\put(8,-16){\line(1,0){32}}
\put(-8,-16){\line(0,-1){16}}
\put(8,-16){\line(0,-1){16}}
\put(0,-30){\circle*{2}}
\put(5,-32){\makebox(0,0){\tiny $F$}}
\qbezier(-4,-11)(-8,-7)(-12,-7)
\qbezier(4,-11)(8,-7)(12,-7)
\qbezier(-4,-11)(0,-15)(4,-11)
\qbezier(-3,-30)(-4,-9)(-12,-9)
\qbezier(3,-30)(2.8,-34)(0,-34)
\qbezier(3,-30)(4,-9)(12,-9)
\qbezier(-3,-30)(-2.8,-34)(0,-34)
\qbezier(-2,-30)(-1.8,-33)(0,-33)
\qbezier(2,-30)(1.8,-33)(0,-33)
\qbezier(-2,-30)(-2.2,-27)(0,-20)
\qbezier(2,-30)(2.2,-27)(0,-20)
\put(-32,-9.1){\makebox(0,0){$\cdots$}}
\put(-32,-7.1){\makebox(0,0){$\cdots$}}
\put(-12,-9){\line(-1,0){16}}
\put(-28,-7){\line(1,0){16}}
\put(32,-9.1){\makebox(0,0){$\cdots$}}
\put(32,-7.1){\makebox(0,0){$\cdots$}}
\put(12,-9){\line(1,0){16}}
\put(28,-7){\line(-1,0){16}}
\put(0,-45){\makebox(0,0){\bf (d)}}
\put(0,7){\makebox(0,0){$0$}}
\color[rgb]{0,1,0}
\put(-50,-8){\makebox(0,0){$(0)$}}
\end{picture}
\caption{\small An example of two geodesic lines intersecting at the
dot-vertex. We present four homotopical types of resolving two
intersections in this pattern (Cases (a)--(d)). A multicurve in
Case (d) contains the loop with only the dot-vertex inside.
This loop is $\tr F=0$, so the whole contribution vanishes in this case. The factors
in brackets pertain to the quantum case in Sec.~\ref{s:q} indicating the weights with
which the corresponding (quantum) geodesic multicurves enter the expression for the quantum
operatorial product $G_1^\hbar G_2^\hbar$.}
\label{fi:dot-skein}
\end{figure}

Given two curves, $\gamma_1$ and $\gamma_2$, with an arbitrary number of crossings,
we now find their Poisson bracket using the following rules:
\begin{itemize}
\item We take a sum of products of geodesic functions
of non(self)intersecting curves obtained when we apply Poisson relation (\ref{Goldman})
at one intersection point and classical skein relation (\ref{skeinclass})
at all the remaining points of intersection; we assume the summation over
all possible cases.
\item If, in the course of calculation, we meet an empty (contractible) loop, then
we associate the factor $-2$ to such a loop; this assignment, as is know~\cite{ChP},
ensures the Reidemeister moves on the set of geodesic lines thus making the bracket
to depend only on the homotopical class of the curve embedding in the surface.
\item If, in the course of calculation, we meet a curve homeomorphic to passing around a
dot-vertex, then we set $\tr F=0$ thus killing the whole corresponding
multicurve function.
\end{itemize}

All these rules are equivalently applicable to the intersections of the geodesic lines depicted
in Fig.~\ref{fi:dot-skein}; the result is presented in the figure.
In obvious notation, $\{G_1,G_2\}=G_\cup-G_\cap$.

Because the Poisson relations are completely determined by homotopy types of curves,
using Lemma~\ref{lem-pending}, we immediately come to the theorem below.

\begin{theorem}\label{th-mark}
Poisson algebras of geodesic functions for two orbifold Riemann surfaces
$\Sigma_{g,s,\{\delta_1,\dots,\delta_s\}}$
and $\Sigma_{g,s,\{\tilde\delta_1,\dots,\tilde\delta_s\}}$ with $\sum_{k=1}^s |\delta_k|=
\sum_{k=1}^s |\tilde\delta_k|$
are isomorphic; the isomorphism is described by Lemma~\ref{lem-pending}.
\end{theorem}

In particular,
it follows from this theorem that we can always collect all the marked points on just one boundary
component.

\newsection{Poisson and braid-group relations for $A_n$ and $D_n$}\label{s:braid}

\subsection{Special algebras of geodesic functions}

We now demonstrate two cases where it is possible to close the Poisson geodesic algebras
on the level of finite number of elements.\footnote{Besides these two cases among which the
case of $A_n$ algebra is equivalent to the Nelson--Regge algebra,
the only other case where it is possible is the case of sphere with four holes.}
In all examples below, we can attain such a closure for a price of introducing nonlinear terms
in the right-hand sides of Poisson relations.

\subsubsection{The $A_n$ algebras}\label{ss:An}

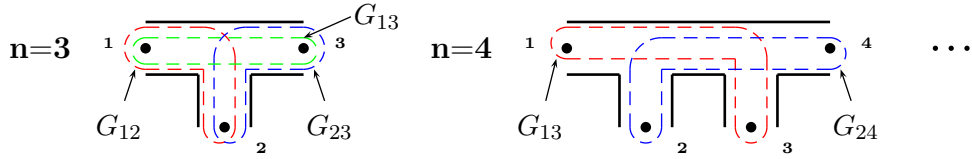
\begin{figure}[tb]
{\psset{unit=0.7}
\begin{pspicture}(-10,-3)(7,1)
\rput(-6,0){\makebox(0,0){$\mathbf{n{=}3}$}}
\pcline[linewidth=1pt](-4,0.5)(-1,0.5)
\pcline[linewidth=1pt](-4,-0.5)(-3,-0.5)
\pcline[linewidth=1pt](-2,-0.5)(-1,-0.5)
\pcline[linewidth=1pt](-3,-0.5)(-3,-1.5)
\pcline[linewidth=1pt](-2,-0.5)(-2,-1.5)
\pscircle*(-4,0){0.1}
\pscircle*(-1,0){0.1}
\pscircle*(-2.5,-1.5){0.1}
\psarc[linecolor=red, linestyle=dashed, linewidth=0.5pt](-4,0){.4}{90}{270}
\psarc[linecolor=red, linestyle=dashed, linewidth=0.5pt](-2.6,-1.5){.3}{180}{0}
\pcline[linecolor=red, linestyle=dashed, linewidth=0.5pt](-2.9,-1.5)(-2.9,-0.4)
\pcline[linecolor=red, linestyle=dashed, linewidth=0.5pt](-2.9,-0.4)(-4,-0.4)
\pcline[linecolor=red, linestyle=dashed, linewidth=0.5pt](-2.3,-1.5)(-2.3,-0.2)
\pcline[linecolor=red, linestyle=dashed, linewidth=0.5pt](-2.9,0.4)(-4,0.4)
\psarc[linecolor=red, linestyle=dashed, linewidth=0.5pt](-2.9,-0.2){.6}{0}{90}
\psarc[linecolor=blue, linestyle=dashed, linewidth=0.5pt](-1,0){.4}{-90}{90}
\psarc[linecolor=blue, linestyle=dashed, linewidth=0.5pt](-2.4,-1.5){.3}{180}{0}
\pcline[linecolor=blue, linestyle=dashed, linewidth=0.5pt](-2.1,-1.5)(-2.1,-0.4)
\pcline[linecolor=blue, linestyle=dashed, linewidth=0.5pt](-2.1,-0.4)(-1,-0.4)
\pcline[linecolor=blue, linestyle=dashed, linewidth=0.5pt](-2.7,-1.5)(-2.7,-0.2)
\pcline[linecolor=blue, linestyle=dashed, linewidth=0.5pt](-2.1,0.4)(-1,0.4)
\psarc[linecolor=blue, linestyle=dashed, linewidth=0.5pt](-2.1,-.2){.6}{90}{180}
\psarc[linecolor=green, linestyle=dashed, linewidth=0.5pt](-4,-0.05){.25}{90}{270}
\psarc[linecolor=green, linestyle=dashed, linewidth=0.5pt](-1,-.05){.25}{-90}{90}
\pcline[linecolor=green, linestyle=dashed, linewidth=0.5pt](-4,.2)(-1,.2)
\pcline[linecolor=green, linestyle=dashed, linewidth=0.5pt](-4,-0.3)(-1,-0.3)
\rput(-4.7,0.1){\makebox(0,0){\tiny$\mathbf1$}}
\rput(-1.8,-1.9){\makebox(0,0){\tiny$\mathbf2$}}
\rput(-0.3,0.1){\makebox(0,0){\tiny$\mathbf3$}}
\pcline[linewidth=0.5pt]{->}(-4.4,-1)(-4.2,-0.4)
\rput(-4.5,-1.5){\makebox(0,0){\small$G_{12}$}}
\pcline[linewidth=0.5pt]{->}(-0.6,-1)(-0.8,-0.4)
\rput(-.5,-1.5){\makebox(0,0){\small$G_{23}$}}
\pcline[linewidth=0.5pt]{->}(-0.1,0.6)(-1,0.2)
\rput(0.4,.6){\makebox(0,0){\small$G_{13}$}}
\rput(2,0){\makebox(0,0){$\mathbf{n{=}4}$}}
\pcline[linewidth=1pt](4,0.5)(9,0.5)
\pcline[linewidth=1pt](4,-0.5)(5,-0.5)
\pcline[linewidth=1pt](6,-0.5)(7,-0.5)
\pcline[linewidth=1pt](8,-0.5)(9,-0.5)
\pcline[linewidth=1pt](5,-0.5)(5,-1.5)
\pcline[linewidth=1pt](6,-0.5)(6,-1.5)
\pcline[linewidth=1pt](7,-0.5)(7,-1.5)
\pcline[linewidth=1pt](8,-0.5)(8,-1.5)
\pscircle*(4,0){0.1}
\pscircle*(9,0){0.1}
\pscircle*(5.5,-1.5){0.1}
\pscircle*(7.5,-1.5){0.1}
\psarc[linecolor=red, linestyle=dashed, linewidth=0.5pt](4,0.1){.3}{90}{270}
\psarc[linecolor=red, linestyle=dashed, linewidth=0.5pt](7.5,-1.5){.3}{180}{0}
\pcline[linecolor=red, linestyle=dashed, linewidth=0.5pt](7.2,-1.5)(7.2,-0.2)
\pcline[linecolor=red, linestyle=dashed, linewidth=0.5pt](7.8,-1.5)(7.8,-0.2)
\pcline[linecolor=red, linestyle=dashed, linewidth=0.5pt](4,0.4)(7.2,0.4)
\pcline[linecolor=red, linestyle=dashed, linewidth=0.5pt](4,-0.2)(7.2,-0.2)
\psarc[linecolor=red, linestyle=dashed, linewidth=0.5pt](7.2,-0.2){.6}{0}{90}
\psarc[linecolor=blue, linestyle=dashed, linewidth=0.5pt](9,-0.1){.3}{-90}{90}
\psarc[linecolor=blue, linestyle=dashed, linewidth=0.5pt](5.5,-1.5){.3}{180}{0}
\pcline[linecolor=blue, linestyle=dashed, linewidth=0.5pt](5.2,-1.5)(5.2,-0.4)
\pcline[linecolor=blue, linestyle=dashed, linewidth=0.5pt](5.8,-1.5)(5.8,-0.4)
\pcline[linecolor=blue, linestyle=dashed, linewidth=0.5pt](5.8,-0.4)(9,-0.4)
\pcline[linecolor=blue, linestyle=dashed, linewidth=0.5pt](5.8,0.2)(9,0.2)
\psarc[linecolor=blue, linestyle=dashed, linewidth=0.5pt](5.8,-.4){.6}{90}{180}
\rput(3.3,0.1){\makebox(0,0){\tiny$\mathbf1$}}
\rput(6.2,-1.9){\makebox(0,0){\tiny$\mathbf2$}}
\rput(8.2,-1.9){\makebox(0,0){\tiny$\mathbf3$}}
\rput(9.7,0.1){\makebox(0,0){\tiny$\mathbf4$}}
\pcline[linewidth=0.5pt]{->}(3.6,-1)(3.8,-0.2)
\rput(3.5,-1.5){\makebox(0,0){\small$G_{13}$}}
\pcline[linewidth=0.5pt]{->}(9.4,-1)(9.2,-0.4)
\rput(9.5,-1.5){\makebox(0,0){\small$G_{24}$}}
\pscircle*(11,0){0.05}
\pscircle*(11.3,0){0.05}
\pscircle*(11.6,0){0.05}
\end{pspicture}
}
\caption{\small Generating graphs for $A_n$ algebras for $n=3,4,\dots$. We indicate character
geodesics whose geodesic functions $G_{ij}$ enter bases of the corresponding algebras.}
\label{fi:An}
\end{figure}

Consider the disc with $n$ orbifold points; examples of the corresponding
representing graph $\Gamma_n$ are depicted in Fig.~\ref{fi:An} for $n=3,4,\dots$.
We enumerate the $n$ dot-vertices clockwise, $i,j=1,\dots, n$.
We then let $G_{ij}$ with $i<j$ denote the geodesic function corresponding to the geodesic
line that encircles exactly two dot-vertices
with the indices $i$ and $j$. The corresponding geodesic line is the line connecting the
preimages $s_i$ and $s_j$ in Fig.~\ref{fi:disc} and the distance is double the
geodesic distance between these points. Three algebraic functions,
$G_{12}$, $G_{23}$, and $G_{13}$, correspond to the lines in Fig.~\ref{fi:An}.

Using the skein relation, we can
close the Poisson algebra thus obtaining for $A_3$:
\be
\label{alg-A3}
\left\{G_{12},G_{23}\right\}=G_{12}G_{23}-2G_{13}\hbox{ and cycl. permut.}
\ee
In higher-order algebras (starting with $n=4$), we meet
a more complicate case of the fourth-order crossing (as shown in the case $n=4$ in Fig.~\ref{fi:An}).
The corresponding Poisson brackets are
\be
\label{alg-A4}
\left\{G_{13},G_{24}\right\}=2G_{12}G_{34}-2G_{14}G_{23}
\ee
(note that the r.h.s. is a linear combination of multicurves).

Algebraic relations (\ref{alg-A3}) and (\ref{alg-A4}) are exactly the Nelson--Regge~\cite{NR} relations
in algebras of geodesics on Riemann surfaces with one and two holes~\cite{ChF2}, and we can use the
well-developed machinery for dealing with these algebras.
These algebras also appear in the Frobenius manifold approach~\cite{DM}.

\subsubsection{The $D_n$-algebras}

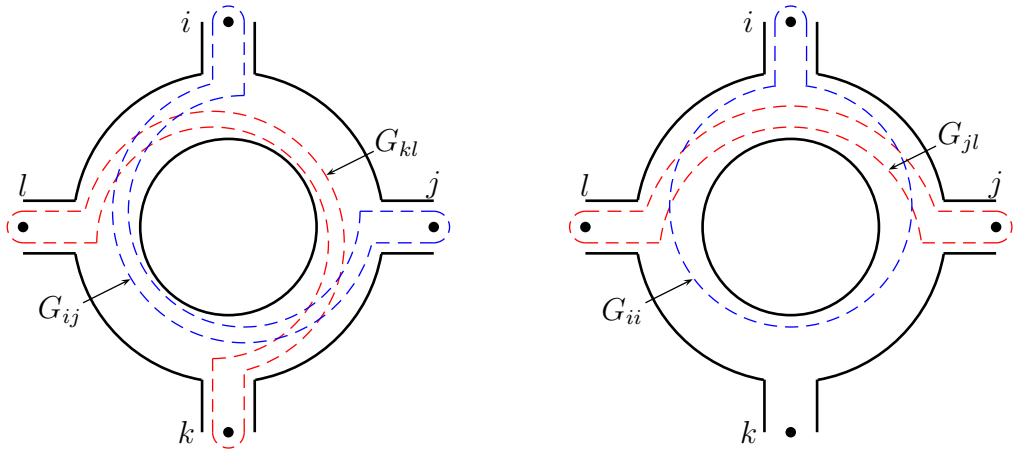
\begin{figure}[tb]
{\psset{unit=0.7}
\begin{pspicture}(-6,-4)(7,4)
\psarc[linewidth=1pt](0,0){2.95}{-80}{-10}
\psarc[linewidth=1pt](0,0){2.95}{10}{80}
\psarc[linewidth=1pt](0,0){2.95}{100}{170}
\psarc[linewidth=1pt](0,0){2.95}{190}{260}
\pscircle[linewidth=1pt](0,0){1.7}
\pcline[linewidth=1pt](-0.5,-2.9)(-0.5,-3.9)
\pcline[linewidth=1pt](0.5,-2.9)(0.5,-3.9)
\pcline[linewidth=1pt](-0.5,2.9)(-0.5,3.9)
\pcline[linewidth=1pt](0.5,2.9)(0.5,3.9)
\pcline[linewidth=1pt](-2.9,-0.5)(-3.9,-0.5)
\pcline[linewidth=1pt](-2.9,0.5)(-3.9,0.5)
\pcline[linewidth=1pt](2.9,-0.5)(3.9,-0.5)
\pcline[linewidth=1pt](2.9,0.5)(3.9,0.5)
\psarc[linecolor=red, linestyle=dashed, linewidth=0.5pt](0,-3.9){.3}{180}{360}
\pscircle*[linewidth=0.5pt](0,-3.9){.1}
\pcline[linecolor=red, linestyle=dashed, linewidth=0.5pt](-0.3,-3.9)(-0.3,-2.5)
\pcline[linecolor=red, linestyle=dashed, linewidth=0.5pt](0.3,-3.9)(0.3,-2.8)
\psarc[linecolor=red, linestyle=dashed, linewidth=0.5pt](-0.3,-0.3){2.2}{-90}{180}
\psarc[linecolor=red, linestyle=dashed, linewidth=0.5pt](-0.3,-0.3){2.5}{-75}{165}
\psarc[linecolor=blue, linestyle=dashed, linewidth=0.5pt](0,3.9){.3}{0}{180}
\pscircle*[linewidth=0.5pt](0,3.9){.1}
\pcline[linecolor=blue, linestyle=dashed, linewidth=0.5pt](-0.3,3.9)(-0.3,2.8)
\pcline[linecolor=blue, linestyle=dashed, linewidth=0.5pt](0.3,3.9)(0.3,2.5)
\psarc[linecolor=blue, linestyle=dashed, linewidth=0.5pt](0.3,0.3){2.2}{90}{360}
\psarc[linecolor=blue, linestyle=dashed, linewidth=0.5pt](0.3,0.3){2.5}{105}{345}
\psarc[linecolor=red, linestyle=dashed, linewidth=0.5pt](-3.9,0){.3}{90}{270}
\pscircle*[linewidth=0.5pt](-3.9,0){.1}
\pcline[linecolor=red, linestyle=dashed, linewidth=0.5pt](-3.9,0.3)(-2.8,0.3)
\pcline[linecolor=red, linestyle=dashed, linewidth=0.5pt](-3.9,-0.3)(-2.5,-0.3)
\psarc[linecolor=blue, linestyle=dashed, linewidth=0.5pt](3.9,0){.3}{-90}{90}
\pscircle*[linewidth=0.5pt](3.9,0){.1}
\pcline[linecolor=blue, linestyle=dashed, linewidth=0.5pt](3.9,0.3)(2.5,0.3)
\pcline[linecolor=blue, linestyle=dashed, linewidth=0.5pt](3.9,-0.3)(2.8,-0.3)
\rput(3.2,1.6){\makebox(0,0){$G_{kl}$}}
\rput(-3.2,-1.6){\makebox(0,0){$G_{ij}$}}
\pcline[linewidth=0.5pt]{->}(2.76,1.44)(1.86,0.99)
\pcline[linewidth=0.5pt]{->}(-2.76,-1.44)(-1.86,-0.99)
\rput(-3.9,0.8){\makebox(0,0){$l$}}
\rput(-0.8,3.9){\makebox(0,0){$i$}}
\rput(3.9,0.8){\makebox(0,0){$j$}}
\rput(-0.8,-3.9){\makebox(0,0){$k$}}
\end{pspicture}
\begin{pspicture}(-3.5,-4)(6,0)
\psarc[linewidth=1pt](0,0){2.95}{-80}{-10}
\psarc[linewidth=1pt](0,0){2.95}{10}{80}
\psarc[linewidth=1pt](0,0){2.95}{100}{170}
\psarc[linewidth=1pt](0,0){2.95}{190}{260}
\pscircle[linewidth=1pt](0,0){1.7}
\pcline[linewidth=1pt](-0.5,-2.9)(-0.5,-3.9)
\pcline[linewidth=1pt](0.5,-2.9)(0.5,-3.9)
\pcline[linewidth=1pt](-0.5,2.9)(-0.5,3.9)
\pcline[linewidth=1pt](0.5,2.9)(0.5,3.9)
\pcline[linewidth=1pt](-2.9,-0.5)(-3.9,-0.5)
\pcline[linewidth=1pt](-2.9,0.5)(-3.9,0.5)
\pcline[linewidth=1pt](2.9,-0.5)(3.9,-0.5)
\pcline[linewidth=1pt](2.9,0.5)(3.9,0.5)
%
\pscircle*[linewidth=0.5pt](0,-3.9){.1}
\psarc[linecolor=red, linestyle=dashed, linewidth=0.5pt](0,-0.6){2.5}{8}{172}
\psarc[linecolor=red, linestyle=dashed, linewidth=0.5pt](0,-0.6){2.9}{20}{160}
\psarc[linecolor=blue, linestyle=dashed, linewidth=0.5pt](0,3.9){.3}{0}{180}
\pscircle*[linewidth=0.5pt](0,3.9){.1}
\pcline[linecolor=blue, linestyle=dashed, linewidth=0.5pt](-0.3,3.9)(-0.3,2.7)
\pcline[linecolor=blue, linestyle=dashed, linewidth=0.5pt](0.3,3.9)(0.3,2.7)
\psarc[linecolor=blue, linestyle=dashed, linewidth=0.5pt](0,0.4){2.3}{97}{83}
\psarc[linecolor=red, linestyle=dashed, linewidth=0.5pt](-3.9,0){.3}{90}{270}
\pscircle*[linewidth=0.5pt](-3.9,0){.1}
\pcline[linecolor=red, linestyle=dashed, linewidth=0.5pt](-3.9,0.3)(-2.8,0.3)
\pcline[linecolor=red, linestyle=dashed, linewidth=0.5pt](-3.9,-0.3)(-2.5,-0.3)
\psarc[linecolor=red, linestyle=dashed, linewidth=0.5pt](3.9,0){.3}{-90}{90}
\pscircle*[linewidth=0.5pt](3.9,0){.1}
\pcline[linecolor=red, linestyle=dashed, linewidth=0.5pt](3.9,0.3)(2.8,0.3)
\pcline[linecolor=red, linestyle=dashed, linewidth=0.5pt](3.9,-0.3)(2.5,-0.3)
\rput(3.2,1.7){\makebox(0,0){$G_{jl}$}}
\rput(-3.2,-1.6){\makebox(0,0){$G_{ii}$}}
\pcline[linewidth=0.5pt]{->}(2.76,1.54)(1.86,1.09)
\pcline[linewidth=0.5pt]{->}(-2.76,-1.44)(-1.86,-0.99)
\rput(-3.9,0.8){\makebox(0,0){$l$}}
\rput(-0.8,3.9){\makebox(0,0){$i$}}
\rput(3.9,0.8){\makebox(0,0){$j$}}
\rput(-0.8,-3.9){\makebox(0,0){$k$}}
\end{pspicture}
}
\caption{\small Typical geodesics corresponding to the geodesic functions
constituting a set of generators of the $D_n$ algebra. We let $G_{ij}$,
$i,j=1,\dots,n$, denote these functions. The order of subscripts
indicates the direction of encompassing the hole (the second boundary component of
the annulus). The most involved pattern of intersection is on the left part of the figure:
the geodesics have there eight-fold intersection; in the right part we present also the
geodesic function $G_{ii}$ corresponding to the geodesic that encircles exactly one orbifold point
and the central hole.}
\label{fi:Dn}
\end{figure}

We now consider the case of annulus with $n$ orbifold points associated to
one of the boundary component (see the example in Fig.~\ref{fi:center}.
In this case, a finite set of geodesic functions closed w.r.t. the Poisson brackets
is given by geodesic functions corresponding to geodesics in Fig.~\ref{fi:Dn}

We therefore describe a set of geodesic functions by the matrix $G_{ij}$ with $i,j=1,\dots, n$ where
the order of indices indicates the direction of encompassing the second boundary component of the annulus.
The corresponding geodesics for $i\ne j$ connect pairwise the points $s_i$ and $s_j$ encompassing the
``central'' hole from one or another side. Counterintuitively, the geodesic function $G_{ii}$
does not correspond to
the geodesic line that starts and terminates at the point $s_i$ encompassing the hole; instead we must take
a smooth (closed) geodesic line that encircles the hole and the given orbifold point.

\begin{lemma} \label{lem-Dn-Poisson}
The set of geodesic functions $G_{ij}$ corresponding to geodesics in Fig.~\ref{fi:Dn} is
Poisson closed.
\end{lemma}

We present below all the nontrivial Poisson relations in the graphical way; because of the rotational
symmetry, only the cyclic order w.r.t. the central hole matters; the starting vertex (with the number one)
can be any among the dot-vertices around the hole; the Roman numbers I and II indicate which geodesic
function occupies the first and which---the second place in the corresponding Poisson bracket.

We have six basic nontrivial brackets (ordered by increasing complexity)

\be
{\psset{unit=0.4}
\begin{pspicture}(-7,-5)(7,3)
\newcommand{\MOLD}{%
\rput(-2,-2){\pscircle*{0.1}}
\rput(-2,2){\pscircle*{0.1}}
\rput(2,-2){\pscircle*{0.1}}
\rput(2,2){\pscircle*{0.1}}
\pscircle[fillstyle=solid, fillcolor=gray](0,0){0.5}
}
\newcommand{\GSHORT}{%
\psarc[linewidth=0.5pt](-2,-2){.5}{120}{300}
\psarc[linewidth=0.5pt](2,-2){.5}{-120}{60}
\psbezier[linewidth=0.5pt](-2.25,-1.567)(-1.817,-1.317)(1.817,-1.317)(2.25,-1.567)
\psbezier[linewidth=0.5pt](-1.75,-2.433)(-1.317,-2.183)(1.317,-2.183)(1.75,-2.433)
}
\newcommand{\GMED}{%
\psarc[linewidth=0.5pt](-2,-2){.5}{90}{270}
\psarc[linewidth=0.5pt](2,2){.5}{0}{180}
\psbezier[linewidth=0.5pt](-2,-1.5)(.3,-1.5)(1.5,-.3)(1.5,2)
\psbezier[linewidth=0.5pt](-2,-2.5)(1,-2.5)(2.5,-1)(2.5,2)
}
\newcommand{\GLONG}{%
\psarc[linewidth=0.5pt](-2,-2){.5}{90}{270}
\psarc[linewidth=0.5pt](-2,2){.5}{90}{270}
\psbezier[linewidth=0.5pt](-2,-1.5)(1.5,-1.5)(1.5,1.5)(-2,1.5)
\psbezier[linewidth=0.5pt](-2,-2.5)(2.1,-2.5)(2.1,2.5)(-2,2.5)
}
\rput(-14,2.5){\makebox(0,0){\bf a.}}
\rput(-11,2){\makebox(0,0){I}}
\rput(-5,2){\makebox(0,0){II}}
\rput(-8,0){\MOLD}
\rput(-8,0){\GMED}
\rput{270}(-8,0){\GMED}
\rput(-4,0){\makebox(0,0){$=2$}}
\rput(0,0){\MOLD}
\rput{90}(0,0){\GSHORT}
\rput{270}(0,0){\GSHORT}
\rput(4,0){\makebox(0,0){$-2$}}
\rput(8,0){\MOLD}
\rput(8,0){\GSHORT}
\rput{270}(8,0){\GLONG}
\end{pspicture}
}
\ee

\be
{\psset{unit=0.4}
\begin{pspicture}(-7,-5)(7,3)
\newcommand{\MOLD}{%
\rput(-2,-2){\pscircle*{0.1}}
\rput(-2,2){\pscircle*{0.1}}
\rput(2,-2){\pscircle*{0.1}}
\rput(2,2){\pscircle*{0.1}}
\pscircle[fillstyle=solid, fillcolor=gray](0,0){0.5}
}
\newcommand{\GSHORT}{%
\psarc[linewidth=0.5pt](-2,-2){.5}{120}{300}
\psarc[linewidth=0.5pt](2,-2){.5}{-120}{60}
\psbezier[linewidth=0.5pt](-2.25,-1.567)(-1.817,-1.317)(1.817,-1.317)(2.25,-1.567)
\psbezier[linewidth=0.5pt](-1.75,-2.433)(-1.317,-2.183)(1.317,-2.183)(1.75,-2.433)
}
\newcommand{\GMED}{%
\psarc[linewidth=0.5pt](-2,-2){.5}{90}{270}
\psarc[linewidth=0.5pt](2,2){.5}{0}{180}
\psbezier[linewidth=0.5pt](-2,-1.5)(.3,-1.5)(1.5,-.3)(1.5,2)
\psbezier[linewidth=0.5pt](-2,-2.5)(1,-2.5)(2.5,-1)(2.5,2)
}
\newcommand{\GLONG}{%
\psarc[linewidth=0.5pt](-2,-2){.5}{90}{270}
\psarc[linewidth=0.5pt](-2,2){.5}{90}{270}
\psbezier[linewidth=0.5pt](-2,-1.5)(1.5,-1.5)(1.5,1.5)(-2,1.5)
\psbezier[linewidth=0.5pt](-2,-2.5)(2.1,-2.5)(2.1,2.5)(-2,2.5)
}
\rput(-13,2.5){\makebox(0,0)[rb]{reduction ${\mathbf a}_1$}}
\rput(-11,2){\makebox(0,0){I}}
\rput(-5,2){\makebox(0,0){II}}
\rput(-8,0){\MOLD}
\rput(-8,0){\GMED}
\rput{270}(-8,0){\GSHORT}
\rput(-4,0){\makebox(0,0){$=$}}
\rput(0,0){\MOLD}
\rput(0,0){\GMED}
\rput{270}(0,0){\GSHORT}
\rput(4,0){\makebox(0,0){$-2$}}
\rput(8,0){\MOLD}
\rput{270}(8,0){\GLONG}
\end{pspicture}
}
\nonumber
\ee

\be
{\psset{unit=0.4}
\begin{pspicture}(-7,-5)(7,3)
\newcommand{\MOLD}{%
\rput(-2,-2){\pscircle*{0.1}}
\rput(-2,2){\pscircle*{0.1}}
\rput(2,-2){\pscircle*{0.1}}
\rput(2,2){\pscircle*{0.1}}
\pscircle[fillstyle=solid, fillcolor=gray](0,0){0.5}
}
\newcommand{\GSHORT}{%
\psarc[linewidth=0.5pt](-2,-2){.5}{120}{300}
\psarc[linewidth=0.5pt](2,-2){.5}{-120}{60}
\psbezier[linewidth=0.5pt](-2.25,-1.567)(-1.817,-1.317)(1.817,-1.317)(2.25,-1.567)
\psbezier[linewidth=0.5pt](-1.75,-2.433)(-1.317,-2.183)(1.317,-2.183)(1.75,-2.433)
}
\newcommand{\GMED}{%
\psarc[linewidth=0.5pt](-2,-2){.5}{90}{270}
\psarc[linewidth=0.5pt](2,2){.5}{0}{180}
\psbezier[linewidth=0.5pt](-2,-1.5)(.3,-1.5)(1.5,-.3)(1.5,2)
\psbezier[linewidth=0.5pt](-2,-2.5)(1,-2.5)(2.5,-1)(2.5,2)
}
\newcommand{\GLONG}{%
\psarc[linewidth=0.5pt](-2,-2){.5}{90}{270}
\psarc[linewidth=0.5pt](-2,2){.5}{90}{270}
\psbezier[linewidth=0.5pt](-2,-1.5)(1.5,-1.5)(1.5,1.5)(-2,1.5)
\psbezier[linewidth=0.5pt](-2,-2.5)(2.1,-2.5)(2.1,2.5)(-2,2.5)
}
\newcommand{\GEAR}{%
\psarc[linewidth=0.5pt](0,0){.7}{90}{270}
\psbezier[linewidth=0.5pt](0,0.7)(2.5,0.7)(3.3,0.5)(3.3,0)
\psbezier[linewidth=0.5pt](0,-0.7)(2.5,-0.7)(3.3,-0.5)(3.3,0)
}
\rput(-14,2.5){\makebox(0,0){\bf b.}}
\rput(-11,2){\makebox(0,0){I}}
\rput(-5,2){\makebox(0,0){II}}
\rput(-8,0){\MOLD}
\rput{45}(-8,0){\GEAR}
\rput{90}(-8,0){\GMED}
\rput(-4,0){\makebox(0,0){$=2$}}
\rput(0,0){\MOLD}
\rput{90}(0,0){\GSHORT}
\rput{135}(0,0){\GEAR}
\rput(4,0){\makebox(0,0){$-2$}}
\rput(8,0){\MOLD}
\rput{315}(8,0){\GEAR}
\rput{180}(8,0){\GSHORT}
\end{pspicture}
}
\ee

\be
{\psset{unit=0.4}
\begin{pspicture}(-7,-5)(7,3)
\newcommand{\MOLD}{%
\rput(-2,-2){\pscircle*{0.1}}
\rput(-2,2){\pscircle*{0.1}}
\rput(2,-2){\pscircle*{0.1}}
\rput(2,2){\pscircle*{0.1}}
\pscircle[fillstyle=solid, fillcolor=gray](0,0){0.5}
}
\newcommand{\GSHORT}{%
\psarc[linewidth=0.5pt](-2,-2){.5}{120}{300}
\psarc[linewidth=0.5pt](2,-2){.5}{-120}{60}
\psbezier[linewidth=0.5pt](-2.25,-1.567)(-1.817,-1.317)(1.817,-1.317)(2.25,-1.567)
\psbezier[linewidth=0.5pt](-1.75,-2.433)(-1.317,-2.183)(1.317,-2.183)(1.75,-2.433)
}
\newcommand{\GMED}{%
\psarc[linewidth=0.5pt](-2,-2){.5}{90}{270}
\psarc[linewidth=0.5pt](2,2){.5}{0}{180}
\psbezier[linewidth=0.5pt](-2,-1.5)(.3,-1.5)(1.5,-.3)(1.5,2)
\psbezier[linewidth=0.5pt](-2,-2.5)(1,-2.5)(2.5,-1)(2.5,2)
}
\newcommand{\GLONG}{%
\psarc[linewidth=0.5pt](-2,-2){.5}{90}{270}
\psarc[linewidth=0.5pt](-2,2){.5}{90}{270}
\psbezier[linewidth=0.5pt](-2,-1.5)(1.5,-1.5)(1.5,1.5)(-2,1.5)
\psbezier[linewidth=0.5pt](-2,-2.5)(2.1,-2.5)(2.1,2.5)(-2,2.5)
}
\newcommand{\GEAR}{%
\psarc[linewidth=0.5pt](0,0){.7}{90}{270}
\psbezier[linewidth=0.5pt](0,0.7)(2.5,0.7)(3.3,0.5)(3.3,0)
\psbezier[linewidth=0.5pt](0,-0.7)(2.5,-0.7)(3.3,-0.5)(3.3,0)
}
\rput(-13,2.5){\makebox(0,0)[rb]{reduction ${\mathbf b}_1$}}
\rput(-5,-2){\makebox(0,0){I}}
\rput(-10,0){\makebox(0,0){II}}
\rput(-8,0){\MOLD}
\rput{45}(-8,0){\GEAR}
\rput{90}(-8,0){\GSHORT}
\rput(-4,0){\makebox(0,0){$=$}}
\rput(0,0){\MOLD}
\rput{90}(0,0){\GSHORT}
\rput{45}(0,0){\GEAR}
\rput(4,0){\makebox(0,0){$-2$}}
\rput(8,0){\MOLD}
\rput{315}(8,0){\GEAR}
\end{pspicture}
}
\nonumber
\ee

\be
{\psset{unit=0.4}
\begin{pspicture}(-7,-5)(7,3)
\newcommand{\MOLD}{%
\rput(-2,-2){\pscircle*{0.1}}
\rput(-2,2){\pscircle*{0.1}}
\rput(2,-2){\pscircle*{0.1}}
\rput(2,2){\pscircle*{0.1}}
\pscircle[fillstyle=solid, fillcolor=gray](0,0){0.5}
}
\newcommand{\GSHORT}{%
\psarc[linewidth=0.5pt](-2,-2){.5}{120}{300}
\psarc[linewidth=0.5pt](2,-2){.5}{-120}{60}
\psbezier[linewidth=0.5pt](-2.25,-1.567)(-1.817,-1.317)(1.817,-1.317)(2.25,-1.567)
\psbezier[linewidth=0.5pt](-1.75,-2.433)(-1.317,-2.183)(1.317,-2.183)(1.75,-2.433)
}
\newcommand{\GMED}{%
\psarc[linewidth=0.5pt](-2,-2){.5}{90}{270}
\psarc[linewidth=0.5pt](2,2){.5}{0}{180}
\psbezier[linewidth=0.5pt](-2,-1.5)(.3,-1.5)(1.5,-.3)(1.5,2)
\psbezier[linewidth=0.5pt](-2,-2.5)(1,-2.5)(2.5,-1)(2.5,2)
}
\newcommand{\GLONG}{%
\psarc[linewidth=0.5pt](-2,-2){.5}{90}{270}
\psarc[linewidth=0.5pt](-2,2){.5}{90}{270}
\psbezier[linewidth=0.5pt](-2,-1.5)(1.5,-1.5)(1.5,1.5)(-2,1.5)
\psbezier[linewidth=0.5pt](-2,-2.5)(2.1,-2.5)(2.1,2.5)(-2,2.5)
}
\newcommand{\GEAR}{%
\psarc[linewidth=0.5pt](0,0){.7}{90}{270}
\psbezier[linewidth=0.5pt](0,0.7)(2.5,0.7)(3.3,0.5)(3.3,0)
\psbezier[linewidth=0.5pt](0,-0.7)(2.5,-0.7)(3.3,-0.5)(3.3,0)
}
\rput(-13,2.5){\makebox(0,0)[rb]{reduction ${\mathbf b}_2$}}
\rput(-11,2){\makebox(0,0){I}}
\rput(-8,-2){\makebox(0,0){II}}
\rput(-8,0){\MOLD}
\rput{45}(-8,0){\GEAR}
\rput{180}(-8,0){\GSHORT}
\rput(-4,0){\makebox(0,0){$=-$}}
\rput(0,0){\MOLD}
\rput{180}(0,0){\GSHORT}
\rput{45}(0,0){\GEAR}
\rput(4,0){\makebox(0,0){$+2$}}
\rput(8,0){\MOLD}
\rput{135}(8,0){\GEAR}
\end{pspicture}
}
\nonumber
\ee

\be
{\psset{unit=0.4}
\begin{pspicture}(-7,-5)(7,3)
\newcommand{\MOLD}{%
\rput(-2,-2){\pscircle*{0.1}}
\rput(-2,2){\pscircle*{0.1}}
\rput(2,-2){\pscircle*{0.1}}
\rput(2,2){\pscircle*{0.1}}
\pscircle[fillstyle=solid, fillcolor=gray](0,0){0.5}
}
\newcommand{\GSHORT}{%
\psarc[linewidth=0.5pt](-2,-2){.5}{120}{300}
\psarc[linewidth=0.5pt](2,-2){.5}{-120}{60}
\psbezier[linewidth=0.5pt](-2.25,-1.567)(-1.817,-1.317)(1.817,-1.317)(2.25,-1.567)
\psbezier[linewidth=0.5pt](-1.75,-2.433)(-1.317,-2.183)(1.317,-2.183)(1.75,-2.433)
}
\newcommand{\GMED}{%
\psarc[linewidth=0.5pt](-2,-2){.5}{90}{270}
\psarc[linewidth=0.5pt](2,2){.5}{0}{180}
\psbezier[linewidth=0.5pt](-2,-1.5)(.3,-1.5)(1.5,-.3)(1.5,2)
\psbezier[linewidth=0.5pt](-2,-2.5)(1,-2.5)(2.5,-1)(2.5,2)
}
\newcommand{\GLONG}{%
\psarc[linewidth=0.5pt](-2,-2){.5}{90}{270}
\psarc[linewidth=0.5pt](-2,2){.5}{90}{270}
\psbezier[linewidth=0.5pt](-2,-1.5)(1.5,-1.5)(1.5,1.5)(-2,1.5)
\psbezier[linewidth=0.5pt](-2,-2.5)(2.1,-2.5)(2.1,2.5)(-2,2.5)
}
\newcommand{\GEAR}{%
\psarc[linewidth=0.5pt](0,0){.7}{90}{270}
\psbezier[linewidth=0.5pt](0,0.7)(2.5,0.7)(3.3,0.5)(3.3,0)
\psbezier[linewidth=0.5pt](0,-0.7)(2.5,-0.7)(3.3,-0.5)(3.3,0)
}
\rput(-14,2.5){\makebox(0,0){\bf c.}}
\rput(-11,2){\makebox(0,0){I}}
\rput(-5,-2){\makebox(0,0){II}}
\rput(-8,0){\MOLD}
\rput{135}(-8,0){\GEAR}
\rput{315}(-8,0){\GEAR}
\rput(-4,0){\makebox(0,0){$=$}}
\rput(0,0){\MOLD}
\rput{270}(0,0){\GMED}
\rput(4,0){\makebox(0,0){$-$}}
\rput(8,0){\MOLD}
\rput{90}(8,0){\GMED}
\end{pspicture}
}
\ee

\be
{\psset{unit=0.4}
\begin{pspicture}(-7,-8)(7,3)
\newcommand{\MOLD}{%
\rput(-2,-2){\pscircle*{0.1}}
\rput(-2,2){\pscircle*{0.1}}
\rput(2,-2){\pscircle*{0.1}}
\rput(2,2){\pscircle*{0.1}}
\pscircle[fillstyle=solid, fillcolor=gray](0,0){0.5}
}
\newcommand{\GSHORT}{%
\psarc[linewidth=0.5pt](-2,-2){.5}{120}{300}
\psarc[linewidth=0.5pt](2,-2){.5}{-120}{60}
\psbezier[linewidth=0.5pt](-2.25,-1.567)(-1.817,-1.317)(1.817,-1.317)(2.25,-1.567)
\psbezier[linewidth=0.5pt](-1.75,-2.433)(-1.317,-2.183)(1.317,-2.183)(1.75,-2.433)
}
\newcommand{\GMED}{%
\psarc[linewidth=0.5pt](-2,-2){.5}{90}{270}
\psarc[linewidth=0.5pt](2,2){.5}{0}{180}
\psbezier[linewidth=0.5pt](-2,-1.5)(.3,-1.5)(1.5,-.3)(1.5,2)
\psbezier[linewidth=0.5pt](-2,-2.5)(1,-2.5)(2.5,-1)(2.5,2)
}
\newcommand{\GLONG}{%
\psarc[linewidth=0.5pt](-2,-2){.5}{90}{270}
\psarc[linewidth=0.5pt](-2,2){.5}{90}{270}
\psbezier[linewidth=0.5pt](-2,-1.5)(1.5,-1.5)(1.5,1.5)(-2,1.5)
\psbezier[linewidth=0.5pt](-2,-2.5)(2.1,-2.5)(2.1,2.5)(-2,2.5)
}
\newcommand{\GEAR}{%
\psarc[linewidth=0.5pt](0,0){.7}{90}{270}
\psbezier[linewidth=0.5pt](0,0.7)(2.5,0.7)(3.3,0.5)(3.3,0)
\psbezier[linewidth=0.5pt](0,-0.7)(2.5,-0.7)(3.3,-0.5)(3.3,0)
}
\newcommand{\GEARTHICK}{%
\psarc[linewidth=0.5pt](0,0){.9}{90}{270}
\psbezier[linewidth=0.5pt](0,0.9)(2.7,0.9)(3.5,0.6)(3.5,0)
\psbezier[linewidth=0.5pt](0,-0.9)(2.7,-0.9)(3.5,-0.6)(3.5,0)
}
\rput(-19,3.5){\makebox(0,0){\bf d.}}
\rput(-19,2){\makebox(0,0){I}}
\rput(-13,2){\makebox(0,0){II}}
\rput(-16,0){\MOLD}
\rput(-16,0){\GLONG}
\rput{180}(-16,0){\GLONG}
\rput(-12,0){\makebox(0,0){$=2$}}
\rput(-8,0){\MOLD}
\rput(-8,0){\GMED}
\rput{270}(-8,0){\GMED}
\rput(-4,0){\makebox(0,0){$-2$}}
\rput(0,0){\MOLD}
\rput{90}(0,0){\GMED}
\rput{180}(0,0){\GMED}
\rput(4,0){\makebox(0,0){$-2$}}
\rput(8,0){\MOLD}
\rput(8,0){\GSHORT}
\rput{270}(8,0){\GLONG}
\rput(12,0){\makebox(0,0){$+2$}}
\rput(16,0){\MOLD}
\rput{180}(16,0){\GSHORT}
\rput{90}(16,0){\GLONG}
\rput(-11,-6){\makebox(0,0){$+4$}}
\rput(-7,-6){\MOLD}
\rput(-7,-6){\GSHORT}
\rput{45}(-7,-6){\GEAR}
\rput{135}(-7,-6){\GEAR}
\rput(-3,-6){\makebox(0,0){$-4$}}
\rput(1,-6){\MOLD}
\rput{180}(1,-6){\GSHORT}
\rput{225}(1,-6){\GEAR}
\rput{315}(1,-6){\GEAR}
\end{pspicture}
}
\ee

\be
{\psset{unit=0.4}
\begin{pspicture}(-7,-8)(7,3)
\newcommand{\MOLD}{%
\rput(-2,-2){\pscircle*{0.1}}
\rput(-2,2){\pscircle*{0.1}}
\rput(2,-2){\pscircle*{0.1}}
\rput(2,2){\pscircle*{0.1}}
\pscircle[fillstyle=solid, fillcolor=gray](0,0){0.5}
}
\newcommand{\GSHORT}{%
\psarc[linewidth=0.5pt](-2,-2){.5}{120}{300}
\psarc[linewidth=0.5pt](2,-2){.5}{-120}{60}
\psbezier[linewidth=0.5pt](-2.25,-1.567)(-1.817,-1.317)(1.817,-1.317)(2.25,-1.567)
\psbezier[linewidth=0.5pt](-1.75,-2.433)(-1.317,-2.183)(1.317,-2.183)(1.75,-2.433)
}
\newcommand{\GMED}{%
\psarc[linewidth=0.5pt](-2,-2){.5}{90}{270}
\psarc[linewidth=0.5pt](2,2){.5}{0}{180}
\psbezier[linewidth=0.5pt](-2,-1.5)(.3,-1.5)(1.5,-.3)(1.5,2)
\psbezier[linewidth=0.5pt](-2,-2.5)(1,-2.5)(2.5,-1)(2.5,2)
}
\newcommand{\GLONG}{%
\psarc[linewidth=0.5pt](-2,-2){.5}{90}{270}
\psarc[linewidth=0.5pt](-2,2){.5}{90}{270}
\psbezier[linewidth=0.5pt](-2,-1.5)(1.5,-1.5)(1.5,1.5)(-2,1.5)
\psbezier[linewidth=0.5pt](-2,-2.5)(2.1,-2.5)(2.1,2.5)(-2,2.5)
}
\newcommand{\GEAR}{%
\psarc[linewidth=0.5pt](0,0){.7}{90}{270}
\psbezier[linewidth=0.5pt](0,0.7)(2.5,0.7)(3.3,0.5)(3.3,0)
\psbezier[linewidth=0.5pt](0,-0.7)(2.5,-0.7)(3.3,-0.5)(3.3,0)
}
\newcommand{\GEARTHICK}{%
\psarc[linewidth=0.5pt](0,0){.9}{90}{270}
\psbezier[linewidth=0.5pt](0,0.9)(2.7,0.9)(3.5,0.6)(3.5,0)
\psbezier[linewidth=0.5pt](0,-0.9)(2.7,-0.9)(3.5,-0.6)(3.5,0)
}
\rput(-19,3.5){\makebox(0,0)[lb]{reduction ${\mathbf d}_1$}}
\rput(-19,-2){\makebox(0,0){I}}
\rput(-13,-2){\makebox(0,0){II}}
\rput(-16,0){\MOLD}
\rput(-16,0){\GLONG}
\rput{270}(-16,0){\GMED}
\rput(-12,0){\makebox(0,0){$=$}}
\rput(-8,0){\MOLD}
\rput(-8,0){\GLONG}
\rput{270}(-8,0){\GMED}
\rput(-4,0){\makebox(0,0){$-2$}}
\rput(0,0){\MOLD}
\rput{90}(0,0){\GMED}
\rput{270}(0,0){\GSHORT}
\rput(4,0){\makebox(0,0){$+2$}}
\rput(8,0){\MOLD}
\rput{90}(8,0){\GLONG}
\rput(-11,-6){\makebox(0,0){$+2$}}
\rput(-7,-6){\MOLD}
\rput(-7,-6){\GSHORT}
\rput{135}(-7,-6){\GEAR}
\rput{135}(-7,-6){\GEARTHICK}
\rput(-3,-6){\makebox(0,0){$-4$}}
\rput(1,-6){\MOLD}
\rput{225}(1,-6){\GEAR}
\rput{315}(1,-6){\GEAR}
\end{pspicture}
}
\nonumber
\ee

\be
{\psset{unit=0.4}
\begin{pspicture}(-7,-5)(7,3)
\newcommand{\MOLD}{%
\rput(-2,-2){\pscircle*{0.1}}
\rput(-2,2){\pscircle*{0.1}}
\rput(2,-2){\pscircle*{0.1}}
\rput(2,2){\pscircle*{0.1}}
\pscircle[fillstyle=solid, fillcolor=gray](0,0){0.5}
}
\newcommand{\GSHORT}{%
\psarc[linewidth=0.5pt](-2,-2){.5}{120}{300}
\psarc[linewidth=0.5pt](2,-2){.5}{-120}{60}
\psbezier[linewidth=0.5pt](-2.25,-1.567)(-1.817,-1.317)(1.817,-1.317)(2.25,-1.567)
\psbezier[linewidth=0.5pt](-1.75,-2.433)(-1.317,-2.183)(1.317,-2.183)(1.75,-2.433)
}
\newcommand{\GMED}{%
\psarc[linewidth=0.5pt](-2,-2){.5}{90}{270}
\psarc[linewidth=0.5pt](2,2){.5}{0}{180}
\psbezier[linewidth=0.5pt](-2,-1.5)(.3,-1.5)(1.5,-.3)(1.5,2)
\psbezier[linewidth=0.5pt](-2,-2.5)(1,-2.5)(2.5,-1)(2.5,2)
}
\newcommand{\GLONG}{%
\psarc[linewidth=0.5pt](-2,-2){.5}{90}{270}
\psarc[linewidth=0.5pt](-2,2){.5}{90}{270}
\psbezier[linewidth=0.5pt](-2,-1.5)(1.5,-1.5)(1.5,1.5)(-2,1.5)
\psbezier[linewidth=0.5pt](-2,-2.5)(2.1,-2.5)(2.1,2.5)(-2,2.5)
}
\newcommand{\GEAR}{%
\psarc[linewidth=0.5pt](0,0){.7}{90}{270}
\psbezier[linewidth=0.5pt](0,0.7)(2.5,0.7)(3.3,0.5)(3.3,0)
\psbezier[linewidth=0.5pt](0,-0.7)(2.5,-0.7)(3.3,-0.5)(3.3,0)
}
\newcommand{\GEARTHICK}{%
\psarc[linewidth=0.5pt](0,0){.9}{90}{270}
\psbezier[linewidth=0.5pt](0,0.9)(2.7,0.9)(3.5,0.6)(3.5,0)
\psbezier[linewidth=0.5pt](0,-0.9)(2.7,-0.9)(3.5,-0.6)(3.5,0)
}
\rput(-13,2.5){\makebox(0,0)[rb]{reduction ${\mathbf d}_2$}}
\rput(-5,1){\makebox(0,0){I}}
\rput(-11,-1){\makebox(0,0){II}}
\rput(-8,0){\MOLD}
\rput{90}(-8,0){\GMED}
\rput{270}(-8,0){\GMED}
\rput(-4,0){\makebox(0,0){$=2$}}
\rput(0,0){\MOLD}
\rput{135}(0,0){\GEAR}
\rput{135}(0,0){\GEARTHICK}
\rput(4,0){\makebox(0,0){$-2$}}
\rput(8,0){\MOLD}
\rput{315}(8,0){\GEAR}
\rput{315}(8,0){\GEARTHICK}
\end{pspicture}
}
\nonumber
\ee

\subsection{Braid group relations for geodesic algebras}\label{ss:braid}

The mapping class group transformation that is a generator of a braid group was presented in
Example~\ref{Ex2} and pertains to interchanging two neighbor orbifold points. It turns out that
such transformations generate the whole mapping class group in the case of $A_n$ and $D_n$ algebras
(in the latter case, we must also add the transformation interchanging the $n$th and the first
orbifold points as an independent generator).

Braid group relations on the level of $Z$-variables were presented in \cite{Ch1}; we skip them here and
proceed forward to the braid-group relations in terms of the geodesic function variables $G_{ij}$.

\subsubsection{Braid group relations for geodesic functions of $A_n$-algebras}

Here we, following Bondal~\cite{Bondal}, propose another, simpler way to derive
the braid group relations using the construction of the groupoid of upper-triangular matrices.
(It was probably first used in \cite{DM} to prove the braid group relations in the case of $A_3$ algebra.)
In the case of $A_n$ algebras for general $n$, let us construct the upper-triangular matrix ${\mathcal A}$
\be
\label{A-matrix}
{\mathcal A}=\left(\begin{array}{ccccc}
                     1 & G_{1,2} & G_{1,3} & \dots & G_{1,n} \\
                     0 & 1 & G_{2,3} & \dots & G_{2,n} \\
                     0 & 0 & 1 & \ddots & \vdots \\
                     \vdots & \vdots & \ddots & \ddots & G_{n-1,n} \\
                     0 & 0 & \dots & 0 & 1 \\
                   \end{array}
\right)
\ee
associating the entries $G_{i,j}$ with the geodesic functions. Using the skein relation,
we can then present the action of the braid group element $R_{i,i+1}$ with $i=1,\dots,n-1$
exclusively in terms of the geodesic functions from this, fixed, set:
\be
\label{braid-A}
R_{i,i+1}{\mathcal A}={\tilde{\mathcal A}},\ \hbox{where} \
\left\{
\begin{array}{ll}
  {\tilde G}_{i+1,j}=G_{i,j} & j>i+1,\\
  {\tilde G}_{j,i+1}=G_{j,i} & j<i, \\
  {\tilde G}_{i,j}=G_{i,j}G_{i,i+1}-G_{i+1,j} & j>i+1, \\
  {\tilde G}_{j,i}=G_{j,i}G_{i,i+1}-G_{j,i+1} & j<i, \\
  {\tilde G}_{i,i+1}=G_{i,i+1} &  \\
\end{array}%
\right. .
\ee

\begin{lemma} \label{lem-braid}
For any $n\ge3$, we have the {\em braid group relation} for $R_{i,i+1}$ in (\ref{braid-A}):
\be
\label{RRR}
R_{i-1,i}R_{i,i+1}R_{i,i-1}=R_{i,i+1}R_{i-1,i}R_{i,i+1},\quad 2\le i\le n-1.
\ee
\end{lemma}

We can conveniently present the braid-group transformation using the special matrices
$B_{i,i+1}$ of the block-diagonal form,
\be
\label{Bii+1}
B_{i,i+1}=\begin{array}{c}
            \vdots \\
            i \\
            i+1 \\
            \vdots \\
          \end{array}
          \left(
          \begin{array}{cccccccc}
            1 &  &  &  &  &  &  &  \\
             & \ddots &  &  &  &  &  &  \\
             &  & 1 &  &  &  &  &  \\
             &  &  &   G_{i,i+1} & -1 &  & &  \\
            &  &  & 1 & 0 &  &  &  \\
             &  &  &  &  & 1 &  &  \\
             &  &  &  &  &  & \ddots &  \\
             &  &  &  &  &  &  & 1 \\
          \end{array}
          \right).
\ee
Then, the action of the braid group generator $R_{i,i+1}$ on ${\mathcal A}$
can be presented in the form of usual matrix product:
\be
\label{BAB}
R_{i,i+1}{\mathcal A}=B_{i,i+1}{\mathcal A}B^{T}_{i,i+1}
\ee
with $B^{T}_{i,i+1}$ the matrix transposed to $B_{i,i+1}$.

We now consider the action of the chain of transformations
$R_{n-1,n}R_{n-2,n-1}\dots R_{2,3}R_{1,2}{\mathcal A}$.
Note that, on each step, the item $G^{(i-1)}_{i,i+1}$ in the corresponding matrix $B_{i,i+1}$
is the transformed quantity
(we assume $G^{(0)}_{ij}$ to coincide with the initial $G_{ij}$ in ${\mathcal A}$).
However, it is easy to see that for just this chain of transformations,
$G^{(i-1)}_{i,i+1}=G^{(0)}_{1,i+1}=G_{1,i+1}$,
and the whole chain of matrices $B$ can be then expressed in terms of the initial variables $G_{i,j}$ as
\be
\label{B...B}
{\mathcal B}\equiv B_{n-1,n}B_{n-2,n-1}\dots B_{2,3}B_{1,2}=\left(%
\begin{array}{ccccc}
  G_{1,2} & -1 & 0 & \dots & 0 \\
  G_{1,3} & 0 & -1 &  & \vdots \\
  \vdots & \vdots & \ddots & \ddots & 0 \\
  G_{1,n} & 0 & \dots & 0 & -1 \\
  1 & 0 & \dots & 0 & 0 \\
\end{array}%
\right),
\ee
whereas its action on ${\mathcal A}$ gives
\be
\label{tildeA}
{\tilde {\mathcal A}}\equiv {\mathcal B}{\mathcal A}{\mathcal B}^{T}=\left(%
\begin{array}{cccccc}
  1 & G_{2,3} & G_{2,4} & \dots & G_{2,n} & G_{1,2} \\
  0 & 1 & G_{3,4} & \dots & G_{3,n} & G_{1,3} \\
  0 & 0 & 1 &  & G_{4,n} & G_{1,4} \\
  \vdots & \vdots &  & \ddots &  & \vdots \\
  0 & 0 & \dots & 0 & 1 & G_{1,n} \\
  0 & 0 & \dots & \dots & 0 & 1 \\
\end{array}%
\right),
\ee
which is a mere permutation of the elements of the initial matrix ${\mathcal A}$.
It is easy to see that the $n$th power of this permutation gives the identical transformation, so we
obtain the last {\em braid group relation}.

\begin{lemma} \label{lem-braid-2}
For any $n\ge3$, we have the second {\em braid group relation} for the $A_n$ algebra:
\be
\label{Rn}
\bigl(R_{n-1,n}R_{n-2,n-1}\cdots R_{2,3}R_{1,2}\bigr)^n=\hbox{Id}.
\ee
\end{lemma}

\subsubsection{Central elements of Poisson/braid group transformations for $A_n$ algebras}

From (\ref{BAB}), we immediately obtain that the same braid-group transformation holds for the transposed
matrix ${\mathcal A}^T$ and therefore for any combination $\lambda{\mathcal A}+\lambda^{-1}{\mathcal A}^T$:
\be
\label{A-lambda}
R_{i,i+1}(\lambda{\mathcal A}+\lambda^{-1}{\mathcal A}^T)=
B_{i,i+1}(\lambda{\mathcal A}+\lambda^{-1}{\mathcal A}^T)B^{T}_{i,i+1},
\ee
and the determinant
\be
\label{det-A}
\det (\lambda{\mathcal A}+\lambda^{-1}{\mathcal A}^T)
\ee
is therefore the generating function for the central elements of the Poisson algebra and, simultaneously,
the invariants of the braid group in the $A_n$ case. There are $\left[\frac{n}{2}\right]$ such independent
elements (due to the symmetry $\lambda\leftrightarrow \lambda^{-1}$, so the maximum Poisson dimension of the
corresponding $A_n$ algebra is $\frac{n(n-1)}{2}-\left[\frac{n}{2}\right]$ (an even number for all $n$).

\subsubsection{Braid group relations for geodesic functions of $D_n$-algebras}

It is possible to express readily the action of the braid group on
the level of the geodesic functions  $G_{i,j}$, $i,j=1,\dots,n$, interpreted also as
entries of the $n\times n$-matrix ${\mathcal D}$
(the elements that are not indicated remain invariant):
\be
R_{i,i+1}{\mathcal D}={\tilde{\mathcal D}},\ \hbox{where} \
\left\{
\begin{array}{ll}
  {\tilde G}_{i+1,k}=G_{i,k} & k\ne i,i+1,\\
  {\tilde G}_{i,k}=G_{i,k}G_{i,i+1}-G_{i+1,k} &  k\ne i,i+1,\\
  {\tilde G}_{k,i+1}=G_{k,i} &  k\ne i,i+1, \\
  {\tilde G}_{k,i}=G_{k,i}G_{i,i+1}-G_{k,i+1} &  k\ne i,i+1, \\
  {\tilde G}_{i,i+1}=G_{i,i+1} &  \\
  {\tilde G}_{i+1,i+1}=G_{i,i} &  \\
  {\tilde G}_{i,i}=G_{i,i}G_{i,i+1}-G_{i+1,i+1} &  \\
  {\tilde G}_{i+1,i}=G_{i+1,i}+G_{i,i+1}G_{i,i}^2-2G_{i,i}G_{i+1,i+1} &  \\
\end{array}%
\right. .
\label{Dn-braid-cl}
\ee

To obtain the full mapping class group, we must complete this set of transformations by the
new element $R_{n,1}$ interchanging $s_1$ and $s_n$:
\be
R_{n,1}{\mathcal D}={\tilde{\mathcal D}},\ \hbox{where} \
\left\{
\begin{array}{ll}
  {\tilde G}_{1,k}=G_{n,k} & k\ne n,1,\\
  {\tilde G}_{n,k}=G_{n,k}G_{n,1}-G_{1,k} &  k\ne n,1,\\
  {\tilde G}_{k,1}=G_{k,n} &  k\ne n,1, \\
  {\tilde G}_{k,n}=G_{k,n}G_{n,1}-G_{k,1} &  k\ne n,1, \\
  {\tilde G}_{n,1}=G_{n,1} &  \\
  {\tilde G}_{1,1}=G_{n,n} &  \\
  {\tilde G}_{n,n}=G_{n,n}G_{n,1}-G_{1,1} &  \\
  {\tilde G}_{1,n}=G_{1,n}+G_{n,1}G_{n,n}^2-2G_{n,n}G_{1,1} &  \\
\end{array}%
\right. .
\label{Dn-braid-cl-n1}
\ee

The first braid group relation follows in this case as well from the three-step process,
but it can be also verified explicitly that the following lemma holds just on the level of
elements $G_{i,j}$.

\begin{lemma} \label{lem-braid-Dn}
For any $n\ge3$, we have the {\em braid group relation} for transformations (\ref{Dn-braid-cl}),
(\ref{Dn-braid-cl-n1}):
\be
\label{RRR-Dn}
R_{i-1,i}R_{i,i+1}R_{i-1,i}{\mathcal D}=R_{i,i+1}R_{i-1,i}R_{i,i+1}{\mathcal D}, i=1,\dots,n\ \mod n.
\ee
\end{lemma}

Note that the second braid-group relation (see Lemma~\ref{lem-braid-2}) is lost
in the case of $D_n$-algebras.

Presenting the braid-group action in the matrix-action (covariant) form (\ref{BAB}) turned out to be
a nontrivial problem. First, it was noted already in \cite{Ch1} that special combinations of $G_{ij}$
admit similar transformation laws under the subgroup of braid-group transformations generated by
relations (\ref{Dn-braid-cl}) alone (omitting the last transformation (\ref{Dn-braid-cl-n1}).

Consider two new $n\times n$ matrices composed from $G_{ij}$:
\bea\label{R.cl}
({\mathcal R})_{i,j}&=&\left\{ \begin{array}{cc}
                        -G_{j,i}-G_{i,j}+G_{i,i}G_{j,j} &\quad j>i \\
                        G_{j,i}+G_{i,j}-G_{i,i}G_{j,j} &\quad  j<i \\
                               0 &\quad  j=i \\
                             \end{array}
                             \right.,
\\
\label{S.cl}
({\mathcal S})_{i,j}&=&G_{i,i}G_{j,j}\quad \hbox{for all}\quad 1\le i,j\le n;
\eea
here ${\mathcal R}$ is skewsymmetric (${\mathcal R}^T=-{\mathcal R}$) and ${\mathcal S}$ is symmetric
(${\mathcal S}^T={\mathcal S}$). Then, together with $\mathcal A$ given by the old formula (\ref{A-matrix}),
we have the following statement.

\begin{lemma} \label{lem-braid-Dn-matrix}
Any linear combination $w_1{\mathcal A}+w_2{\mathcal A}^T+\rho {\mathcal R}+\sigma {\mathcal S}$
with complex $w_1$, $w_2$, $\rho$, and $\sigma$ transforms by formula (\ref{BAB})
under the subgroup (\ref{Dn-braid-cl}) of the braid group.
\end{lemma}

However, incorporating the last generator of the braid group took a long time. Here, we present the new
result obtained recently in collaboration with M.~Mazzocco~\cite{ChM}\footnote{The way to come to this
result is interesting by itself; it involves some new insight on the whole topic. Here we, however,
confine ourself to a mere presentation, the reader can find details in \cite{ChM}.}

Let us introduce the $(4n)\times (4n)$ upper triangular block matrix ${\mathbb A}$:
\be
\label{mathbbA}
{\mathbb A}=\left[\begin{array}{cccc}
            {\mathcal A}& B & C & B^T \\
            0 & {\mathcal A}& B & C  \\
            0 & 0 &{\mathcal A}& B \\
            0 & 0 & 0 & {\mathcal A}
            \end{array}
            \right],
\quad\hbox{where}\ \left\{\begin{array}{l}
B={\mathcal S}+{\mathcal R}+{\mathcal A}-{\mathcal A}^T,\\
C=2{\mathcal S}-{\mathcal A}-{\mathcal A}^T, \ C^T=C.
\end{array}
\right.
\ee

The matrices of braid-group transformations $R_{i,i+1}$ with $i=1,\dots,n-1$, ${\mathbb B}_{i,i+1}$,
have a simple block-diagonal structure:
\be
\label{B-D-ii+1}
{\mathbb B}_{i,i+1}=\left[\begin{array}{cccc}
            B_{i,i+1}& 0 & 0 & 0 \\
            0 & B_{i,i+1} & 0 & 0  \\
            0 & 0 & B_{i,i+1} & 0 \\
            0 & 0 & 0 & B_{i,i+1}
            \end{array}\right],
\ee
whereas a new matrix ${\mathcal B}_{n,1}(\lambda)$ must depend on the parameter $\lambda$ and has the form

\be
\label{Bn1}
{\mathbb B}_{n,1}(\lambda)=  \left(
          \begin{tabular}{ccc|ccc|ccc|ccc}
               $0$ &&& &&& &&&   && \hbox{\color{red}$\lambda^2$}  \\
               &$\mathcal I$&& &&& &&&  &&  \\
               &  &   $G_{1,n}$ &  $-1$ && &&&&  &&   \\
            \hline
              &  & $1$ &  $0$ &  && &&&  &&    \\
               &&&  &$\mathcal I$&&  &&&  &&  \\
              &&&   && $G_{1,n}$  &  $-1$ &&& &&     \\
              \hline
              &  &  &   &  &$1$& $0$&&& &&    \\
               &&& &&& &$\mathcal I$&& &&  \\
              &&& &&&   && $G_{1,n}$  &  $-1$ &&     \\
              \hline
              &&& &&&    && $1$ &  $0$ &  &    \\
               &&& &&&  &&& &$\mathcal I$&  \\
               \hbox{\color{red}$-\lambda^{-2}$} &  &  & &&& &&&  & & $G_{1,n}$   \\
          \end{tabular}
          \right),
\ee
with $\mathcal I$ being the $(n-2)\times (n-2)$ unit matrices on the diagonal.

We then have the theorem.

\begin{theorem}\label{th-braid}
The braid group relations  (\ref{Dn-braid-cl}) and (\ref{Dn-braid-cl-n1}) can be presented
in the form of matrix relations for the matrix $\lambda{\mathbb A}+\lambda^{-1}{\mathbb A}^T$
with the matrix ${\mathbb A}$ defined in (\ref{mathbbA}):
\bea
R_{i,i+1}\bigl(\lambda{\mathbb A}+\lambda^{-1}{\mathbb A}^T\bigr)
&=&{\mathbb B}_{i,i+1}\bigl(\lambda{\mathbb A}+\lambda^{-1}{\mathbb A}^T\bigr){\mathbb B}_{i,i+1}^T,
\quad i=1,\dots,n-1
\label{braid-D-1}
\\
R_{n,1}(\lambda {\mathbb A}+\lambda^{-1}{\mathbb A}^T)&=&{\mathbb B}_{n,1}(\lambda)
(\lambda {\mathbb A}+\lambda^{-1}{\mathbb A}^T){\mathbb B}_{n,1}^T(\lambda^{-1}),
\label{braid-D-2}
\eea
with ${\mathbb B}_{i,i+1}$ from (\ref{B-D-ii+1}) and ${\mathbb B}_{n,1}(\lambda)$ from
(\ref{Bn1}).
\end{theorem}

\subsubsection{Central elements of Poisson/braid group transformations for $D_n$ algebras}

Proceeding with analogy from the case of the $A_n$ algebra, we take the determinant
\be
\label{det-D}
\det (\lambda{\mathbb A}+\lambda^{-1}{\mathbb A}^T)
\ee
as the generating function for the central elements of the $D_n$ algebra. This is a $(4n)\times(4n)$-matrix,
so one could expect the existence of $\left[\frac{4n}{2}\right]=2n$ central elements. However, this matrix
has a special structure, and if we take it for $\lambda=1$, that is, consider the symmetric matrix
${\mathbb A}+{\mathbb A}^T$ (of a bilinear form), then, in terms of original matrices $\mathcal A$,
$\mathcal S$, and $\mathcal R$, we have for ${\mathbb A}+{\mathbb A}^T$ the expression
\be
\left[\begin{tabular}{c|c|c|c}
        ${\mathcal A}+{\mathcal A}^T$ & ${\mathcal S}+{\mathcal R}+{\mathcal A}-{\mathcal A}^T$  &
        $2{\mathcal S}-{\mathcal A}-{\mathcal A}^T$  &
        ${\mathcal S}-{\mathcal R}-{\mathcal A}+{\mathcal A}^T$
         \\
        \hline
        ${\mathcal S}-{\mathcal R}-{\mathcal A}+{\mathcal A}^T$  &${\mathcal A}+{\mathcal A}^T$ &
        ${\mathcal S}+{\mathcal R}+{\mathcal A}-{\mathcal A}^T$  &
        $2{\mathcal S}-{\mathcal A}-{\mathcal A}^T$
         \\
        \hline
        $2{\mathcal S}-{\mathcal A}-{\mathcal A}^T$ &
        ${\mathcal S}-{\mathcal R}-{\mathcal A}+{\mathcal A}^T$  &${\mathcal A}+{\mathcal A}^T$ &
        ${\mathcal S}+{\mathcal R}+{\mathcal A}-{\mathcal A}^T$
         \\
        \hline
        ${\mathcal S}+{\mathcal R}+{\mathcal A}-{\mathcal A}^T$  &
        $2{\mathcal S}-{\mathcal A}-{\mathcal A}^T$ &
        ${\mathcal S}-{\mathcal R}-{\mathcal A}+{\mathcal A}^T$  &${\mathcal A}+{\mathcal A}^T$
         \\
        \end{tabular}
            \right],
\ee
and adding first line of blocks to the third line and second to the fourth, we obtain the matrix in which
two last lines of blocks are composed from the same matrix $2{\mathcal S}$. By its structure, the matrix
${\mathcal S}$ has rank one being the outer product of two vectors, so we conclude that the matrix
${\mathbb A}+{\mathbb A}^T$ has at most rank $2n+1$, that is, taking into account the symmetry
$\lambda\leftrightarrow\lambda^{-1}$, we have
\bea
\det \bigl(\lambda{\mathbb A}+\lambda^{-1}{\mathbb A}^T\bigr)&=&(\lambda-\lambda^{-1})^{2n-1}\Bigl[
\lambda^{2n+1}+S_1\lambda^{2n-1}+\cdots\Bigr.
\nonumber
\\
&{}&\quad \Bigl.\cdots + S_n\lambda - S_n\lambda^{-1} -\cdots - S_1\lambda^{1-2n}-\lambda^{-2n-1}\Bigr],
\label{central-D}
\eea
so, in total, we have exactly $n$ independent central elements $S_i$, $i=1,\dots,n$, and the highest
Poisson leaf dimension is $n^2-n=n(n-1)$.

\newsection{Quantum Teichm\"uller spaces of orbifold Riemann surfaces}\label{s:q}

\subsection{Canonical quantization of the Poisson algebra}

A quantization of a Poisson manifold, which is equivariant under the action of a discrete
group $\cal D$,
is a family of $*$-algebras ${\cal A}^\hbar$ depending on a positive
real parameter $\hbar$ with
$\cal D$ acting by outer automorphisms and having the
following properties:

\begin{itemize}

\item [{\bf 1.}] (Flatness.)  All algebras are isomorphic (noncanonically)
as linear spaces.

\item [{\bf 2.}]
(Correspondence.) For $\hbar=0$, the algebra is isomorphic as a $\cal
D$-module to the $*$-algebra of complex-valued functions ${\cal A}^0$ on the Poisson
manifold.

\item [{\bf 3.}]  (Classical Limit.) The Poisson bracket on ${\cal A}^0$ given
by $\{a_1, a_2\} = \lim_{\hbar \rightarrow 0}\frac{[a_1,a_2]}{\hbar}$ coincides with the
Poisson bracket given by the Poisson structure of the manifold.

\end{itemize}

Fix a three-valent fatgraph $\Gamma_{g,\delta}$ as a spine of $\Sigma_{g,\delta}$,
and let ${\cal T}^\hbar={\cal T}^\hbar(\Gamma_{g,\delta})$
be the algebra generated by the operators $Z_\alpha^\hbar$, one generator for each
unoriented edge $\alpha$ of $\Gamma_{g,\delta}$, with relations
\be
\label{comm}
[Z^\hbar_\alpha, Z^\hbar_\beta ] = 2\pi i\hbar\{Z_\alpha, Z_\beta\}
\ee
(cf.\ (\ref{WP-PB})) and the
$*$-structure
\be
(Z^\hbar_\alpha)^*=Z^\hbar_\alpha,
\ee
where $Z_\alpha$  and
$\{\cdot,\cdot\}$ denotes the respective coordinate functions
and the Poisson bracket on the classical Teichm\"uller space.
Because of (\ref{WP-PB}), the right-hand side of (\ref{comm}) is a constant
taking only five values $0$, \ $\pm 2\pi i \hbar$, and $\pm 4 \pi i \hbar$
depending upon five variants of identifications of endpoints of edges labelled $\alpha$ and $\beta$.

All the standard statements that we have in the case of Teichm\"uller spaces of
Riemann surfaces with holes are transferred to the case of orbifold Riemann surfaces.

\begin{lemma}\label{nondeg}
The center ${\cal Z}^\hbar$ of the algebra ${\cal T}^\hbar$ is generated by the sums
$\sum_{\alpha\in I}{Z^\hbar_\alpha}$ over all edges $\alpha\in I$ surrounding
a given boundary component, the center has dimension $s$, and the quantum structure is
nondegenerate on the quotient ${\cal T}^\hbar/{\cal Z}^\hbar$.
\end{lemma}

The examples of the {\em boundary-parallel} curves whose quantum lengths are the Casimir operators
are again in Fig.~\ref{fi:center}. Of course,
those are the same curves that provide the center of the Poisson algebra.

\begin{cor}\label{corr11}
There is a basis for ${\cal T}^\hbar/{\cal Z}^\hbar$
given by operators $p_i,q_i$, for $i=1,\ldots ,3g-3+s+\sum_{j=1}^s|\delta_j|$
satisfying the standard commutation relations $[p_i,q_j]=2\pi i \hbar\delta_{ij}$.
\end{cor}

\subsection{Quantum flip transformations}

The Whitehead move becomes now a morphism of (quantum) algebras.
The {\it quantum Whitehead moves} or quantum {\it flips} along an edge $Z$
of $\Gamma$ are now described by Eq. (\ref{abc}), Fig.~\ref{fi:mcg-pending}, and Eq.~\ref{braid1}
with the (quantum) function~\cite{Faddeev},~\cite{ChF}
\begin{equation} \label{phi}
\phi(z)\equiv \phi^\hbar(z) =
-\frac{\pi\hbar}{2}\int_{\Omega} \frac{e^{-ipz}}{\sinh(\pi p)\sinh(\pi \hbar
p)}dp,
\end{equation}
where the contour $\Omega$ goes along
the real axis bypassing the origin from above.
For each unbounded self-adjoint operator $Z^\hbar$ on the Hilbert space
${\cal H}$ of $L^2$-functions, $\phi^\hbar (Z^\hbar)$ is a well-defined
unbounded self-adjoint operator on ${\cal H}$.

The function $\phi^\hbar(Z)$ satisfies the relations (see~\cite{ChF})
\bea
&{}&\phi^\hbar(Z)-\phi^\hbar(-Z)=Z,
\nonumber
\\
&{}&
\phi^\hbar(Z+i\pi\hbar)-\phi^\hbar(Z-i\pi\hbar)=\frac{2\pi i\hbar}{1+\e^{-Z}},
\nonumber
\\
&{}&
\phi^\hbar(Z+i\pi)-\phi^\hbar(Z-i\pi)=\frac{2\pi i}{1+\e^{-Z/\hbar}}
\nonumber
\eea
and is meromorphic in the complex plane with the poles at the
points $\{\pi i(m+n\hbar),\ m,n\in {\Bbb Z}_+\}$ and
$\{-\pi i(m+n\hbar),\ m,n\in {\Bbb Z}_+\}$.

The function $\phi^\hbar(Z)$ is therefore holomorphic in the strip
$|\hbox{Im\,}Z|<\pi\hbox{\,min\,}(1,\hbox{Re\,}\hbar)-\epsilon$ for any $\epsilon>0$,
so we need only its asymptotic behavior as
$Z\in{\Bbb R}$ and $|Z|\to\infty$, for which we have (see, e.g., \cite{Kashaev3})
\be
\biggr.\phi^\hbar(Z)\biggl|_{|Z|\to\infty}=(Z+|Z|)/2+O(1/|Z|).
\label{phih-as}
\ee

We then have the following theorem (\cite{ChF},~\cite{Kashaev})

\begin{theorem}\label{th-Q}
The family of algebras ${\cal T}^\hbar={\cal T}^\hbar(\Gamma_{g,\delta} )$ is a
quantization of ${\cal T}^H_{g,\delta}$ for any three-valent fatgraph spine
$\Gamma_{g,\delta}$ of $\Sigma_{g,\delta}$, that is,
\begin{itemize}
\item In the limit $\hbar \mapsto 0$, morphism {\rm(\ref{abc})} using
(\ref{phi}) coincides with classical morphism {\rm(\ref{abc})} with $\phi(Z)=\log(1+\e^Z)$.
\item Morphism {\rm(\ref{abc})} using {\rm(\ref{phi})} is indeed a morphism of $*$-algebras.
\item A flip $W_Z$ satisfies $W_Z^2=I$.
\item Flips on inner edges having exactly one common vertex satisfy the pentagon relation.
\end{itemize}
\end{theorem}

\subsection{Quantum geodesic functions}

We next embed the algebra of geodesic functions
(\ref{G}) into a suitable
completion of the constructed algebra ${\cal T}^\hbar$.
For any geodesic $\gamma$,
the geodesic function $G_\gamma$ can be expressed  in terms of
shear coordinates on ${\cal T}^H$:
\begin{equation}
G_\gamma \equiv \tr P_{Z_1\cdots Z_n}=
\sum_{j\in J}\exp\left\{{\frac{1}{2} \sum_{\alpha \in E(\Gamma)}
m_j(\gamma,\alpha) Z_\alpha}\right\},  \label{clen}
\end{equation}
where $m_j(\gamma,\alpha)$ are
integers and $J$ is a finite set of indices.

In general, sets of integers $\left\{m_j(\gamma,\alpha)\right\}_{\alpha=1}^{6g-6+3s+2|\delta|}$
may coincide for different $j_1,j_2\in J$; we however distinguish between them as soon
as they come from different products of exponentials $e^{\pm Z_i/2}$ in
traces of matrix products in (\ref{clen}).

For any closed path $\gamma$ on $\Sigma_{g,\delta}$, define the {\it quantum geodesic}
operator $G^\hbar_\gamma \in {\cal T}^\hbar$ to be
\begin{equation} \label{qlen}
G^\hbar_\gamma \equiv
\ORD{\tr P_{Z_1\dots Z_n}}\equiv
\sum_{j\in J}
\exp\left\{{\frac{1}{2} \sum_{\alpha \in E(\Gamma_{g,\delta})}
\bigl(m_j(\gamma,\alpha) Z^\hbar_\alpha
+2\pi i\hbar
c_j(\gamma,\alpha)
\bigr)}\right\},
\end{equation}
where the {\it quantum ordering} $\ORD{\cdot}$ implies that we vary the classical
expression (\ref{clen}) by introducing additional
integer coefficients $c_j(\gamma,\alpha)$,
which must be determined from the conditions below.

That is, we assume that each term in the classical expression (\ref{clen}) can get
multiplicative corrections only of the form $q^n$, $n\in {\Bbb Z}$, with
\be
\label{qdef}
q\equiv e^{-i\pi\hbar}.
\ee

We now formulate the defining properties of quantum geodesics.
\begin{itemize}
\item[{\bf 1.}]  {\it Commutativity.} If closed paths $\gamma$ and $\gamma^\prime$ do not
intersect, then the operators $G^\hbar_\gamma$ and
$G^\hbar_{\gamma^\prime}$ commute.
\item[{\bf 2.}]  {\it Naturality.} The mapping class group
(\ref{abc}) acts naturally on the set $\{G^\hbar_\gamma\}$, i.e., for any transformation
$W^\hbar$ from the mapping class groupoid and for a closed path $\gamma$ in a spine $\Gamma_{g,\delta}$ of
$\Sigma_{g,\delta}$, we have
$$
W^\hbar(G^\hbar_\gamma) = G^\hbar_{W(\gamma)}.
$$
\item[{\bf 3.}]  {\it Quantum geodesic algebra}. The product of two quantum geodesics
is a linear combination of quantum multicurves governed by the (quantum) skein
relation below.
\item[{\bf 4.}] {\it Orientation invariance.}
As in the classical case, the quantum geodesic operator does not depend on the
orientation of the corresponding geodesic line.
\item[{\bf 5.}] {\it Exponents of geodesics.}
A quantum geodesic $G^\hbar_{n\gamma}$ corresponding to the $n$-fold concatenation of $\gamma$
is expressed via $G^\hbar_{\gamma}$ exactly as in the classical case, namely,
\be
\label{cheb}
G^\hbar_{n\gamma}=2T_n\bigl(G^\hbar_{\gamma}/2\bigr),
\ee
where $T_n(x)$ are Chebyshev's polynomials.
\item[{\bf 6.}] {\it Hermiticity.} A quantum geodesic is a Hermitian
operator having by definition a real spectrum.
\end{itemize}

We let  the standard normal ordering symbol ${:}\ {:}$ denote the {\it Weyl
ordering}, ${:}\e^{a_1}\e^{a_2}\cdots\e^{a_n}{:}\equiv\e^{a_1+\cdots+a_n}$,
for any set of exponents with $a_i\neq -a_j$ for $i\neq j$.

\begin{defin}
{\rm
For a spine $\Gamma_{g,\delta}$, we call a geodesic {\it graph simple} if it does not pass twice
through any of inner edges of the graph and {\em undergoes at most one inversion} at any of the
dot vertices.
}
\end{defin}

\begin{proposition}\label{lem31}
For any graph simple geodesic $\gamma$ with respect to any spine $\Gamma$,
the coefficients $c_j(\gamma,\alpha)$ in {\rm(\ref{qlen})} are identically zero, i.e.,
the quantum ordering is the Weyl ordering.
\end{proposition}

\subsection{Quantum skein relations}

We now formulate the {\em general} rules that allow one to disentangle the product of
any two quantum geodesics.

Let $G^\hbar_1$ and $G^\hbar_2$ be two quantum geodesic operators corresponding to
geodesics $\gamma_1$ and $\gamma_2$ where all the inversion relations are resolved using
the dot-vertex construction. Then
\begin{itemize}
\item
We must apply the {\em quantum skein relation}
\footnote{Here the order of crossing lines corresponding to
$G^\hbar_1$ and $G^\hbar_2$ depends on which quantum geodesic
occupies the first place in the product;
the rest of the graph remains unchanged for all items in (\ref{skein}).}
\be
\setlength{\unitlength}{.8mm}%
\begin{picture}(90,40)(0,45)
\thicklines
\put(-10,50){\line(1, 1){9}}
\put( 1, 61){\line(1, 1){9}}
\put(-10,70){\line(1,-1){20}}
\put(18,60){\makebox(0,0){$=$}}
\put(30,61){\makebox(0,0){$\e^{-i\pi\hbar/2}$}}
\put(40,60){\oval(20,20)[r]}
\put(63,60){\oval(20,20)[l]}
\put(70,60){\makebox(0,0){$+$}}
\put(78,61){\makebox(0,0){$\e^{i\pi\hbar/2}$}}
\put(95,48){\oval(20,20)[t]}
\put(95,72){\oval(20,20)[b]}
\put(-11,68){\makebox(0,0)[rb]{\Large$G^\hbar_1$}}
\put(-11,52){\makebox(0,0)[rt]{\Large$G^\hbar_2$}}
\put(52,74){\makebox(0,0)[cb]{\Large$G^\hbar_Z$}}
\put(95,74){\makebox(0,0)[cb]{\Large$\wtd G^\hbar_Z$}}
\end{picture}
\label{skein}
\ee
{\em simultaneously} at {\em all} intersection points.
\item After the application of the quantum skein relation we can
obtain empty (contractible) loops; we assign the factor $-q-q^{-1}$
to each such loop and this suffices to ensure the quantum Reidemeister moves.
\item We can also obtain loops that are homeomorphic to going around a dot-vertex;
as in the classical case, we claim the corresponding geodesic functions to vanish, $\tr F=0$,
so we erase all such cases of geodesic laminations in the quantum case as well.
\end{itemize}

The main lemma is in order.

\begin{lemma}\label{lem34}{\rm \cite{ChF},~\cite{Teshner}}
There exists
a unique quantum ordering $\ORD{\dots}$ {\rm (\ref{qlen})}, which is
generated by the quantum geodesic
algebra (\ref{skein}) and is consistent with the quantum
transformations {\rm(\ref{abc})}, i.e., so that the quantum
geodesic algebra is invariant under the action of the quantum mapping
class groupoid.
\end{lemma}

\begin{example}\label{ex3}
{\rm
For the pattern of the quantum geodesic functions in Fig.~\ref{fi:dot-skein}, we have
(provided $\gamma_1$ and $\gamma_2$ have no more intersections)
$$
G_1^\hbar\cdot G_2^\hbar=qG_\cup^\hbar+q^{-1}G_\cap^\hbar+G_\supset^\hbar\cdot G_\subset^\hbar.
$$
This algebra simplifies further in the case where $G_1^\hbar=\ORD{\tr F_{i_1}F_j}:=G_{i_1,j}^\hbar$ and
$G_2^\hbar=\ORD{\tr F_{j}F_{i_2}}:=G_{j,i_2}^\hbar$ with $i_1<j<i_2$ because in this case
$G_\supset^\hbar=\ORD{\tr F_{i_1}}=0$ and $G_\subset^\hbar=\ORD{\tr F_{i_2}}=0$. Then, taking into
account that the product in the opposite order gives
$$
G_2^\hbar\cdot G_1^\hbar=q^{-1}G_\cup^\hbar+qG_\cap^\hbar+G_\supset^\hbar\cdot G_\subset^\hbar,
$$
and that, for $G_1^\hbar=G_{i_1,j}^\hbar$ and $G_2^\hbar=G_{j,i_2}^\hbar$,
$G_\cap^\hbar=\ORD{\tr F_{i_1}F_{i_2}}:=G_{i_1,i_2}^\hbar$
we obtain for the {\em $q$-commutator}
$$
qG_{i_1,j}^\hbar\cdot G_{j,i_2}^\hbar-q^{-1}G_{j,i_2}^\hbar\cdot G_{i_1,j}^\hbar
=(q^2-q^{-2})G_{i_1,i_2}^\hbar.
$$
This relation is among basic relations for the quantum Nelson--Regge algebras~\cite{NR}.
}
\end{example}

\begin{figure}[tb]
{\psset{unit=0.8}
\begin{pspicture}(-8,-8)(8,7)
\newcommand{\CROSS}[1]{%
{\psset{unit=#1}
\psarc[linewidth=0.5pt](-3,0){1.}{90}{270}
\psarc[linewidth=0.5pt](3,0){1.}{-90}{90}
\psarc[linewidth=0.5pt](0,3){1.}{0}{180}
\psarc[linewidth=0.5pt](0,-3){1.}{-180}{0}
\rput(-3,0){\makebox(0,0){\tiny $\bullet$}}
\rput(3,0){\makebox(0,0){\tiny $\bullet$}}
\rput(0,3){\makebox(0,0){\tiny $\bullet$}}
\rput(0,-3){\makebox(0,0){\tiny $\bullet$}}
\pcline[linewidth=0.5pt](-3,1)(-1.3,1)
\pcline[linewidth=0.5pt](-3,-1)(-1.3,-1)
\pcline[linewidth=0.5pt](3,1)(1.3,1)
\pcline[linewidth=0.5pt](3,-1)(1.3,-1)
\pcline[linewidth=0.5pt](-0.7,1)(0.7,1)
\pcline[linewidth=0.5pt](-0.7,-1)(0.7,-1)
\pcline[linewidth=0.5pt](-1,-3)(-1,3)
\pcline[linewidth=0.5pt](1,-3)(1,3)
}
}
\newcommand{\POND}[1]{%
{\psset{unit=#1}
\psarc[linewidth=0.5pt](-3,0){1.}{90}{270}
\psarc[linewidth=0.5pt](3,0){1.}{-90}{90}
\psarc[linewidth=0.5pt](0,3){1.}{0}{180}
\psarc[linewidth=0.5pt](0,-3){1.}{-180}{0}
\rput(-3,0){\makebox(0,0){\tiny $\bullet$}}
\rput(3,0){\makebox(0,0){\tiny $\bullet$}}
\rput(0,3){\makebox(0,0){\tiny $\bullet$}}
\rput(0,-3){\makebox(0,0){\tiny $\bullet$}}
\pcline[linewidth=0.5pt](-3,1)(-2,1)
\pcline[linewidth=0.5pt](-3,-1)(-2,-1)
\pcline[linewidth=0.5pt](3,1)(2,1)
\pcline[linewidth=0.5pt](3,-1)(2,-1)
\pcline[linewidth=0.5pt](-1,2)(-1,3)
\pcline[linewidth=0.5pt](1,2)(1,3)
\pcline[linewidth=0.5pt](-1,-2)(-1,-3)
\pcline[linewidth=0.5pt](1,-2)(1,-3)
}
}
\newcommand\IZVIV[1]{%
{\psset{unit=#1}
\psarc[linewidth=0.5pt](-0.75,1){1}{-90}{0}
\psarc[linewidth=0.5pt](1.25,-1){1}{90}{180}}
}
\newcommand{\Comm}[5]{%
{\psset{unit=#1}
\POND{#1}
\rput{#2}(-#1,#1){\IZVIV{#1}}
\rput{#3}(#1,#1){\IZVIV{#1}}
\rput{#4}(#1,-#1){\IZVIV{#1}}
\rput{#5}(-#1,-#1){\IZVIV{#1}}}
}
\rput(-5.8,7.5){\makebox(0,0){\small I}}
\rput(-4,6.7){\makebox(0,0){\small II}}
\rput(-5,6){\CROSS{0.35}}
\rput(-3,6){\makebox(0,0){\small$=q^2$}}
\rput(-1,6){\Comm{0.6}{0}{0}{0}{0}}
\rput(1,6){\makebox(0,0){\small$+\ q$}}
\rput(3,6){\Comm{0.6}{90}{0}{0}{0}}
\rput(5,6){\makebox(0,0){\small$+\ q$}}
\rput(7,6){\Comm{0.6}{0}{90}{0}{0}}
\rput(-6,3){\makebox(0,0){\small$+\ q$}}
\rput(-4,3){\Comm{0.6}{0}{0}{90}{0}}
\rput(-2,3){\makebox(0,0){\small$+\ q$}}
\rput(0,3){\Comm{0.6}{0}{0}{0}{90}}
\rput(2,3){\makebox(0,0){\small$+\ 1$}}
\rput(4,3){\Comm{0.6}{90}{90}{0}{0}}
\rput(6,3){\makebox(0,0){\small$+\ 1$}}
\rput(8,3){\Comm{0.6}{90}{0}{90}{0}}
\rput(-6.5,0){\makebox(0,0){\small$+\ 1$}}
\rput(-4.5,0){\Comm{0.6}{90}{0}{0}{90}}
\rput(-2.5,0){\makebox(0,0){\small$+\ 1$}}
\rput(-0.5,0){\Comm{0.6}{0}{90}{90}{0}}
\rput(1.5,0){\makebox(0,0){\small$+\ 1$}}
\rput(3.5,0){\Comm{0.6}{0}{90}{0}{90}}
\rput(5.5,0){\makebox(0,0){\small$+\ 1$}}
\rput(7.5,0){\Comm{0.6}{0}{0}{90}{90}}
\rput(-6,-3){\makebox(0,0){\small$+ q^{-1}$}}
\rput(-3.8,-3){\Comm{0.6}{90}{90}{90}{0}}
\rput(-1.7,-3){\makebox(0,0){\small$+ q^{-1}$}}
\rput(0.4,-3){\Comm{0.6}{90}{90}{0}{90}}
\rput(2.5,-3){\makebox(0,0){\small$+ q^{-1}$}}
\rput(4.6,-3){\Comm{0.6}{90}{0}{90}{90}}
\rput(6.7,-3){\makebox(0,0){\small$+ q^{-1}$}}
\rput(8.8,-3){\Comm{0.6}{0}{90}{90}{90}}
\rput(-6.5,-6){\makebox(0,0){\small$+ q^{-2}$}}
\rput(-4.3,-6){\Comm{0.6}{90}{90}{90}{90}}
\end{pspicture}
}
\caption{\small The quantum skein relation for the quantum geodesics functions $G^\hbar_{ij}$
(indicated by I) and $G^\hbar_{kl}$ (indicated by II) at $i<k<j<l$.
Multicurves containing components homeomorphic to passing around
the orbifold points vanish. Taking into account the symmetry
w.r.t. changing the order of $G_{ij}$ and $G_{kl}$ in the product, we find that only
the first (with $q^2$) and the last (with $q^{-2}$) terms contribute to the commutator.}
\label{fi:POND}
\end{figure}
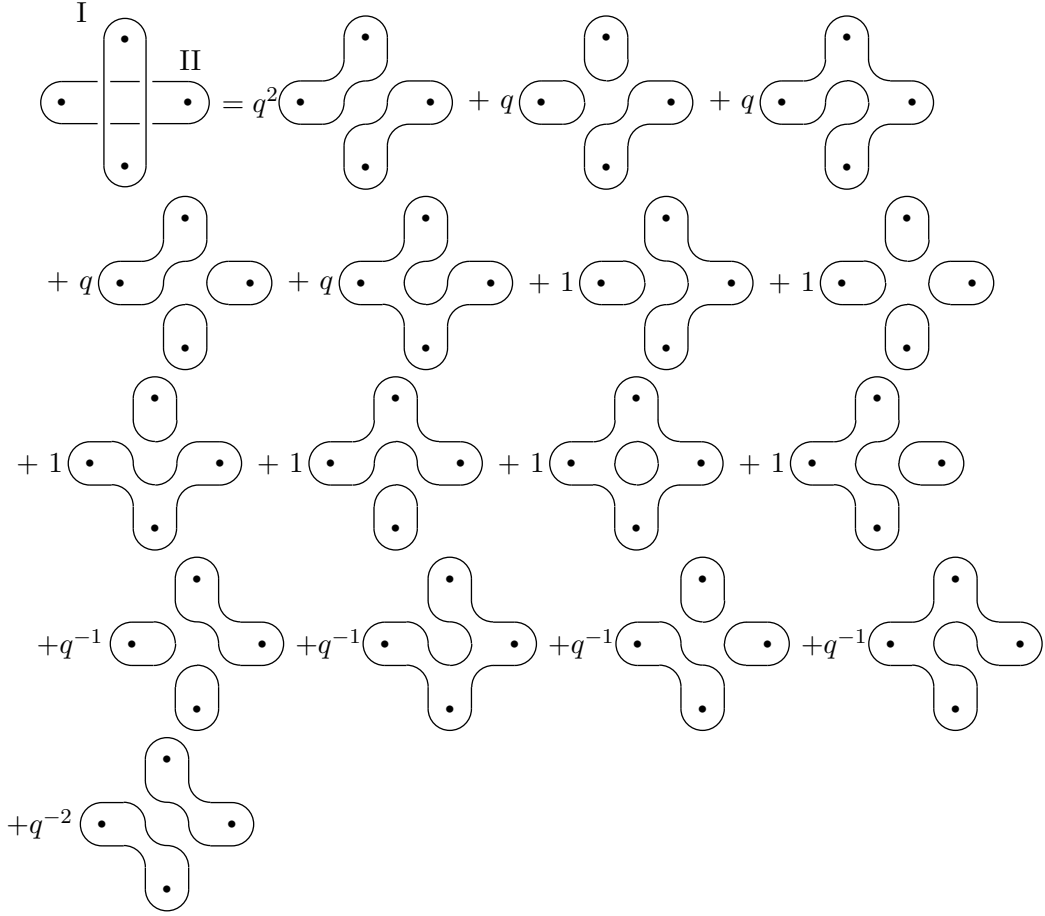

In Fig.~\ref{fi:POND}, we present the quantum skein relation for quadruple intersection of
geodesic functions (note that we must assign $-q-q^{-1}$ to every contractible loop).

\subsection{Quantum braid group relation}\label{sss:q-braid}

\subsubsection{Quantum $A_n$-algebra}\label{sss:q-braid-An}

We now consider the quantum geodesic functions associated with paths in the $A_n$-algebra
pattern in Fig.~\ref{fi:An}.

We first generalize Example~\ref{ex3} to the case of general $A_n$ algebras.
For the quantum geodesic functions $G^\hbar_{i,j}$ \ $(i<j)$ we have (assuming $j<i<l<k$)
\bea
&{}&[G^\hbar_{ik},G^\hbar_{jl}]=\xi\biggl(G^\hbar_{jk}G^\hbar_{il}-G^\hbar_{ji}G^\hbar_{lk}\biggr);
\nonumber
\\
&{}&qG^\hbar_{il}G^\hbar_{ji}-q^{-1}G^\hbar_{ji}G^\hbar_{il}=\xi G^\hbar_{jl};\qquad \xi=q^2-q^{-2}.
\label{An-q}
\\
&{}&qG^\hbar_{jl}G^\hbar_{il}-q^{-1}G^\hbar_{il}G^\hbar_{jl}=\xi G^\hbar_{ji};
\nonumber
\eea
and, apparently, quantum geodesic functions corresponding to nonintersecting geodesics commute.

From the quantum skein relation, it is easy to obtain quantum transformations for
$G^\hbar_{i,j}$. We introduce the ${\mathcal A}^\hbar$-matrix
\be
\label{A-matrix-q}
{\mathcal A}^\hbar=\left(\begin{array}{ccccc}
                     q^{-1} & G^\hbar_{1,2} & G^\hbar_{1,3} & \dots & G^\hbar_{1,n} \\
                     0 & q^{-1} & G^\hbar_{2,3} & \dots & G^\hbar_{2,n} \\
                     0 & 0 & q^{-1} & \ddots & \vdots \\
                     \vdots & \vdots & \ddots & \ddots & G^\hbar_{n-1,n} \\
                     0 & 0 & \dots & 0 & q^{-1} \\
                   \end{array}
\right)
\ee
associating the Hermitian operators $G^\hbar_{i,j}$ with the quantum
geodesic functions. Using the skein relation,
we can then present the action of the braid group element $R^\hbar_{i,i+1}$ exclusively in terms
of the geodesic functions from this, fixed set:
$R^\hbar_{i,i+1}{\mathcal A}^\hbar={\tilde{\mathcal A}^\hbar}$, where
\be
\label{R-matrix-q}
\begin{array}{ll}
  {\tilde G}^\hbar_{i+1,j}=G^\hbar_{i,j} & j>i+1,\\
  {\tilde G}^\hbar_{j,i+1}=G^\hbar_{j,i} & j<i, \\
  {\tilde G}^\hbar_{i,j}=q^{-1}G^\hbar_{i,j}G^\hbar_{i,i+1}-q^{-2}G^\hbar_{i+1,j}
  =qG^\hbar_{i,i+1}G^\hbar_{i,j}-q^{2}G^\hbar_{i+1,j} & j>i+1, \\
  {\tilde G}^\hbar_{j,i}=q^{-1}G^\hbar_{j,i}G^\hbar_{i,i+1}-q^{-2}G^\hbar_{j,i+1}
  =qG^\hbar_{i,i+1}G^\hbar_{j,i}-q^{2}G^\hbar_{j,i+1} & j<i, \\
  {\tilde G}^\hbar_{i,i+1}=G^\hbar_{i,i+1} &  \\
\end{array}%
 .
\ee
We can again present this transformation via the special matrices
$B^\hbar_{i,i+1}$ of the block-diagonal form
\be
\label{Bii+1-q}
B^\hbar_{i,i+1}=\begin{array}{c}
            \vdots \\
            i \\
            i+1 \\
            \vdots \\
          \end{array}
          \left(
          \begin{array}{cccccccc}
            1 &  &  &  &  &  &  &  \\
             & \ddots &  &  &  &  &  &  \\
             &  & 1 &  &  &  &  &  \\
             &  &  &   qG^\hbar_{i,i+1} & -q^{2} &  & &  \\
            &  &  & 1 & 0 &  &  &  \\
             &  &  &  &  & 1 &  &  \\
             &  &  &  &  &  & \ddots &  \\
             &  &  &  &  &  &  & 1 \\
          \end{array}
          \right).
\ee
Then, the action of the quantum braid group generator $R^\hbar_{i,i+1}$ on ${\mathcal A}^\hbar$
can be expressed as the matrix product (taking into account the noncommutativity of quantum matrix entries)
\be
\label{BAB-q}
R^\hbar_{i,i+1}{\mathcal A}^\hbar=B^\hbar_{i,i+1}{\mathcal A}^\hbar\bigl(B^\hbar_{i,i+1}\bigr)^{\dagger}
\ee
with $\bigl(B^{\hbar}_{i,i+1}\bigr)^{\dagger}$ the matrix Hermitian conjugate to
$B^\hbar_{i,i+1}$ (its nontrivial $2\times 2$-block has the form $\left(%
\begin{array}{cc}
  q^{-1}G^\hbar_{i,i+1} & 1 \\
  -q^{-2} & 0 \\
\end{array}%
\right)$).
Using the same technique as above, it is then straightforward to prove the following lemma.

\begin{lemma} \label{lem-q-braid}
For any $n\ge3$, we have the {\em quantum braid group relations}
\bea
\label{RRR-q}
&&R^\hbar_{i-1,i}R^\hbar_{i,i+1}R^\hbar_{i-1,i}=R^\hbar_{i,i+1}R^\hbar_{i-1,i}R^\hbar_{i,i+1},
\quad 2\le i\le n-1,
\\
\label{Rn-q}
&&\bigl(R^\hbar_{n-1,n}R^\hbar_{n-2,n-1}\cdots R^\hbar_{2,3}R^\hbar_{1,2}\bigr)^n=\hbox{Id}.
\eea
\end{lemma}

\subsubsection{Quantum $D_n$-algebra}\label{sss:q-Dn}

We now quantize the Poisson algebra of geodesic functions $G_{ij}$ corresponding to paths
as shown in Fig.~\ref{fi:Dn}. For each Poisson geodesic relation for
generators of the $D_n$ algebra in Sec.~\ref{s:braid}, we have the corresponding quantum
counterpart.

We assume the following cyclic ordering of indices in formulas below:
$$
{\psset{unit=0.4}
\begin{pspicture}(-7,-3)(7,3)
\newcommand{\ARC}{%
\psarc[linewidth=0.5pt]{->}(0,0){2}{30}{60}
}
\pscircle[fillstyle=solid, fillcolor=gray](0,0){0.5}
\rput(2,0){\makebox(0,0){$j$}}
\rput(0,2){\makebox(0,0){$i$}}
\rput(-2,0){\makebox(0,0){$l$}}
\rput(0,-2){\makebox(0,0){$k$}}
\rput(0,0){\ARC}
\rput{90}(0,0){\ARC}
\rput{180}(0,0){\ARC}
\rput{270}(0,0){\ARC}
\end{pspicture}
}
$$

The quantum permutation relations read ($q=e^{-i\pi\hbar}$, $\xi\equiv q^2-q^{-2}$)
\bea
&{}&\hbox{Case ${\mathbf a}$}\qquad
[G^\hbar_{ik},G^\hbar_{jl}]=\xi\biggl(G^\hbar_{jk}G^\hbar_{il}-G^\hbar_{ji}G^\hbar_{lk}\biggr);
\nonumber
\\
&{}&\hbox{Case ${\mathbf a}_1$}\qquad
qG^\hbar_{jl}G^\hbar_{kj}-q^{-1}G^\hbar_{kj}G^\hbar_{jl}=\xi G^\hbar_{kl};
\nonumber
\\
&{}&\hbox{Case ${\mathbf b}$}\qquad
[G^\hbar_{jl},G^\hbar_{ii}]=\xi\biggl(G^\hbar_{ji}G^\hbar_{ll}-G^\hbar_{il}G^\hbar_{jj}\biggr);
\nonumber
\\
&{}&\hbox{Cases ${\mathbf b}_{1,2}$}\ \
qG^\hbar_{jj}G^\hbar_{kj}-q^{-1}G^\hbar_{kj}G^\hbar_{jj}=\xi G^\hbar_{kk},\qquad
qG^\hbar_{jk}G^\hbar_{jj}-q^{-1}G^\hbar_{jj}G^\hbar_{jk}=\xi G^\hbar_{kk};
\nonumber
\\
&{}&\hbox{Case ${\mathbf c}$}\qquad
[G^\hbar_{ii},G^\hbar_{kk}]=(q-q^{-1})\bigl(G^\hbar_{ik}-G^\hbar_{ki}\bigr).
\label{Dn-q}
\\
&{}&\hbox{Case $\mathbf d$}\qquad [G^\hbar_{ij},G^\hbar_{kl}]=\xi\biggl(G^\hbar_{kj}G^\hbar_{li}
-G^\hbar_{jk}G^\hbar_{il}
+G^\hbar_{jl}G^\hbar_{ik}-G^\hbar_{lj}G^\hbar_{ki}\biggr.
\nonumber
\\
&{}&\qquad\qquad\qquad\qquad\qquad\qquad\qquad\qquad\biggl.
+(q+q^{-1})(G^\hbar_{il}G^\hbar_{jj}G^\hbar_{kk}-G^\hbar_{kj}G^\hbar_{ll}G^\hbar_{ii})
\biggr);
\nonumber
\\
&{}&\hbox{Case ${\mathbf d}_1$}\qquad qG^\hbar_{jl}G^\hbar_{ij}-q^{-1}G^\hbar_{ij}G^\hbar_{jl}=
\xi\biggl(q^{-1}G^\hbar_{lj}G^\hbar_{ji}+qG^\hbar_{ii}G^\hbar_{ll}+q^{-1}G^\hbar_{ll}G^\hbar_{ii}\biggr.
\nonumber
\\
&{}&\qquad\qquad\qquad\qquad\qquad\qquad\qquad\qquad
\biggl.-q^{-2}G^\hbar_{li}-G^\hbar_{il}(G^\hbar_{jj})^2\biggr);
\nonumber
\\
&{}&\hbox{Case ${\mathbf d}_2$}\qquad
[G^\hbar_{jl},G^\hbar_{lj}]=\xi\biggl((G^\hbar_{ll})^2-(G^\hbar_{jj})^2\biggr);
\nonumber
\eea

Although these relations contain not only triple terms in the r.h.s. but also noncommuting terms
(this is the price for closing the algebra), they nevertheless establish the lexicographic ordering
on the corresponding set of quantum variables $\{G^\hbar_{ij}\}$.

\begin{lemma}\label{lem-Dn-quantum}
Permutation relations postulated by (\ref{Dn-q}) satisfy the (quantum) Jacobi identities.
\end{lemma}

The proof is tedious but straightforward calculations. Note that
algebra (\ref{Dn-q}) is consistent even without relation to geometry
of modular spaces; the similar phenomenon was already observed in
the case of $A_n$-algebras.

We now provide the quantum version of the braid group transformations (\ref{Dn-braid-cl})
and (\ref{Dn-braid-cl-n1}). For $R_{i,i+1}^\hbar$ with $1\le i\le n-1$, we have
\be
\label{R-matrix-Dn-q}
\begin{array}{ll}
  {\tilde G}^\hbar_{i+1,k}=G^\hbar_{i,k} & k\ne i,i+1,\\
  {\tilde G}^\hbar_{i,k}=qG^\hbar_{i,i+1}G^\hbar_{i,k}-q^{2}G^\hbar_{i+1,k}
  =q^{-1}G^\hbar_{i,k}G^\hbar_{i,i+1}-q^{-2}G^\hbar_{i+1,k} & k\ne i,i+1, \\
  {\tilde G}^\hbar_{k,i+1}=G^\hbar_{k,i} & k\ne i,i+1,\\
  {\tilde G}^\hbar_{k,i}=qG^\hbar_{i,i+1}G^\hbar_{k,i}-q^{2}G^\hbar_{k,i+1}
  =q^{-1}G^\hbar_{k,i}G^\hbar_{i,i+1}-q^{-2}G^\hbar_{k,i+1} & k\ne i,i+1, \\
  {\tilde G}^\hbar_{i,i+1}=G^\hbar_{i,i+1}, &  \\
  {\tilde G}^\hbar_{i+1,i+1}=G^\hbar_{i,i}, &  \\
  {\tilde G}^\hbar_{i,i}=qG^\hbar_{i,i+1}G^\hbar_{i,i}-q^{2}G^\hbar_{i+1,i+1}
  =q^{-1}G^\hbar_{i,i}G^\hbar_{i,i+1}-q^{-2}G^\hbar_{i+1,i+1}, &  \\
  {\tilde G}^\hbar_{i+1,i}=G^\hbar_{i+1,i}+G^\hbar_{i,i}G^\hbar_{i,i+1}G^\hbar_{i,i}
  -q^{-1}G^\hbar_{i+1,i+1}G^\hbar_{i,i}-qG^\hbar_{i,i}G^\hbar_{i+1,i+1}, &  \\
\end{array}%
\ee
and for $R_{n,1}^\hbar$, we have
\be
\label{R-matrix-Dn-q-n1}
\begin{array}{ll}
  {\tilde G}^\hbar_{1,k}=G^\hbar_{n,k} & k\ne n,1,\\
  {\tilde G}^\hbar_{n,k}=qG^\hbar_{n,1}G^\hbar_{n,k}-q^{2}G^\hbar_{1,k}
  =q^{-1}G^\hbar_{n,k}G^\hbar_{n,1}-q^{-2}G^\hbar_{1,k} & k\ne n,1, \\
  {\tilde G}^\hbar_{k,1}=G^\hbar_{k,n} & k\ne n,1,\\
  {\tilde G}^\hbar_{k,n}=qG^\hbar_{n,1}G^\hbar_{k,n}-q^{2}G^\hbar_{k,1}
  =q^{-1}G^\hbar_{k,n}G^\hbar_{n,1}-q^{-2}G^\hbar_{k,1} & k\ne n,1, \\
  {\tilde G}^\hbar_{n,1}=G^\hbar_{n,1}, &  \\
  {\tilde G}^\hbar_{1,1}=G^\hbar_{n,n}, &  \\
  {\tilde G}^\hbar_{n,n}=qG^\hbar_{n,1}G^\hbar_{n,n}-q^{2}G^\hbar_{1,1}
  =q^{-1}G^\hbar_{n,n}G^\hbar_{n,1}-q^{-2}G^\hbar_{1,1}, &  \\
  {\tilde G}^\hbar_{1,n}=G^\hbar_{1,n}+G^\hbar_{n,n}G^\hbar_{n,1}G^\hbar_{n,n}
  -q^{-1}G^\hbar_{1,1}G^\hbar_{n,n}-qG^\hbar_{n,n}G^\hbar_{1,1}, &  \\
\end{array}%
\ee

\begin{lemma} \label{lem-q-braid-Dn}
For any $n\ge3$, we have the {\em quantum braid group relations}
\be
\label{RRR-Dn-q}
R^\hbar_{i-1,i}R^\hbar_{i,i+1}R^\hbar_{i-1,i}=R^\hbar_{i,i+1}R^\hbar_{i-1,i}R^\hbar_{i,i+1},
\quad i=1,\dots, n \ \mod n
\ee
for transformations (\ref{R-matrix-Dn-q}),(\ref{R-matrix-Dn-q-n1}) of quantum operators subject to
quantum algebra (\ref{Dn-q}).
\end{lemma}

Again, the second identity (\ref{Rn-q}) is lost in the case of $D_n$ algebras.

\subsubsection{Matrix representation for $D_n$-algebra and invariants}\label{sss:q-Dn-mat}

We now construct the quantum version of Theorem~\ref{th-braid}. For this, we first need a
preparatory lemma.

\begin{lemma} \label{lem-q-Dn-matrix}
The following four matrices with operatorial entries, together with all their linear combinations,
transform in accordance with the quantum braid-group action (\ref{BAB-q}): ${\mathcal A}^\hbar$
(\ref{A-matrix-q}), $\bigl({\mathcal A}^\hbar\bigr)^\dagger$, ${\mathcal R}^\hbar$, and ${\mathcal S}^\hbar$,
where
\bea\label{R.q}
({\mathcal R}^\hbar)_{i,j}&=&\left\{ \begin{array}{cc}
                        -G^\hbar_{j,i}-q^2G^\hbar_{i,j}+qG^\hbar_{i,i}G^\hbar_{j,j} &\quad j>i \\
                        G^\hbar_{i,j}+q^{-2}G^\hbar_{j,i}-q^{-1}G^\hbar_{i,i}G^\hbar_{j,j} &\quad  j<i \\
                               0 &\quad  j=i \\
                             \end{array}
                             \right.;\qquad
\bigl({\mathcal R}^\hbar\bigr)^\dagger=-{\mathcal R}^\hbar,
\\
\label{S.q}
({\mathcal S}^\hbar)_{i,j}&=&G^\hbar_{i,i}G^\hbar_{j,j}\quad \hbox{for all}\quad 1\le i,j\le n,
\qquad \bigl({\mathcal S}^\hbar\bigr)^\dagger={\mathcal S}^\hbar.
\eea
\end{lemma}

The quantum $(4n)\times(4n)$ matrix ${\mathbb B}^\hbar_{i,i+1}$ for $1\le i\le n-1$ has the
block-diagonal form (\ref{B-D-ii+1}) with diagonal entries being $n\times n$-matrices $B_{i,i+1}^\hbar$
(\ref{Bii+1-q}), whereas the remaining matrix ${\mathbb B}^\hbar_{n,1}(\lambda)$ reads
\be
\label{Bn1-q}
{\mathbb B}^\hbar_{n,1}(\lambda)=  \left(
          \begin{tabular}{ccc|ccc|ccc|ccc}
               \small $0$ &&& &&& &&&   && \hbox{\small \color{red}$\lambda^2$}  \\
               &\small $\!\!\! \mathcal I\!\!\!$&& &&& &&&  &&  \\
               &  & \small   $\!\!\! qG^\hbar_{1,n}$ & \small  $-q^{2}$ && &&&&  &&   \\
            \hline
              &  &\small  $1$ & \small  $0$ &  && &&&  &&    \\
               &&&  &\small $\!\!\! \mathcal I\!\!\!$&&  &&&  &&  \\
              &&&   &&\small  $\!\!\! qG^\hbar_{1,n}$  & \small  $-q^{2}$ &&& &&     \\
              \hline
              &  &  &   &  &\small $1$&\small $0$&&& &&    \\
               &&& &&& &\small $\!\!\! \mathcal I\!\!\!$&& &&  \\
              &&& &&&   &&\small  $\!\!\! qG^\hbar_{1,n}$  & \small  $-q^{2}$ &&     \\
              \hline
              &&& &&&    &&\small  $1$ & \small  $0$ &  &    \\
               &&& &&&  &&&&\small $\!\!\! \mathcal I\!\!\!$&  \\
               \hbox{\small \color{red}$-\lambda^{-2}$}\small $q^{2}\!\!\! \!\!\!\!\!\! $
               &  &  & &&& &&&  & &\small  $\!\!\! qG^\hbar_{1,n}  $   \\
          \end{tabular}
          \right),
\ee
and introducing the quantum matrix
\be
\label{mathbbA-q}
{\mathbb A}^\hbar=\left[\begin{array}{cccc}
            {\mathcal A}^\hbar& B^\hbar & C^\hbar & \bigl(B^\hbar\bigr)^\dagger \\
            0 & {\mathcal A}^\hbar& B^\hbar & C^\hbar  \\
            0 & 0 &{\mathcal A}^\hbar& B^\hbar \\
            0 & 0 & 0 & {\mathcal A}^\hbar
            \end{array}
            \right],
\quad\hbox{where}\ \left\{\begin{array}{l}
B^\hbar={\mathcal R}^\hbar+q^{-1}{\mathcal S}^\hbar+q^2{\mathcal A}^\hbar
-q^{-2}\bigl({\mathcal A}^\hbar\bigr)^\dagger,\\
C^\hbar=(q+q^{-1}){\mathcal S}^\hbar-{\mathcal A}^\hbar-\bigl({\mathcal A}^\hbar\bigr)^\dagger,
\end{array}
\right.,
\ee
we come to the main theorem about quantum braid-group representation.

\begin{theorem}\label{th-braid-q}
The quantum braid group relations  (\ref{R-matrix-Dn-q}) and (\ref{R-matrix-Dn-q-n1}) can be presented
in the form of matrix relations for the matrix
$\lambda{\mathbb A}^\hbar+\lambda^{-1}\bigl({\mathbb A}^\hbar\bigr)^\dagger$
with the matrix ${\mathbb A}^\hbar$ defined in (\ref{mathbbA-q}):
\bea
R^\hbar_{i,i+1}\Bigl(\lambda{\mathbb A}^\hbar+\lambda^{-1}\bigl({\mathbb A}^\hbar\bigr)^\dagger\Bigr)
&=&{\mathbb B}^\hbar_{i,i+1}
\Bigl(\lambda{\mathbb A}^\hbar+\lambda^{-1}\bigl({\mathbb A}^\hbar\bigr)^\dagger\Bigr)
\bigl({\mathbb B}^\hbar_{i,i+1}\bigr)^\dagger,
\quad i=1,\dots,n-1
\label{braid-D-1-q}
\\
R^\hbar_{n,1}\Bigl(\lambda{\mathbb A}^\hbar+\lambda^{-1}\bigl({\mathbb A}^\hbar\bigr)^\dagger\Bigr)
&=&{\mathbb B}^\hbar_{n,1}(\lambda)
\Bigl(\lambda{\mathbb A}^\hbar+\lambda^{-1}\bigl({\mathbb A}^\hbar\bigr)^\dagger\Bigr)
\bigl({\mathbb B}^\hbar_{n,1}(\lambda^{-1})\bigr)^\dagger,
\label{braid-D-2-q}
\eea
where ${\mathbb B}^\hbar_{i,i+1}$ has the form (\ref{B-D-ii+1}) with $B_{i,i+1}$
replaced by $B^\hbar_{i,i+1}$ from
(\ref{Bii+1-q}) and with ${\mathbb B}^\hbar_{n,1}(\lambda)$ of the form (\ref{Bn1-q}).
\end{theorem}

\begin{example}\label{Ex4}
{\rm
In the case $n=2$, the combination
$$
G^\hbar_{1,1}G^\hbar_{2,2}-qG^\hbar_{1,2}-q^{-1}G^\hbar_{2,1}=
G^\hbar_{2,2}G^\hbar_{1,1}-q^{-1}G^\hbar_{1,2}-qG^\hbar_{2,1}
$$
is a central element of the (quantum) algebra $D_2$; the other central element is
$$
G^\hbar_{1,2}G^\hbar_{2,1}-q^2(G^\hbar_{2,2})^2-q^{-2}(G^\hbar_{1,1})^2=
G^\hbar_{2,1}G^\hbar_{1,2}-q^{-2}(G^\hbar_{2,2})^2-q^2(G^\hbar_{1,1})^2.
$$
}
\end{example}

\begin{example}\label{Ex5}
{\rm
A cyclic permutation of indices $P:\,i\mapsto i+1 \mod n;\ j\mapsto j+1 \mod n$
destroys the structure of the matrix
${\mathcal A}^\hbar$ and results in the following transformations for ${\mathcal R}^\hbar$
and ${\mathcal S}^\hbar$:
\bea
\label{R.cycle}
P:\ {\mathcal R}^\hbar&\mapsto&
\left(%
\begin{array}{cccc}
  0 & 1 &  &  \\
   & \ddots & \ddots &  \\
   &  & \ddots & 1 \\
  -q^{-2} &  &  & 0 \\
\end{array}%
\right)
{\mathcal R}^\hbar
\left(%
\begin{array}{cccc}
  0 &  &  & -q^2 \\
  1 & \ddots & &  \\
   & \ddots & \ddots &  \\
   &  & 1 & 0 \\
\end{array}%
\right),
\\
\label{S.cycle}
P:\ {\mathcal S}^\hbar&\mapsto&
\left(%
\begin{array}{cccc}
  0 & 1 &  &  \\
   & \ddots & \ddots &  \\
   &  & \ddots & 1 \\
  1 &  &  & 0 \\
\end{array}%
\right)
{\mathcal S}^\hbar
\left(%
\begin{array}{cccc}
  0 &  &  & 1 \\
  1 & \ddots & &  \\
   & \ddots & \ddots &  \\
   &  & 1 & 0 \\
\end{array}%
\right).
\eea
These transformations together with (\ref{BAB-q}) must also
generate a full modular group. From this, we find that $\det {\mathcal R}$ must be itself
the mapping-class group invariant lying therefore in the
center of the Poisson algebra.
Same is true for ${\mathcal S}$, but $\det {\mathcal S}\equiv 0$ whereas
$\det {\mathcal R}$ is nonzero for even $n=2m$ (and vanishes for odd $n$):
denoting $Q_{i,j}:=({\mathcal R})_{i,j}$
for $i<j$, we have $\det {\mathcal R}=\hbox{Pf}_{2m}^2$,
where the Pfaffian $\hbox{Pf}_{2m}$
is given by the parity-signed sum over all possible pairings in the set of indices
$1,2,\dots,2m-1,2m$.

For example, for $m=2$, we have
$$
\hbox{Pf}_4=Q_{1,2}Q_{3,4}+Q_{1,4}Q_{2,3}-Q_{1,3}Q_{2,4}
$$
(recall that $Q_{i,j}=G_{i,j}+G_{j,i}-G_{i,i}G_{j,j}$ in the classical case). In the quantum case,
these elements acquire $q$-corrections.
}
\end{example}

\newsection*{Acknowledgments}

The author acknowledges useful discussions with V.~V.~Fock, S.~Fomin, M.~Mazzocco,
S.~N.~Natanzon, R.~C.~Penner, M.~Shapiro, and D.~Thurston.
The author thanks the referee for the useful remarks and for the
careful reading of the manuscript.

The work was partially financially supported by the Russian Foundation for Basic Research
(Grant Nos.~06-02-17383 and 09-01-92433-CE), Grants of Support for the Scientific
Schools 795.2008.1, by the Program Mathematical Methods of
Nonlinear Dynamics,
by the European Community through the FP6
Marie Curie RTN {\em ENIGMA} (Contract number MRTN-CT-2004-5652), and
by EPSRC Grant EP/D071895/1.

\end{document}